%% file: main.tex
\documentclass{article}

\input{includes}

\title{Quantum Correlations in Frustrated Three-Body Systems}
\newif\ifuniqueAffiliation

\ifuniqueAffiliation 
\author{Chaitali ~Shah  \\
	Centre for High Energy Physics\\
	Indian Institute of Science\\
	Bengaluru 560012\\
	\texttt{chaitalishah@iisc.ac.in} \\
	\And
	Apoorva ~Patel \\
	Centre for High Energy Physics\\
	Indian Institute of Science\\
	Bengaluru 560012\\
	\texttt{adpatel@iisc.ac.in} \\

}
\else
\usepackage{authblk}

\author[1,2,3]{%
Chaitali ~Shah %
}
\author[1]{%
Apoorva D.~Patel %
}
\affil[1]{Centre for High Energy Physics, Indian Institute of Science, Bengaluru 560012}
\affil[2]{Department of Instrumentation and Applied Physics, Indian Institute of Science, Bengaluru 560012}
\affil[3]{Department of Chemistry, University of California, Berkeley, Berkeley, CA 94720, USA}
\fi

\hypersetup{
pdftitle={Quantum Correlations in Frustrated Three-Body Systems},
pdfsubject={quant-ph},
pdfauthor={Chaitali ~Shah, Apoorva D.~Patel},
pdfkeywords={Quantum Correlations, Frustration, Three Body Problem, Entanglement, Variational Method},
}

\begin{document}
\maketitle

\input{abstract}
\input{keywords}
\input{introduction}
\input{helium}
\input{dihydrogencation}
\input{hbond}
\input{conclusion}
\input{acknowledgement}

\bibliographystyle{unsrt}
\bibliography{references}  





\newpage
\appendix
\input{appendix}
\input{appendix2}

\end{document}

%% file: includes.tex
\usepackage{arxiv}

\usepackage[utf8]{inputenc} 
\usepackage[T1]{fontenc}    
\usepackage{amssymb}
\usepackage{amsmath}
\usepackage{hyperref}       
\usepackage{url}            
\usepackage{booktabs}       
\usepackage{amsfonts}       
\usepackage{nicefrac}       
\usepackage{microtype}      
\usepackage{cleveref}       
\usepackage{lipsum}         
\usepackage{graphicx}
\usepackage{doi}
\usepackage{latexsym}
\usepackage{multicol}
\usepackage{multirow}
\usepackage{booktabs}
\usepackage{multirow}
\usepackage{fancyhdr}
\usepackage[numbers]{natbib}
\usepackage{wrapfig}
\usepackage{algorithm}
\usepackage{algorithmic}
\usepackage{enumerate}
\usepackage{cancel}
\usepackage{mathrsfs}
\usepackage{color}
\usepackage[compact]{titlesec}
\usepackage{mdwlist}
\usepackage{tikz}
\usepackage{varwidth}
\usepackage{datetime}
\usepackage{pdfpages}
\usepackage{ragged2e}
\usetikzlibrary{shapes,arrows, trees}
\usepackage{tcolorbox}
\usepackage{listings}
\usepackage{float}
\usepackage{caption}
\usepackage{subcaption}
\usepackage{braket}
\usepackage{array}
\usepackage{tabularx}
\usepackage{svg}
\usepackage{wrapfig}
\usepackage{longtable}
\usepackage{mathabx}

\newcolumntype{L}[1]{>{\raggedright\let\newline\\\arraybackslash\hspace{0pt}}m{#1}}
\newcolumntype{C}[1]{>{\centering\let\newline\\\arraybackslash\hspace{0pt}}m{#1}}
\newcolumntype{R}[1]{>{\raggedleft\let\newline\\\arraybackslash\hspace{0pt}}m{#1}}
\newcolumntype{M}[1]{>{\centering\arraybackslash}m{#1}}

\newcommand{\dilog}{\Lambda_2}
\newcommand{\mc}[1]{\mathcal{#1}}
\newcommand{\mr}[1]{\mathrm{#1}}
\newcommand{\til}[1]{\widetilde{#1}}
\newcommand{\al}{\alpha}
\newcommand{\bkt}[2]{\braket{#1}_{#2}}
\newcommand{\rab}{r_{12}}
\newcommand{\ra}{r_{1}}
\newcommand{\rb}{r_{2}}

\newcommand{\cst}{\cos(\theta)}
\newcommand{\snt}{\sin(\theta)}
\newcommand{\pdv}[2]{\frac{\partial{#1}}{\partial{#2}}}
\renewcommand{\arraystretch}{1.8}

%% file: abstract.tex
\begin{abstract}
	Many physical systems are not understood from first principles due to the presence of multi-body quantum correlations. The effort taken to simulate such systems on classical computers increases exponentially with increase in the system size. We study simple, yet non-trivial, three-body frustrated systems that can help in understanding the underlying quantum correlations. First we investigate the ground state of helium-like atoms using the variational method and physically meaningful ans\"atze, revealing how quantum entanglement arises due to frustration in the system. Next we consider another frustrated three-body system, the hydrogen molecular ion, and analytically demonstrate the nature of its wavefunction arising from the quantum tunneling phenomenon. Finally, we extend these results to model yet another physically important system, the hydrogen bond.
\end{abstract}

%% file: keywords.tex
\keywords{Quantum Correlations \and Frustration \and Three Body Problem \and Entanglement\and Variational Method}

%% file: introduction.tex
\section{Introduction}
Many theoretically interesting and practically important physical systems have not been understood in their entirety despite many advances in the field of physics, because they have not been solved in closed form. Usually, such systems are studied using simulations and numerical computations that model them. Attempts that model real physical systems classically, fail to explain certain unique features about them. The deviations in the classical simulation results from the experimental data are labeled `anomalies'. They are really not `anomalies' but quantum mechanical features making their presence felt. Many physical systems have multi-body quantum correlations, and their modeling is essential to understand the observed properties of the systems. In this work, we study three three-body systems and the nature of the quantum correlations that arise between their constituents under frustrating interactions. 

Helium-like atoms are the simplest multi-electron systems that can be used to understand the concept of quantum entanglement. We investigate the ground state wavefunction of helium-like atoms using the variational method. The variational method is an extremely useful technique in quantum mechanics, employed to find the ground state energy of a system. It forms the basis for powerful quantum algorithms like the Variational Quantum Eigensolver, which can potentially solve many practical problems in quantum chemistry and biology \cite{nisq,vqa,vqe}. Helium-like atoms consist of a nucleus with two electrons. There is frustration in the system because there are two attractive Coulomb forces (nucleus-electron) and one repulsive force (electron-electron). Numerous calculations have obtained the ground state energy to sufficient accuracy using multi-parameter variational trial wavefunctions \cite{hyl1,hyl2,pek,henrich,schwartz}. This study aims to obtain a better understanding of the ground state structure in terms of quantum entanglement using a small number of variational parameters.  

The hydrogen molecular ion H$_2^+$ is another fascinating three-body system with frustration. It is the most elementary molecule, but the closed form solution for its energy eigenstates and eigenvalues has not yet been discovered. It is also the simplest system that can be studied to understand the nature of a chemical bond \cite{genesis}. The Schrodinger equation for this system is separable \cite{arfken}, and there has been a lot of work to find its series solutions using various expansions \cite{burrau, hylleraas, wilson, bates, aubert}. A more general approach to find its ground state energy is to use the variational method \cite{pauling}, with the ground state wavefunction being chosen as a linear combination of atomic orbitals. Our work focuses on determining the ground state wave function \cite{mitin,grivet} that describes the 3-dimensional probability distribution of the electron in the bond. The detailed understanding of this system can be extended to the concept of tunnelling and the formation of bonds in more complex systems. 

The third system we consider is the hydrogen bond between two oxygen atoms. It has structural similarity with the H$_2^+$ system, with two attractive (proton-oxygen) and one repulsive (oxygen-oxygen) interactions. So the ground state wavefunction for H$_2^+$ can be used to model the wavefunction for the hydrogen bond. Such a first-principle attempt to model the hydrogen bond has limitations, because the oxygen atoms in the hydrogen bond are not point objects and have filled core orbitals. Still, we can model the experimental data reasonably, and we obtain a sensible wavefunction for the hydrogen bond with a small number of phenomenological parameters.

%% file: helium.tex
\section{Helium-like Atoms}

The helium atom is the simplest multi-electron system and consists of two electrons bound to a nucleus. The closed form solution to its Schr{\"o}dinger equation has not yet been discovered, while various approximations have been used to estimate its ground state energy and wavefunction.

The Schr{\"o}dinger equation for helium-like atoms is
\begin{equation} 
\mc{H} \psi(r_1,r_2) = E \psi(r_1,r_2) ~,
\end{equation}
with the Hamiltonian $\mc{H}$ given by
\begin{equation}
\mc{H} = -\frac{\hbar^2}{2m}\nabla_{r_1}^2-\frac{\hbar^2}{2m}\nabla_{r_2}^2-\frac{Ze^2}{r_1}-\frac{Ze^2}{r_2} + \frac{e^2}{r_{12}} ~.
\label{H_helium}
\end{equation}
Here $m$ is the mass of the electron, $Z$ is the atomic number, $\vec{r}_{1}$ and $\vec{r}_{2}$  are the distance vectors between the electrons and the nucleus, and $r_{12}=|\vec{r}_1-\vec{r}_2|$.
We have assumed the nucleus to be infinitely heavy, and ignored the magnetic moments of the electrons and the nucleus. Tiny corrections due to finite nuclear mass and interaction between magnetic moments of the electrons and the nucleus can be estimated perturbatively, when comparing the eigenvalues of the above Hamiltonian with experimental data.

Choosing atomic units (a.u.), where the length scale $\lambda=a_0$ is the Bohr radius and the energy scale $\epsilon=E_h$ is the Hartree, the Hamiltonian becomes
\begin{equation}
\mc{H} = -\frac{1}{2}\nabla_{r_1}^2-\frac{1}{2}\nabla_{r_2}^2-\frac{Z}{r_1}-\frac{Z}{r_2}+\frac{1}{r_{12}} ~.
\label{H_helium_au}
\end{equation}
The first two terms are the kinetic energies of the two electrons, and the other terms comprise the potential energy from Coulomb interactions between the three particles. The electron-electron repulsion term (the last term in the Hamiltonian) prevents a solution to this eigenvalue equation by separation of variables. Hence, we resort to approximation techniques to obtain the ground state energy of the system.

\subsection{The Variational Method}

The Variational Method is extremely useful in finding approximate estimates of the ground state energy of quantum systems.
The Rayleigh-Ritz variational principle provides an upper bound to the ground state energy \cite{griffiths}, i.e. for any wavefunction $|\psi\rangle$, 
\begin{equation}
\braket{\mc{H}} \equiv \frac{\braket{\psi|\mc{H}|\psi}}{\braket{\psi|\psi}} \geq E_0 ~,
\label{var_upper}
\end{equation}
where $E_0$ is the ground state energy of the system and $braket{\mc{H}}$ is the expectation value of the Hamiltonian.

A lower bound on the ground state energy can be obtained in terms of the expectation value and variance of the Hamiltonian $\mc{H}$ \cite{mathews}:
\begin{equation}
\braket{\mc{H}} - \braket{\mc{H}^2-\braket{\mc{H}}^2}^{1/2} \leq E_0 ~,
\label{var_lower}
\end{equation}
provided that $\braket{\mc{H}}$ is closer to $E_0$ than to any other eigenvalue. It can be noted that when $|\psi\rangle$ approximates the ground state wavefunction to $O(\epsilon)$, the upper bound is accurate to $O(\epsilon^2)$, while the lower bound is accurate to $O(\epsilon)$. So the size of the variance of the Hamiltonian gives an estimate of how accurately $|\psi\rangle$ approximates the ground state wavefunction.

In the variational method, the left-hand sides of the inequalities \eqref{var_upper} and \eqref{var_lower} are evaluated using a trial wavefunction, a \emph{variational ansatz} $|\tilde{\psi}\rangle$ that depends on a number of parameters. Both inequalities reduce to equalities when the chosen ansatz is the actual ground state. The bounds on the ground state energy are then obtained by varying the parameters, until the upper bound is minimized, and the lower bound is maximized (both these extrema need not occur for the same parameter values).

\subsection{The Variational Ans\"atze}

We now apply the variational method using various ans\"atze to estimate the ground state energies of two-electron atoms with $Z=1,2,3,4$. The two electrons are in an antisymmetric spin-singlet in the ground state. So their spatial wavefunction must be exchange-symmetric, as dictated by the statistics for identical fermions. We use four few-parameter ans\"atze to understand the contributions of various wavefunction properties to the ground state energy.

\begin{enumerate}
\item \textbf{Ansatz 1:  $|\tilde{\psi}_1\rangle$}

If the electron-electron interaction term is omitted from the Hamiltonian in \eqref{H_helium_au}, then it becomes the sum of two hydrogen-like Hamiltonians. Then the ground state wavefunction is the product of two hydrogenic wavefunctions (in a.u.) \cite{griffiths}:
\begin{equation}
    \psi(r_1,r_2) = \psi_g(r_1) \times \psi_g(r_2) = \frac{Z^3}{\pi} e^{-Z(r_1+r_2)} ~.
    \label{hel_wf}
\end{equation}

Hence, our first choice for the variational ansatz is:
\begin{equation}
    \tilde{\psi}_1(r_1,r_2) = e^{-\alpha(r_1+r_2)} ~,
    \label{ansatz-1}
\end{equation}
where $\alpha$ is the variational parameter.

\item \textbf{Ansatz 2: $|\tilde{\psi}_2\rangle$}

We improve the ansatz $\tilde{\psi}_1$ by adding an extra parameter that allows the two electrons to be at different distances from the nucleus. The spatial symmetry of the wavefunction requires it to have the form of a $2\times2$ \textit{Permanent} (in contrast to the \textit{Slater determinant} form of the spin-singlet wavefunction).

Our unnormalised variational wavefunction ansatz is therefore \cite{chandrasekhar}:
\begin{equation}
    \tilde{\psi}_2(r_1,r_2)= e^{-(\alpha r_1 + \beta r_2)} + e^{-(\alpha r_2 + \beta r_1)} ~, 
    \label{ansatz-2}
\end{equation}
with $\alpha$ and $\beta$ as the variational parameters.

\item \textbf{Ansatz 3: $|\tilde{\psi}_3\rangle$}

The previous variational ansatzes do not account the electron-electron Coulomb repulsion. In an attempt to understand the contribution of this repulsive interaction, we modify $\tilde{\psi}_1$ to:
\begin{equation}
     \tilde{\psi}_3(r_1,r_2) = e^{-\alpha(r_1+r_2)} (1+c r_{12}) ~,
     \label{ansatz-3}
\end{equation}  
where $\alpha$ and $c$ are the variational parameters, and $1+c r_{12}$ can be considered an approximation to $e^{c r_{12}}$.

\item \textbf{Ansatz 4: $|\tilde{\psi}_4\rangle$}
A still better approximation to the ground state wavefunction, which takes into account both the permanent structure and the repulsive interaction, was used in \cite{chandrasekhar}:
\begin{equation}
    \tilde{\psi}_4(r_1,r_2) = ( e^{-(\alpha r_1 + \beta r_2)} + e^{-(\alpha r_2 + \beta r_1)} )(1+c r_{12}) ~.
    \label{ansatz-4}
\end{equation}
Note that Ans\"atze 1-3 can be considered as special cases of Ansatz 4 with particular choices for $\beta$ and $c$.

\end{enumerate}

\subsection{Results: Upper and Lower Bounds}

High accuracy estimates of the ground state energies, as well as their experimental values, available in the literature for atomic numbers Z=1 to 4, are listed in Table \ref{tab:energy_ref}. The details of our variational calculation are provided in the Supplementary Information.

\begin{table}[htbp]
\renewcommand{\arraystretch}{1}
 \caption{The ground state energy values for atomic numbers Z=1 to 4 (rounded to 6 decimal places), from high accuracy solutions to Eq.\eqref{H_helium_au} \cite{osti}, and experimental data \cite{nist} , as per available literature.}
    \centering
    \begin{tabular}{|c|c|c|}
    \hline 
        Atomic number Z & $E_0$ from solving Eq.\eqref{H_helium_au} (in a.u.) & $E_0$ from experimental data \\ 
        \hline 
         1 & -0.527751  & -0.527716  \\
         2 & -2.903724 & -2.903385 \\ 
         3 & -7.279913 & -7.279833   \\ 
         4 & -13.655 566 & -13.656585   \\ 
         \hline
    \end{tabular}  
    \label{tab:energy_ref}
\end{table} 

\begin{table}[htbp]
\renewcommand{\arraystretch}{1.2}
    \centering
    \caption{The ground state energy bounds $\widehat{E}_0$ (rounded to 6 decimal places) and the variational parameters (rounded to 4 decimal places) for ans\"atze $\til{\psi}_1$,$\til{\psi}_2$,$\til{\psi}_3$ and $\til{\psi}_4$.}
    \begin{tabular}{|M{6em}|M{3em}|M{3em}|M{5em}|M{5em}|M{5em}|M{5em}|}
    \hline
       \multicolumn{3}{|c|}{Atomic Number Z} & 1 & 2 & 3 & 4 \\
       \hline
       \multirow{13}{4em}{Upper Bounds (Obtained using  \eqref{var_upper})}  & \multirow{2}{3em}{$\til{\psi}_1$} & $\widehat{\alpha}$ & 0.6875 & 1.6875 & 2.6875 & 3.687 \\ 
       \cline{3-7}
       &  & {$\widehat{E}_0$} & -0.472656 & -2.847656 & -7.222656 & -13.597656 \\ 
       \cline{2-7}
        & \multirow{3}{3em}{$\til{\psi}_2$} & $\widehat{\alpha}$ & 1.0392 & 2.1832 & 3.2949 & 4.3897  \\
          \cline{3-7}
          &  & {$\widehat{\beta}$} & 0.2832 & 1.1885 & 2.0790 & 2.9847 \\
            \cline{3-7}
        &  & {$\widehat{E}_0$} & -0.513303 & -2.875661 & -7.248748 & -13.633965  \\
          \cline{2-7}
          & \multirow{4}{3em}{$\til{\psi}_3$} & $\widehat{\alpha}$ & 0.8257 & 1.8497 & 2.8564 & 3.8592  \\
          \cline{3-7}
            &  & {$\widehat{c}$} & 0.4933 & 0.3689 & 0.3354 & 0.3212 \\
            \cline{3-7}
        &  & {$\widehat{E}_0$} & -0.508780 & -2.891121 & -7.268157 & -13.644052 \\
          \cline{2-7}
          & \multirow{4}{3em}{$\til{\psi}_4$} & $\widehat{\alpha}$ & 1.0749 & 2.2084 & 3.2994 & 4.3744  \\
          \cline{3-7}
          &  & {$\widehat{\beta}$} & 0.4774 & 1.4362 & 2.3618 & 3.2934  \\
            \cline{3-7}
            &  & {$\widehat{c}$} & 0.3125 & 0.2927 & 0.2770 & 0.2688  \\
            \cline{3-7}
        &  & {$\widehat{E}_0$} & -0.525919 & -2.901420 & -7.277174 & -13.652545  \\
          \hline
       \multirow{13}{4em}{Lower Bounds (Obtained using  \eqref{var_lower})}  & \multirow{2}{3em}{$\til{\psi}_1$} & $\widecheck{\alpha}$ & - & 1.8529 & 2.8624 & 3.8662 \\
       \cline{3-7}
       &  & {$\widecheck{E}_0$} & - & -3.738875 & -8.609198 & -15.479185 \\  
       \cline{2-7}
        & \multirow{3}{3em}{$\til{\psi}_2$} & $\widecheck{\alpha}$ & - & - & - & 4.4772   \\
          \cline{3-7}
          &  & {$\widecheck{\beta}$} & - & -  & - & 2.8631  \\
            \cline{3-7}
        &  & {$\widecheck{E}_0$} & - & - & - & -15.221156  \\
          \cline{2-7}
          & \multirow{4}{3em}{$\til{\psi}_3$} & $\widecheck{\alpha}$ & 0.8800 & 1.8986 & 2.9028 & 3.9046  \\
          \cline{3-7}
            &  & {$\widecheck{c}$} & 0.2040 & 0.1560 & 0.1442 & 0.1390  \\
            \cline{3-7}
        &  & {$\widecheck{E}_0$} & -0.7644 & -3.5543 & -8.3425 & -15.1301  \\
          \cline{2-7}
          & \multirow{4}{3em}{$\til{\psi}_4$} & $\widecheck{\alpha}$ & - & - & 3.3453 & 4.43119 \\
          \cline{3-7}
          &  & {$\widecheck{\beta}$} & - & - & 2.1912 & 3.1494 \\
          \cline{3-7}
          &  & {$\widecheck{c}$} & - & - & 0.1626 & 0.1604 \\
            \cline{3-7}
        &  & {$\widecheck{E}_0$} & - & - & -8.135617 & - 14.860020  \\
          \hline
        \multirow{13}{4em}{Lower Bounds (Obtained using upper bound parameters in \eqref{var_lower})}  & \multirow{2}{3em}{$\til{\psi}_1$} & $\widehat{\alpha}$ & 0.6875 & 1.6875 & 2.6875 & 3.687 \\ 
       \cline{3-7}
       &  & {$\breve{E}_0$} & -0.858578 & -3.794920 & -8.731261 & -15.667603 \\ 
       \cline{2-7}
        & \multirow{3}{3em}{$\til{\psi}_2$} & $\widehat{\alpha}$ & 1.0392 & 2.1832 & 3.2949 & 4.3897  \\
          \cline{3-7}
          &  & {$\widehat{\beta}$} & 0.2832 & 1.1885 & 2.0790 & 2.9847 \\
            \cline{3-7}
        &  & {$\breve{E}_0$} & -0.716955 & -3.582992 & -8.414869 & -15.244669  \\
          \cline{2-7}
          & \multirow{4}{3em}{$\til{\psi}_3$} & $\widehat{\alpha}$ & 0.8257 & 1.8497 & 2.8564 & 3.8592  \\
          \cline{3-7}
            &  & {$\widehat{c}$} & 0.4933 & 0.3689 & 0.3354 & 0.3212 \\
            \cline{3-7}
        &  & {$\breve{E}_0$} & -0.810935 & -3.743793 & -8.676828 & -15.606900 \\
          \cline{2-7}
          & \multirow{4}{3em}{$\til{\psi}_4$} & $\widehat{\alpha}$ & 1.0749 & 2.2084 & 3.2994 & 4.3744  \\
          \cline{3-7}
          &  & {$\widehat{\beta}$} & 0.4774 & 1.4362 & 2.3618 & 3.2934  \\
            \cline{3-7}
            &  & {$\widehat{c}$} & 0.3125 & 0.2927 & 0.2770 & 0.2688  \\
            \cline{3-7}
        &  & {$\breve{E}_0$} & -0.699406 & -3.4805769 & -8.248090 & -15.012942  \\
          \hline
          
    \end{tabular}
    \label{table:bounds}
\end{table}

Our results for the upper bounds, variances and lower bounds for all four ans\"atze are summarised in Table \ref{table:bounds}. The upper bounds, denoted by $\widehat{E}_0$ are computed by minimising \eqref{var_upper}, and the corresponding variational parameters are denoted by the `\ $\widehat{} \ $' symbol over them. The lower bounds, denoted by $\widecheck{E}_0$ are computed by minimising \eqref{var_lower}, and the corresponding variational parameters are denoted by the `\ $\widecheck{} \ $' symbol over them. In the lower bound calculations, there are instances where we could not maximize the energy (these cases are marked by `-' in the table). In these instances, the lower bounds are evaluated using the less optimal variational parameters obtained from the corresponding upper bound minimisations.

On comparing Tables \ref{table:bounds} and \ref{tab:energy_ref} the following inferences can be made:
\begin{itemize}
    \item The upper bounds of Eq.\eqref{var_upper} are closer to the exact ground state energy than the lower bounds of Eq.\eqref{var_lower}, as expected. Also, the lower bound function fails to produce a proper value for smaller Z.
    \item Ansatz 4 estimates the ground state energy within fractional accuracy of 0.0035 for Z=1, and the fractional accuracy improves with increasing Z. It is therefore a good variational parametrisation for ground state energy estimation.
    \item Even with Ansatz 4, the lower bound has fractional accuracy of 0.09 for Z=4, and the fractional accuracy gets worse with decreasing Z. It means that there is enough room for improving Ansatz 4, for ground state wavefunction determination.
    \item Looking at how the results for Ansatz 1 progress towards those of Ansatz 4, the permanent symmetrization and the Coulomb repulsion have comparable contribution to the ground state energy. Still, the permanent symmetrization is more important for Z=1, while the Coulomb repulsion is more important for larger Z.
    \item All ground state energy estimates improve with increasing Z, with the results for Anzatz 1 and Ansatz 4 moving closer. It implies that the importance of permanent symmetrization and Coulomb repulsion goes down with increasing Z.
    \end{itemize}

\subsection{Frustration and Broken Symmetry}
\label{sec_frust}

Next, we attempt to understand the origin of the permanent structure in the Ansatz 2, and the conditions under which breaking the symmetry ($\alpha \neq \beta$) lowers the ground state energy. To do this, we vary the strength of the Coulomb repulsion term in the Hamiltonian \eqref{H_helium_au} by introducing a parameter $\lambda$:
\begin{equation}
\mc{H}_\lambda = -\frac{1}{2}\nabla_{r_1}^2-\frac{1}{2}\nabla_{r_2}^2-\frac{Z}{r_1}-\frac{Z}{r_2}+\frac{\lambda}{r_{12}}
\label{H_hel_au}
\end{equation}

\begin{itemize}
    \item For $\lambda > 0$, there are two attractive forces (between the electrons and the nucleus), and one repulsive force (between the two electrons). This is a frustrated system, with $\lambda = 1$ corresponding to the Hamiltonian $\mc{H}$.
    \item When $\lambda < 0$, the electron-electron interaction turns attractive. Then there are three attractive forces and no frustration.
    \item When $\lambda = 0$, there is no interaction between the electrons and the ground state wavefunction is exactly given by Eq.\eqref{hel_wf}, which is the product of two hydrogenic wavefunctions. 
\end{itemize}

\begin{figure}[htbp]
    \centering
    \includegraphics[width=0.7\textwidth]{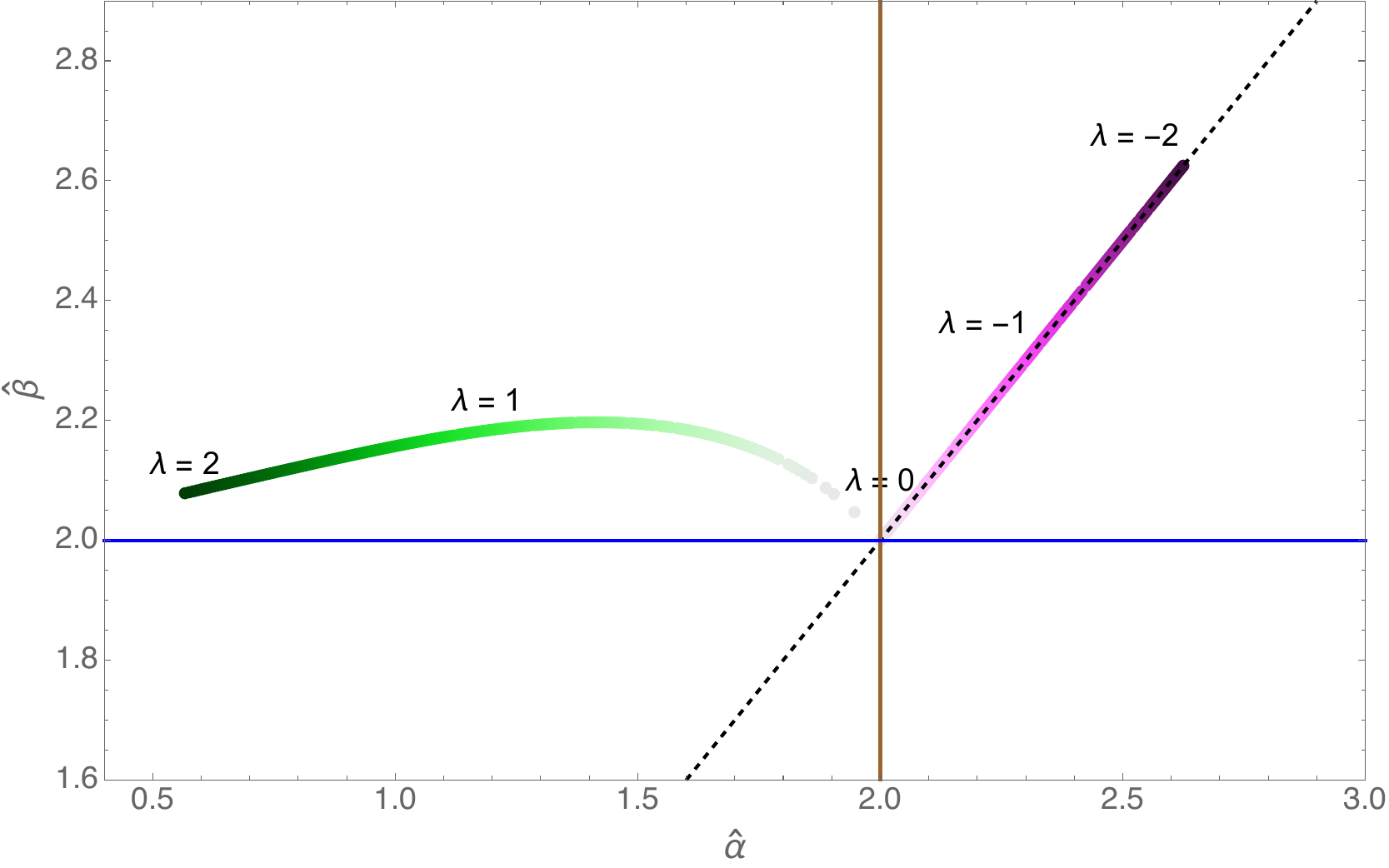}
    \caption{Locus of the point ($\widehat{\alpha}$,$\widehat{\beta}$) for $-2 \leq \lambda \leq 2$ and Z=2}
    \label{locus}
\end{figure}

We perform the variational calculations for $\mc{H}_\lambda$ using Ansatz 2, and find the expectation value of the Hamiltonian to be:
\begin{align}
 \widehat{E}_0 & = (  8(16\alpha\beta - 16Z(\alpha+\beta)+5 \alpha \lambda +5 \beta \lambda) + \frac{1}{\alpha^3\beta^3}(\alpha+\beta)^3(\alpha^5+3\alpha^4\beta+\beta^5\nonumber \\
 &-2Z(\alpha+\beta)^4+2\alpha^3\beta(2\beta+\lambda)+\alpha\beta^3(3\beta+2\lambda)+2\alpha^2\beta^2(2\beta+3\lambda))) \nonumber \\
 & \times \left(\frac{1 }{16(\alpha+\beta)^6\frac{1}{8\alpha^3\beta^3} + \frac{8}{(\alpha+\beta)^6}}\right)
 \label{lambdaexp}
\end{align} 
We then minimize this expression to obtain the ground state energy $\widehat{E}_0$ for Z=2 at $\widehat{\alpha}$ and $\widehat{\beta}$, for $-2 \leq \lambda \leq 2$. The locus of the point ($\widehat{\alpha}$,$\widehat{\beta}$) is plotted in the $\alpha-\beta$ plane in Figure \ref{locus}. The points go from green to pink as $\lambda$ increases from -2 to 2.

We observe that
\begin{itemize}
    \item For $-2 \leq \lambda \leq 0$, the points ($\widehat{\alpha}$, $\widehat{\beta}$) lie on the $\widehat{\alpha}=\widehat{\beta}$ line. This is a symmetric solution, and the permanent does not appear in the wavefunction for the non-frustrated system.
    \item For $0 < \lambda \leq 2$, the points ($\widehat{\alpha}$, $\widehat{\beta}$) deviate from the $\widehat{\alpha}=\widehat{\beta}$ line. For such a frustrated system, breaking the symmetry lowers the energy and the wavefunction is a permanent. In particular, $\alpha \neq \beta$ denotes an \textit{entangled quantum state}. 
\end{itemize}

\subsection*{2.5 Entanglement Structure of Ansatz 2}

Both the symmetric permanent and the antisymmetric determinant describe correlations of entangled
quantum states, though with different complexity class labels. For the electronic ground state of
Helium-like atoms, the spatial wavefunction is described by a permanent, while the spin
wavefunction is described by a determinant. With our parametrisation, the entanglement of the
spatial wavefunction depends on $\alpha\ne\beta$, while the spin-singlet wavefunction is maximally
entangled.

The two-electron entanglement entropy for the Ansatz 2 wavefunction is easily calculated in closed form,
by constructing the one-electron reduced density matrix. The unnormalised spatial wavefunction is
\begin{equation}
\tilde\psi_{2}(r_1,r_2)
= e^{-\alpha r_1 - \beta r_2}
+ e^{-\alpha r_2 - \beta r_1} ~,
\end{equation}
which can be written in terms of normalised one-electron orbitals
\begin{equation}
\phi_{\alpha}(\mathbf r)
= \left(\frac{\alpha^3}{\pi}\right)^{1/2} e^{-\alpha r} ~,
\qquad
\phi_{\beta}(\mathbf r)
= \left(\frac{\beta^3}{\pi}\right)^{1/2} e^{-\beta r} ~,
\end{equation}
with overlap
\begin{equation}
S \equiv \langle \phi_\alpha | \phi_\beta \rangle
= 8\,\frac{(\alpha\beta)^{3/2}}{(\alpha+\beta)^3} ~.
\label{eq:overlaps}
\end{equation}
The normalised two-electron spatial wavefunction is then
\begin{equation}
\Psi_{2}(\mathbf r_1,\mathbf r_2)
= \mathcal{N}\,\bigl(
\phi_{\alpha}(r_1)\phi_{\beta}(r_2)
+
\phi_{\beta}(r_1)\phi_{\alpha}(r_2)
\bigr) ~,
\qquad
\mathcal{N} = \frac{1}{\sqrt{2(1+S^2)}} ~.
\end{equation}

To obtain the entanglement between the two electrons, we perform the Schmidt
decomposition of $\Psi_2$.  Introducing the orthonormal combinations
\begin{equation}
|\varphi_{+}\rangle
= \frac{| \phi_\alpha\rangle + | \phi_\beta\rangle}{\sqrt{2(1+S)}} ~,
\qquad
|\varphi_{-}\rangle
= \frac{| \phi_\alpha\rangle - | \phi_\beta\rangle}{\sqrt{2(1-S)}} ~,
\end{equation}
a straightforward calculation yields
\begin{equation}
\Psi_{2}
= \frac{1}{\sqrt{2(1+S^2)}}
\Big[
(1+S)\,|\varphi_{+}\varphi_{+}\rangle
+
(S-1)\,|\varphi_{-}\varphi_{-}\rangle
\Big] ~.
\label{eq:schmidtform}
\end{equation}
The corresponding eigenvalues of the one-particle reduced density matrix are
\begin{equation}
\lambda_{+}
= \frac{(1+S)^2}{2(1+S^2)} ~,
\qquad
\lambda_{-}
= \frac{(1-S)^2}{2(1+S^2)} ~,
\label{eq:eigenvalues}
\end{equation}
with $\lambda_{+}+\lambda_{-}=1$.  They are also the natural occupation
numbers (divided by 2) of the reduced states.

The spatial entanglement entropy of Ansatz 2 is therefore given exactly by
the von Neumann entropy
\begin{equation}
S_{\mathrm{vN}}
= -\lambda_{+}\ln \lambda_{+}
  -\lambda_{-}\ln \lambda_{-} ~.
\label{eq:vnentropy}
\end{equation}

\begin{figure}[t]
  \centering
  \includegraphics[width=0.7\columnwidth]{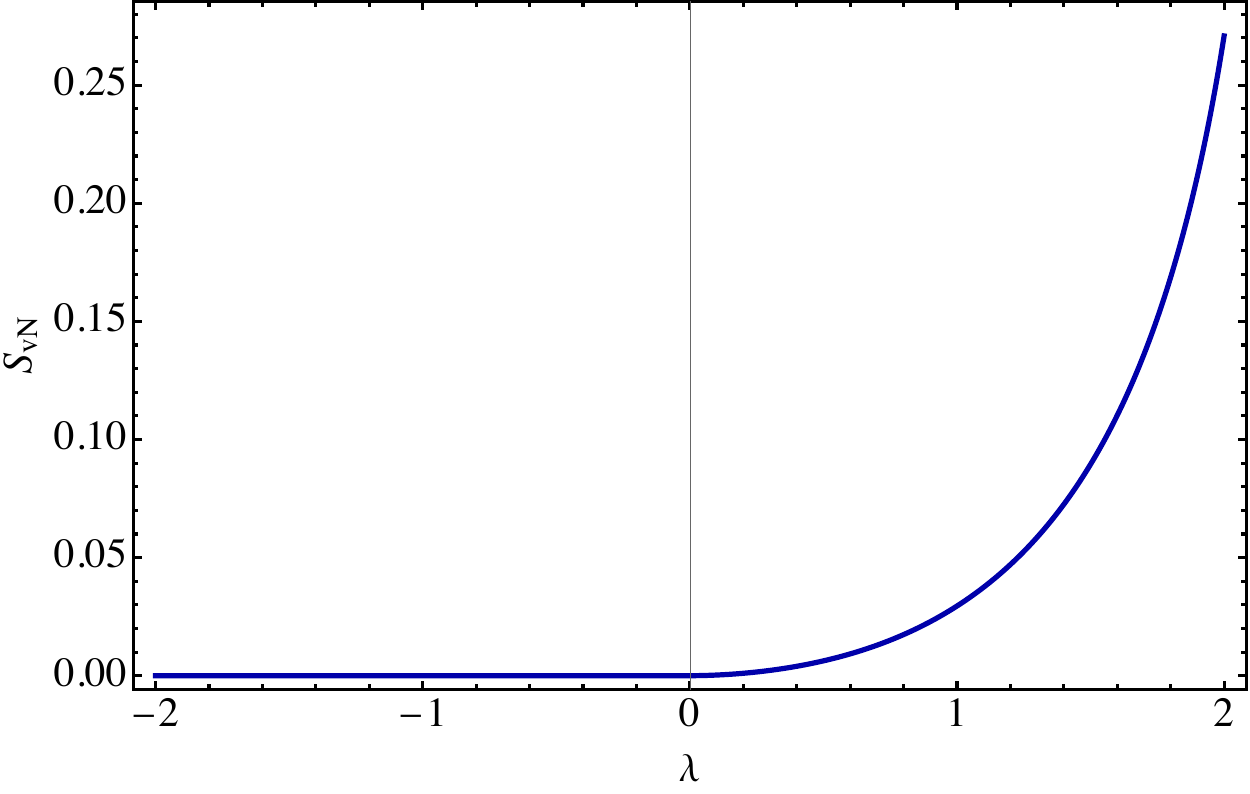}
  \caption{Entropy $S_{\mathrm{vN}}$ of Ansatz~2 as a function of the
  frustration parameter $\lambda$.
  }
  \label{fig:entropy_vs_lambda}
\end{figure}

It is shown in Fig.~\ref{fig:entropy_vs_lambda} as a function of $\lambda$.

The closed-form expression for $\mathrm{S_{vN}}$ allows us to track the entanglement as a
function of frustration in the Hamiltonian. In particular:
(i) For $\lambda\le0$ and $\alpha=\beta$, the two orbitals are identical, $S=1$,
and we obtain $\lambda_{+}=1$, $\lambda_{-}=0$, implying no entanglement.
(ii) With increasing frustration, $\lambda>0$ in Eq.(11) and $\alpha\neq\beta$,
$S$ reduces, generating entanglement entropy. The spatial entanglement, however,
cannot reach the maximum value 0.5 (corresponding to $S=0$), because the two orbitals
cannot become orthogonal.

%% file: dihydrogencation.tex
\section{Hydrogen Molecular Ion}
\label{sec_dihydcat}

The hydrogen molecular ion (or the dihydrogen cation) is the simplest molecule, consisting of two nuclei (protons) and an electron. Its study is important, because historically the semi-classical quantum theory failed to produce results that agreed with experiments \cite{pauli}, while the quantum theory using the Schr\"odinger equation succeeded \cite{burrau}.

The Schr\"odinger equation for the $H_2^+$ system is
\begin{equation}
    \mc{H}\psi = \left(\frac{P_1^2}{2M}+\frac{P_2^2}{2M}+\frac{p^2}{2m}-\frac{e^2}{r_1}-\frac{e^2}{r_2}+\frac{e^2}{r_{12}}   \right)\psi = E' \psi ~,
\end{equation}
where $m$ is the electron mass, $M$ is the nuclear mass, $\vec{r}_1$ and $\vec{r}_2$ are the distance vectors between the electron and the nuclei, $r_{12}$ is the distance between the two nuclei, and $P_1^2$,$P_2^2$ and $p^2$ are the momenta of the two nuclei and the electron.

The nuclei are very heavy compared to the electron, and empirical molecular spectroscopy data show that electronic and vibrational/rotational energy scales of the molecule are well-separated. So we can use the Born-Oppenheimer approximation \cite{cartoon}, and ignore the motion of the nuclei. The distance between the two nuclei is then set to $r_{12} = R$, and the Hamiltonian simplifies to
\begin{equation}
    \mc{H} = \left(-\frac{\hbar^2}{2m_e}\nabla^2-\frac{e^2}{r_1}-\frac{e^2}{r_2}\right) + \frac{e^2}{R} ~.
    \label{ham_dicat}
\end{equation}
Converting to atomic units, with rescaling of length and energy,
\begin{equation}
    \lambda = a_0 = \frac{e^2}{E_h} ~, \quad \epsilon = E_h = \frac{\hbar^2}{m a_0^2} ~,
    \label{h2p_scale}
\end{equation}
the Schr\"odinger equation becomes, with $E = E'-\frac{1}{R}$,
\begin{equation}
    \mc{H}\psi = \left(-\frac{1}{2}\nabla^2-\frac{1}{r_1}-\frac{1}{r_2}\right)\psi = E \psi ~.
    \label{se_h2p}
\end{equation}

This equation is separable in the prolate spheroidal coordinates \cite{arfken}. With z-axis as the inter-nuclear axis, these coordinates are:
\[ \xi = \frac{r_1+r_2}{R} , \quad  1 \leq \xi < \infty \]
\[ \eta = \frac{r_1-r_2}{R} , \quad  -1 \leq \eta \leq 1 \]
\[ \phi = \mr { \ azimuth \ angle}, \quad 0 \leq \phi < 2\pi \]

The Laplacian in these coordinates is
\begin{equation}
\nabla^2 = \frac{4}{R^2(\xi^2-\eta^2)}\left( \frac{\partial}{\partial \xi} \left((\xi^2 -1)\frac{\partial}{\partial \xi}\right) + \frac{\partial}{\partial \eta} \left((1-\eta^2)\frac{\partial}{\partial \eta}\right) +\left( \frac{1}{\xi^2-1} + \frac{1}{1-\eta^2} \right) \frac{\partial^2}{\partial \phi^2} \right) ~.
\label{lap}
\end{equation}
Separation of the coordinates, $\psi(\xi,\eta,\phi) = X(\xi)Y(\eta)\Phi(\phi)$, gives the following linear second order ordinary differential equations, with $A$ and $p^2 = -\frac{ER^2}{2}$ as the separation parameters that link the equations:

\begin{equation}
\frac{d^2\Phi}{d\phi^2}+m^2 \Phi = 0 ~,
\label{azimuth}
\end{equation}

\begin{equation}
\frac{d}{d\eta}\left( (1-\eta^2)\frac{dY}{d\eta}\right) + \left( p^2\eta^2 - \frac{m^2}{1-\eta^2} - A \right)Y = 0 ~,
\label{angular}    
\end{equation}

\begin{equation}
\frac{d}{d\xi}\left( (\xi^2 -1)\frac{dX}{d\xi}\right) + \left(- p^2\xi^2 +2R\xi - \frac{m^2}{\xi^2-1} + A \right)X = 0 ~.
\label{radial}
\end{equation}

\subsection{Solutions of the equations}

The azimuthal equation, Eq.\eqref{azimuth}, has the physical solutions
\begin{equation}
    \Phi(\phi) = \frac{1}{\sqrt{2\pi}} e^{\pm im\phi}, \quad m=0,1,2, ... ~,
\end{equation}
where the integer $m$ corresponds to the angular momentum along the inter-nuclear axis. For the ground state (experimentally denoted as the $^2\Sigma_g^+$ state) of $H_2^+$, $m = 0$. 

\subsubsection{Angular equation}
The angular equation, Eq.\eqref{angular}, can be rewritten as
\begin{equation}
(1-\eta^2)\frac{d^2Y}{d\eta^2} - 2\eta\frac{dY}{d\eta} + (p^2\eta^2-A)Y = 0 ~.
\label{ang}
\end{equation}
We need to obtain the relation between $A$ and $p^2$ to solve this equation. The equation has regular singular points at $\eta=\pm1$, hence according to Fuch's theorem we can obtain at least one solution by the Frobenius series method about an ordinary point like $\eta=0$. The ground state of $H_2^+$ is symmetric in $r_1$ and $r_2$, and the solution is an even function of $\eta$.


Hylleraas suggested an expansion in terms of the associated Legendre polynomial series \cite{hylleraas}:
\begin{equation}
Y(\eta) = \sum_{l=0}^{\infty}a_l P_l^0(\eta) ~.
\label{legendre}
\end{equation}
Substituting it in Eq.\eqref{ang}, and using the equation satisfied by the associated Legendre polynomials, we obtain
\begin{equation}
\sum_{l=0}^{\infty} a_l (p^2\eta^2 -A-l(l+1))P_l^0(\eta) = 0 ~.  
\label{leg}
\end{equation}

Associated Legendre polynomials obey the recurrence relation,
\begin{equation}
  (2l+1)\eta P_l^m(\eta) = (l+m) P_{l-1}^m(\eta)+(l-m+1)P_{l+1}^m(\eta) ~.
  \label{leg_rec}
\end{equation}
It can be used to express $\eta^2P_l^0$ in terms of $P_{l-2}^0(\eta)$, $P_{l}^0(\eta)$ and $P_{l+2}^0(\eta)$, which when substituted in Eq.\ref{leg} yields the following recurrence relation between the coefficients $a_l$: 
\begin{equation}
  a_{l+2}  \frac{p^2(l+2)(l+1)}{(2l+3)(2l+5)}+a_{l} \left( \frac{p^2l^2}{(2l+1)(2l-1)} + \frac{p^2(l+1)(l-1)}{(2l+1)(2l+3)} - A -l(l+1)\right)+ a_{l-2}  \frac{p^2(l-1)l}{(2l-3)(2l-1)} = 0 ~,
\end{equation}
abbreviated as $a_{l-2}\alpha(l) + a_{l}\beta(l) + a_{l+2}\gamma(l) = 0$.

For the even values of $l$ appearing in the expansion of the ground state wavefunction, the coefficients can be represented by the tridiagonal matrix:
\[ M_{even} = \begin{bmatrix}
    \frac{p^2}{3}- A & \frac{2p^2}{15} & 0 & 0 & ...\\
     \frac{2p^2}{3} &  \frac{11p^2}{21} -A -6 &  \frac{4p^2}{21} & 0 & ... \\
     0 &  \frac{12p^2}{35} &  \frac{39p^2}{77}-A-20 &  \frac{30p^2}{143} & ...\\
     : & : & : & : & ^..
\end{bmatrix} \]


A tridiagonal matrix can be decomposed into a lower triangular and upper triangular matrix:
\[ M = LU = \begin{bmatrix}
    1 & 0 & 0 & ...\\
    l_0 & 1 & 0 & ... \\
    : & : & : & ^..
\end{bmatrix}\begin{bmatrix}
    d_0 & u_0 & ... & ...\\
    0 & d_1 & u_1 & ... \\
    : & : & : & ^..
\end{bmatrix} = \begin{bmatrix}
    d_0 & u_0 & ... & ...\\
    l_0d_0 & d_1+l_0u_0 & u_1 & ... \\
    : & : & : & ^..
\end{bmatrix}\]
For a non-trivial solution to exist, the determinant of the matrix must vanish, i.e.
\begin{equation}
|M| = |L||U| = |U| = \prod_0^{\infty}d_n =0 ~,
\end{equation}
where the diagonal elements are specified by
$d_n = \beta(2n)-\frac{\alpha(2n)\gamma(2n-2)}{d_{n-1}}$,
with $d_0 = \beta(0) = \frac{p^2}{3} - A$.

Approximate solutions for the relation between $p^2$ and $A$ are obtained by setting $d_N =0$ for some sufficiently large N, and solving the (N+1)$^\mr{th}$ degree equation for $A$ in terms of $p^2$.
For the ground state, the series representation for $A$ as a function of $p^2$ is found to be:
\begin{equation}
A(p^2) = \frac{p^2}{3}+\frac{2p^4}{135}+\frac{4p^6}{8505}+O(p^8) ~.
\end{equation}

\subsubsection{Radial equation}
Having determined $A$ as a function of $p^2$, the radial Eq.\eqref{radial} has to be solved to obtain the relation between $R$ and $p^2$. The radial equation can be rewritten as
\begin{equation}
(\xi^2-1)\frac{d^2X}{d\xi^2} + 2\xi\frac{dX}{d\eta} +(2R\xi-p^2\xi^2+A)X = 0 ~.
\label{rad}
\end{equation}
It can be simplified by the choice \cite{jaffe}:
\begin{equation}
X(\xi) = e^{-p\xi} (\xi+1)^\sigma f(x) ~,
\quad x = \frac{\xi-1}{\xi+1} ~,
\quad \sigma = \frac{R}{p}-1 ~,
\label{xi_sub}
\end{equation}
leading to the differential equation for $f(x)$:
\begin{equation}
    x(1-x)^2f''+((1-2\sigma)x^2+2(\sigma-2p-1)x+1)f'+(\sigma^2x+\sigma(1+2p)+A-p^2)f = 0 ~.
    \label{pow_xi}
\end{equation}

This equation for $f(x)$ can be solved using a power series expansion about the origin:
\begin{equation}
 f(x) = \sum_{n=0}^{\infty} a_nx^n ~.
 \label{power_xi}
\end{equation} 
Substitution in Eq.\eqref{pow_xi} gives the three-term recurrence relation:
\begin{equation}
    a_{n+1} (n+1)^2 + a_n (2n(\sigma-2p-n)+\sigma(1+2p)+A-p^2) + a_{n-1}(n-1-\sigma)^2 = 0 ~.
\end{equation}
The series solution converges for $0\leq x\leq 1$. For a convergent solution, we can start with $a_{-1}=0$ and finite $a_0$, and then demand $a_n \rightarrow 0$ as $n\rightarrow\infty$. Setting $a_N=0$ for sufficiently large $N$ determines $p^2$ as a function of $R$ with sufficient accuracy.

With accurate expressions for $A$ and $p^2$, minimization of the energy eigenvalue $E'$ as a function of $R$ determines the ground state of $H_2^+$. The most accurate results (in a.u.) are \cite{scott}:
\begin{equation}
p = 1.48501462, \quad A = 0.811729585, \quad R = 2,
\quad E' = -0.60263462 ~.
\label{h2p_values}
\end{equation} 

\subsubsection{Complete solution in terms of Confluent Heun functions}

The substitution $x=1-\eta^2$ in Eq.\eqref{ang} gives the following differential equation for the function $y(x) = Y(\eta)$:
\begin{equation}
    x(x-1)y'' +\left(\frac{3}{2}x-1\right)y' +\frac{1}{4}(p^2x+A-p^2)y = 0 ~.
    \label{heun_eta}
\end{equation}
It is the non-symmetric Confluent Heun equation \cite{heun}, with the solutions (one regular at x=1 and other at x=0) \cite{rufus}:
\begin{align}
    y_0(x) & = HeunC\left( 0,-\frac{1}{2},0,-\frac{p^2}{4},\frac{A+1}{4},1-x \right) ~, \nonumber \\
    y_1(x) & = HeunC\left(0,0,-\frac{1}{2},\frac{p^2}{4},\frac{A-p^2+1}{4},x \right) ~. \nonumber
\end{align}
Transforming back to the variable $\eta$,
\begin{align}
    Y_0(x) & = HeunC\left( 0,-\frac{1}{2},0,-\frac{p^2}{4},\frac{A+1}{4},\eta^2 \right) ~, \\
    Y_1(x) & = HeunC\left(0,0,-\frac{1}{2},\frac{p^2}{4},\frac{A-p^2+1}{4},1-\eta^2\right) ~.
    \label{heunc_eta}
\end{align}

For appropriate values of $A$ and $p^2$, the solutions become linearly dependent, which is what we need to obtain a solution that is regular at both x=1 and x=0. The Wronskian of the two solutions can be used to find these appropriate values of $A$ and $p^2$, because it vanishes for linearly dependent solutions \cite{rufus}.

The radial equation can also be converted to the non-symmetric Confluent Heun equation, with regular singularities at $\xi=\pm1$. The corresponding solutions are \cite{rufus}:
\begin{align}
X_1(\xi) &= e^{-p(\xi-1)} HeunC\left( 4p,0,0,-4R,2R-p^2+A,\frac{1-\xi}{2}\right) ~, \\
X_{-1}(\xi) &= e^{p(\xi+1)}HeunC\left( 4p,0,0,-4R,-2R-p^2+A,\frac{1+\xi}{2}\right) ~.
\end{align}
The point $\xi=-1$ is unphysical, and the corresponding solution is ignored.

\subsection{The Ground State Wavefunction}

The ground state wavefunction of the hydrogen molecular ion is the product of the radial and angular solutions described in the preceding sections. In unnormalized form, it is
\begin{align}
    \psi(\xi,\eta) = &e^{-p(\xi-1)} HeunC\left( 4p,0,0,-4R,2R-p^2+A,\frac{1-\xi}{2}\right)\\ &\times HeunC\left(0,0,-\frac{1}{2},\frac{p^2}{4},\frac{A-p^2+1}{4},1-\eta^2\right) ~.
\end{align}
The coordinates can be changed from prolate spheroidal to cylindrical for ease of visualizing the wavefunction:
\[ \rho = \frac{R}{2}\sqrt{(\xi^2-1)(1-\eta^2)}, ~ \quad 0 \leq \rho < \infty ~, \]
\[ z = \frac{R}{2}\xi\eta ~, \quad -\infty < z < \infty ~. \]

\begin{figure}[htbp]
     \centering
     \begin{subfigure}[b]{0.3\textwidth}
         \centering        \fbox{\includegraphics[width=\textwidth,height=3.7cm]{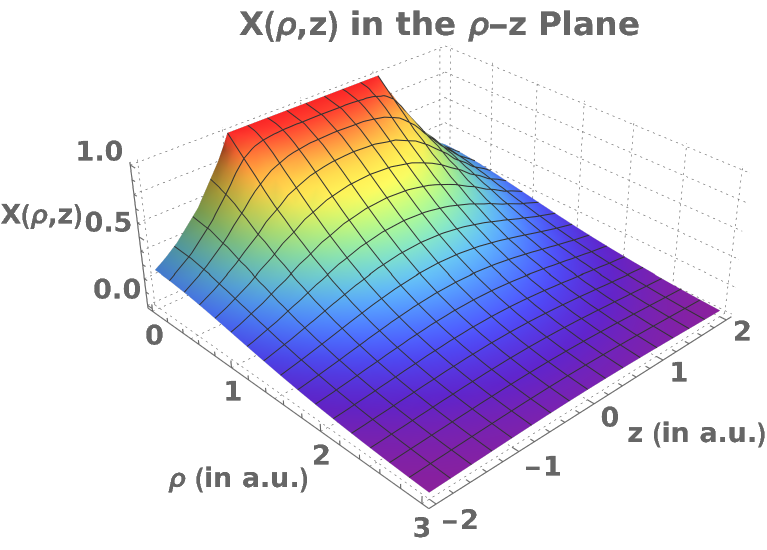}}
         \caption{$X(\rho,z)$}
         \label{xcyl}
     \end{subfigure}
     \hspace{3mm}     
     \begin{subfigure}[b]{0.3\textwidth}
         \centering
        \fbox{\includegraphics[width=\textwidth,height=3.7cm]{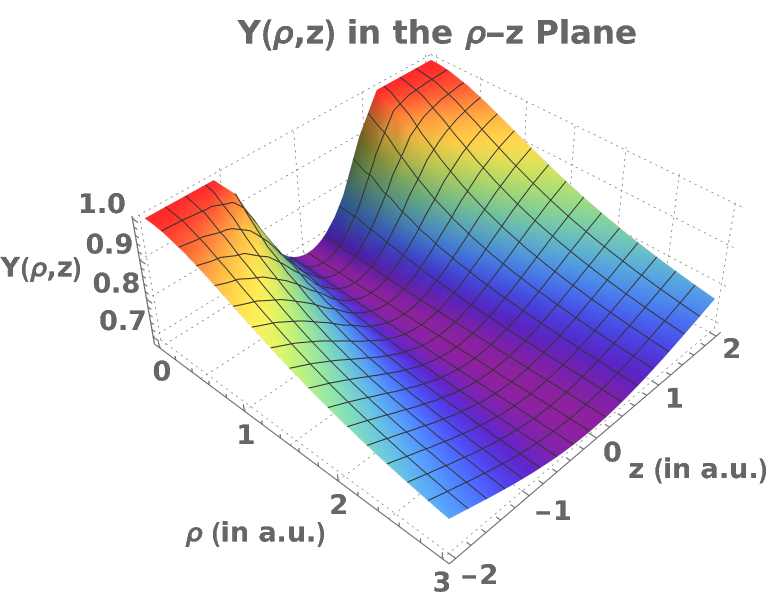}}
         \caption{$Y(\rho,z)$}
         \label{ycyl}
     \end{subfigure}
      \hspace{3mm}    
     \begin{subfigure}[b]{0.3\textwidth}
         \centering
         \fbox{\includegraphics[width=\textwidth,height=3.7cm]{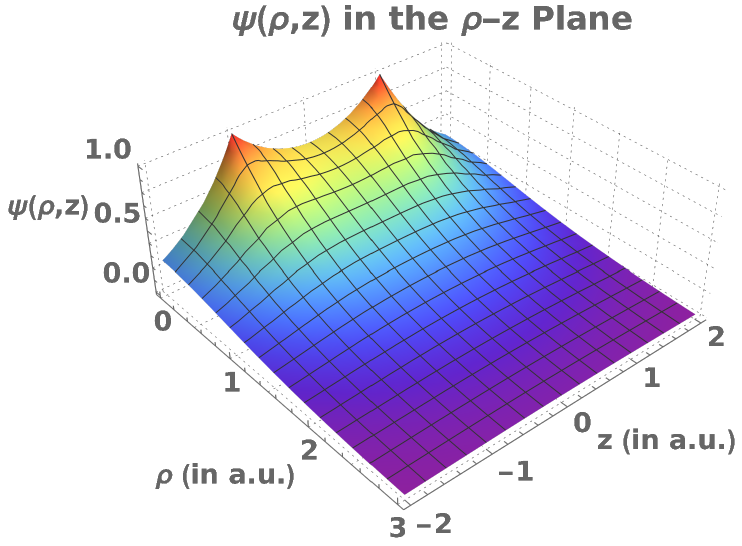}}
         \caption{$\psi(\rho,z)$}
         \label{psicyl}
     \end{subfigure}
     \caption{$X(\rho,z)$,$Y(\rho,z)$ and $\psi(\rho,z)$ in the $\rho-z$ plane}
     \label{fig_xypsi}
\end{figure}

The transformed functions $X(\rho,z)$, $Y(\rho,z)$ and $\psi(\rho,z)$ are plotted in Fig.\ref{fig_xypsi}.

\begin{figure}[htbp]
     \centering
     \begin{subfigure}[b]{0.3\textwidth}
         \centering
         \fbox{\includegraphics[width=\textwidth,height=3.7cm]{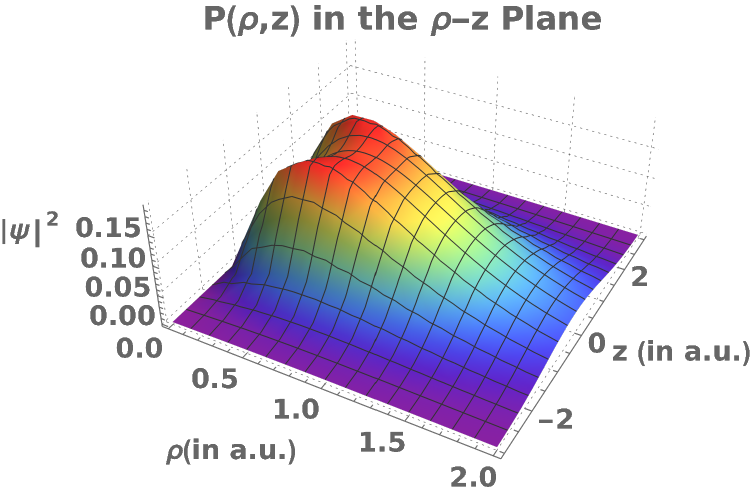}}
         \caption{$P(\rho,z)$}
         \label{def}
     \end{subfigure}
     \hspace{3mm}
     \begin{subfigure}[b]{0.3\textwidth}
         \centering
         \fbox{\includegraphics[width=\textwidth,height=3.7cm]{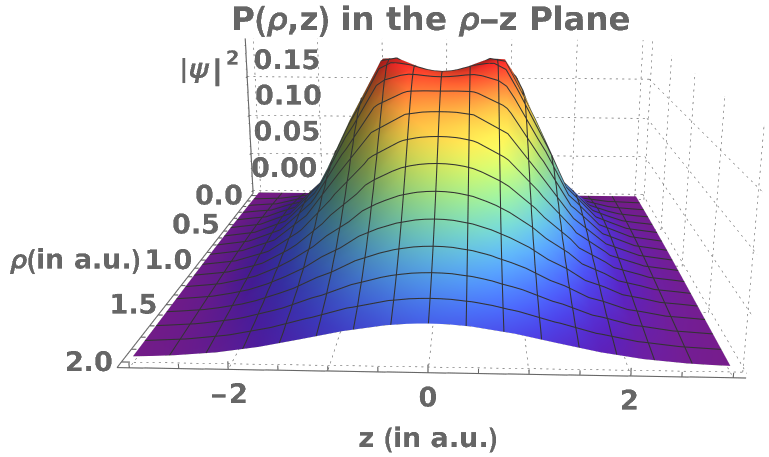}}
         \caption{Front view}
         \label{front}
     \end{subfigure}  
     \caption{Probability distribution of the electron in the ground state of $H_2^+$}
     \label{fig_prob}
\end{figure}

The associated (unnormalized) probability distribution of the electron,
\[ P(\rho,z) =  |\psi(\rho,z)|^2 \rho ~, \]
is displayed in Fig.\ref{fig_prob}. It can be seen that there are two peaks, indicating a high probability of finding the electron in those areas. Their locations are $(\rho,z) = (0.546097,\pm 0.71125)$.

In the ground state, the electron is more likely to be found near one of the nuclei, rather than in the middle. The nuclei are located at $z=\pm 1$ and $\rho = 0$, whereas the peaks occur at $|z|$ less than unity and $\rho$ greater than zero. This is the structure of the chemical bond between the two nuclei. In particular, the double-peaked wavefunction indicates tunnelling of the electron between the two nuclei.

\subsection{Approximate solutions}

We can fit the exact solution for the $H_2^+$ ground state wavefunction to finite order polynomial approximations, with sufficient accuracy and small number of coefficients. For a power series expansion, we choose
\begin{align}
    X_{pow}(\xi) & = e^{-p(\xi-1)}\sum_{n=0}^{N_{\mr{rad}}} x_n \xi^{n} ~, \\
    Y_{pow}(\eta) & = \sum_{n=0}^{N_{\mr{ang}}/2} y_{2n} \eta^{2n} ~. 
    \label{approx_pow}
\end{align}
The angular wavefunction is even, and so only even powers of $\eta$ appear in its expansion. Approximating the ground state wavefunction as $\psi_{pow}(\xi,\eta) = X_{pow}(\xi)Y_{pow}(\eta)$, we calculate the approximate ground state energy by integrating over $\xi$ and $\eta$:
\[ E_{\mr{approx}} = \bkt{\psi_{pow}|\mc{H}|\psi_{pow}}{} / \bkt{\psi_{pow}|\psi_{pow}}{} ~. \]
As detailed in the Supplementary Information, it can match the correct result to 6 decimal point accuracy with $N_{\mr{rad}} = 4$ and $N_{\mr{ang}} = 4$.

The power series can be improved by expansion in terms of the Chebyshev polynomials of the first kind, $T_n(\theta) = \cos(n\theta)$. That requires rescaling of $\xi \in [1,\xi_m]$ to fit it in the Chebyshev polynomial domain [-1,1].
\begin{align}
    X_{cheb}(\xi) & = e^{-p(\xi-1)}\sum_{n=0}^{N_{\mr{rad}}} \tilde{x}_n T_n(\xi') ~,
    \quad \xi' = \frac{2\xi-\xi_m-1}{\xi_m-1} ~, \\
    Y_{cheb}(\eta) & = \sum_{n=0}^{N_{\mr{ang}}/2} \tilde{y}_n T_n(\eta) ~.    
    \label{approx_cheb}
\end{align}
We again calculate the approximate ground state energy, using $\psi_{cheb}(\xi,\eta) = X_{cheb}(\xi)Y_{cheb}(\eta)$. As detailed in the Supplementary Information, it can match the correct result to 6 decimal point accuracy with $N_{\mr{rad}} = 4$ and $N_{\mr{ang}} = 4$, for $\xi_m = 8$. We observe that, compared to the power series expansion, the expansion coefficients fall off faster for the Chebyshev expansion.

%% file: hbond.tex
\section{The Hydrogen Bond}

The hydrogen bond is an exclusively quantum phenomenon produced by tunnelling of a proton between two heavier atoms. Understanding its structure is important for studying many real systems in quantum chemistry and biology. The ubiquitous system with a network of hydrogen bonds is water. Water has many unique properties, considered anomalous because they are unexplained by classical models. Quantum mechanical modeling of the hydrogen bond can aid the analysis of such properties.

\subsection{Proposed Model}

We consider the hydrogen bond involving two oxygen atoms and a proton. Its structure has a formal similarity with that of the hydrogen molecular ion, with the signs of charges flipped. Assigning an effective charge $Z_e$ to the oxygen atoms, the Hamiltonian for this system under the Born-Oppenheimer approximation is:
\begin{equation}
    \mc{H} = -\frac{\hbar^2}{2 m_p} \nabla^2 - \frac{Z_e e^2}{\ra} -\frac{Z_e e^2}{\rb} + \frac{Z_e^2 e^2}{R} ~.
    \label{hb_ham}
\end{equation}
The drawback of this approximation is that the oxygen atoms possess filled core electron orbitals and cannot be really treated as point charges. The additional spectator electron orbitals also contribute to Coulomb repulsion and dipole-dipole interaction. So our model is actually a starting point for a more realistic treatment.

We rescale the Hamiltonian with units related to those used for the hydrogen molecular ion in Eq.\eqref{h2p_scale}:
\begin{equation}
    \lambda' = \frac{a_0}{Z_e \mu} = \frac{\lambda}{Z_e \mu} ~,
    \quad \epsilon' = Z_e^2 \mu E_h = Z_e^2 \mu \epsilon ~.
    \label{rescale_hb}
\end{equation}
Here $\mu = \frac{m_p}{m_e} =1836.15$, and $\lambda'$ and $\epsilon'$ are the new length and energy scales. In units of $\lambda'$ and $\epsilon'$, the Hamiltonian is:
\begin{equation}
    \til{\mc{H}} = -\frac{1}{2}\til{\nabla}^2 -\frac{1}{\til{\ra}}-\frac{1}{\til{\rb}}+\frac{Z_e}{\til{R}} ~,
\end{equation}
and the Schr{\"o}dinger equation simplifies to
\begin{equation}
    \til{\mc{H}}\psi = \left(-\frac{1}{2}\til{\nabla}^2-\frac{1}{\til{r_1}}-\frac{1}{\til{r_2}}\right)\psi = E \psi ~,
    \label{se_hb}
\end{equation}
where $E = E'-\frac{Z_e}{\til{R}}$. This equation is similar to Eq.\eqref{se_h2p}, differing only in the Coulomb repulsion term. It can be analytically solved for various values of $\til{R}$ obtained by varying $Z_e$, similar to what we did in Section~\ref{sec_dihydcat}.

The experimentally determined properties of the $O$-$H$-$O$ hydrogen bond, obtained from dissociation of a water dimer molecule are \cite{rocher}:
\begin{align}
    R & = 2.98 \ \text{\AA} = 10340 \ Z_e \ \lambda'  \\
    \quad E_d & = -0.137 \ \text{eV} = -\left( \frac{2.74 \times 10^{-6}}{Z_e^2} \right) \epsilon' ~.
    \label{e_dissoc}
\end{align}
The hydrogen bond energy $E'$ is the sum of $E_d$ and the energy of the dissociated products (i.e. free $O$ and bound $H$-$O$). That makes $E' = E_d-\frac{1}{2}$ in our rescaled units.

\begin{figure}[htbp]
     \centering
         \centering
         \includegraphics[width=0.5\textwidth]{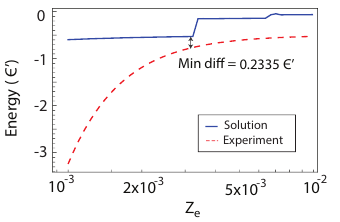}
     \caption{Comparison between the calculated analytical solution and experimental constraints as a function of the parameter $Z_e$. The blue solid line represents the solution to Eq.\ref{se_hb}, while the red dashed line tracks the experimental energy curve under the constraint of Eq.\ref{e_dissoc}. The minimum difference is 0.2335 $\epsilon'$ at $Z_e = 3.3251\times10^{-2}$.}
     \label{fig_energy}
\end{figure}

We carried out a simultaneous fit to the solution to Eq.\eqref{se_hb} and the experimental constraint of Eq.\eqref{e_dissoc}, as a function of $Z_e$. The hydrogen bond is a rather weakly bound system, and so we needed to solve the angular and radial equations to high orders, for the solutions to stabilise with sufficient numerical accuracy. In our symmetric choice for the variational wavefunction, the leading contamination to the ground state wavefunction comes from the second excited state. This contamination becomes visible in the behaviour of the variational wavefunction near $\eta=0$ (or $z=0$), and we attempted to avoid that.   

Our best results, obtained for $N_\mr{ang}=20$ and $N_\mr{rad}=14$, are displayed in Fig. \ref{fig_energy}. The numerical and experimental curves do not intersect, implying that we do not have an exact solution and some truncation effects remain. There are modeling errors too, in approximating the oxygen atoms as point charges. Still, the difference is small enough at the point of closest approach, specified by the parameters:
\begin{equation}
p = 17.6824, \quad A = 278.318, \quad Z_e = 0.00332, \quad E' = -0.528992~.
\label{hbond_values}
\end{equation}

\begin{figure}[htbp]
     \centering
     \begin{subfigure}[b]{0.3\textwidth}
         \centering
         \fbox{\includegraphics[width=\textwidth, height = 3.4cm]{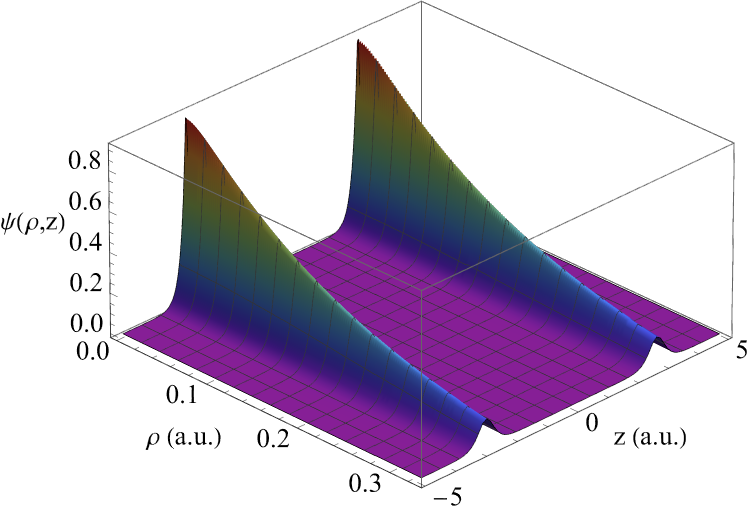}}
         \caption{Wavefunction $\psi(\rho,z)$}
         \label{def}
     \end{subfigure}
     \hspace{3mm}
     \begin{subfigure}[b]{0.3\textwidth}
         \centering
         \fbox{\includegraphics[width=\textwidth, height=3.4cm]{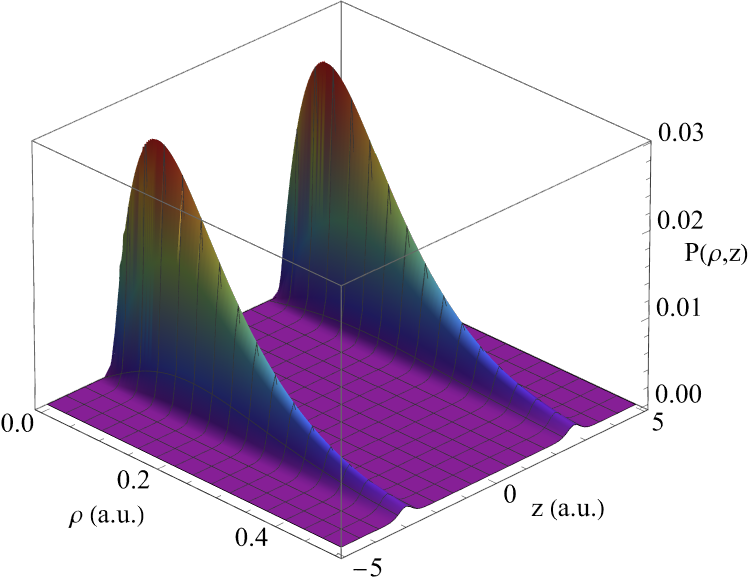}}
         \caption{Probability distribution, $P(\rho,z)$}
         \label{front}
     \end{subfigure}  
     \caption{The structure of the hydrogen bond wavefunction and probability distribution in the $\rho-z$ plane}
     \label{fig_hbondprob}
\end{figure}

The hydrogen bond wavefunction corresponding to these parameters is plotted in physical units in Fig. \ref{fig_hbondprob}, and shows the anticipated two-peak structure similar to that in Fig. \ref{fig_prob}. The peaks are located at $(\rho,z) = (0.0819,\pm 2.81561)$ a.u., while the oxygen atoms reside at $(\rho,z) = (0,\pm 2.81569)$. The noteworthy feature is that the hydrogen bond wavefunction is more sharply
peaked close to the oxygen atoms, compared to the rounded bumps of the $H_2^+$ wavefunction between the nuclei. It means that the proton of the hydrogen bond is much more likely to be close to one of the oxygen atoms than being in the intervening region of the bond.
\subsection{Possible Improvements}

The major drawback of our model is the treatment of oxygen atoms as static point objects. That ignores contribution of filled core electron orbitals, Coulomb and dipole-dipole interactions arising from the spectator electron orbitals, and the vibrational motion of the oxygen atoms. Inclusion of these effects, without introducing too many parameters, would make the model more realistic and tractable.

To an extent, we have bypassed the shortcomings of our model, by fitting it to the experimental data, instead of variationally minimizing the total energy to find the ground state. Our analysis may be looked upon as a reasonable solution, where the detailed molecular ingredients are replaced by a hard core potential.

%% file: conclusion.tex
\section{Summary}

Our analysis of quantum correlations in the ground state of helium-like atoms demonstrates entanglement between the two electrons. Frustrating interactions make the ground state of the system a permanent, and we show that the contributions of the permanent structure and Coulomb repulsion to the ground state energy are comparable in magnitude for small Z. Efficient modeling of such permanent structures in larger quantum systems can help in understanding their properties.

Our analysis of the hydrogen molecular ion exhibits the nature of the chemical bond through the tunnelling phenomenon in its wavefunction. Again frustrating interactions are the reason for the tunnelling. Such tunnelling is a common feature of chemical bonds, and its proper modeling can help in obtaining detailed understanding of many molecular structures.

We also model the $O$-$H$-$O$ hydrogen bond, exploiting its similarity with the hydrogen molecular ion. Our simple model fits the experimental data reasonably, yielding a double-peaked 3-dimensional structure for the wavefunction. It is an improvement over the scenarios that look at the hydrogen bond as tunnelling along a single reaction coordinate. Our model can be made more realistic in a more detailed analysis, with an improved treatment of the oxygen atom structure.

%% file: acknowledgement.tex
\section*{Acknowledgements}
This study was partially funded by the Accenture Project PC53015 at IISc. The numerical calculations in this work were carried out using Mathematica.

%% file: appendix.tex
\section*{Supplementary Information}

Here we provide the complete details of the variational calculations to obtain the upper and the lower bounds on the ground state energy of Helium-like atoms.

\section{Helium-like Atoms}

\subsection{Upper Bound Calculations}
The Hamiltonian for helium-like atoms in atomic units is
\begin{equation*}
\mc{H} = -\frac{1}{2}\nabla_{r_1}^2-\frac{1}{2}\nabla_{r_2}^2-\frac{Z}{r_1}-\frac{Z}{r_2}+\frac{1}{r_{12}} ~,
\end{equation*}
where $r_{12} = \sqrt{r_1^2+r_2^2-2r_1r_2 \cos(\theta)}$ and $\cos(\theta) = \sin(\theta_1)\sin(\theta_2)\cos(\phi_1-\phi_2)+\cos(\theta_1)\cos(\theta_2)$.

Consider the variational ansatz $\til{\psi}_4$ from Eq.\eqref{ansatz-4}:
\begin{align*}
    \til{\psi} & = (e^{-\al r_1 - \beta r_2}+ e^{-\beta r_1 - \al r_2})(1+cr_{12}) ~, \\
    \til{\psi} & =\psi_a + \psi_b = (\psi_{a1}+\psi_{a2}+\psi_{b1}+\psi_{b2}) ~,  
\end{align*}
with
\begin{align*}
\psi_a & = \psi_{a1}+ \psi_{a2} , \quad \psi_{a1} = e^{-\al r_1 -\beta r_2} , \quad \psi_{a2} = cr_{12}\psi_{a1} ~, \\
\psi_b & = \psi_{b1}+ \psi_{b2} , \quad \psi_{b1} = e^{-\beta r_1 - \al r_2} , \quad \psi_{b2} = cr_{12}\psi_{b1} ~.
\end{align*}

Obtaining results for this ansatz is sufficient for obtaining them for the other ansatzes that are special cases of it. Setting $c=0$ gives $\til{\psi}_2$, setting $\al=\beta$ gives $\til{\psi}_3$, and setting both $c=0$ and $\al=\beta$ gives $\til{\psi}_1$ upto a factor of 2. 

The upper bound is obtained by calculating the normalised expectation value of the Hamiltonian:
\begin{align*}
   \til{E} & =  \frac{\bra{\til{\psi}}\mc{H}\ket{\til{\psi}}}{\braket{\til{\psi}|\til{\psi}}} \\
     \til{E}  & = \frac{(\bra{\psi_a}+\bra{\psi_b})\mc{H}(\ket{\psi_a}+\ket{\psi_b})}{(\bra{\psi_a}+\bra{\psi_b})(\ket{\psi_a}+\ket{\psi_b})} \\
    \til{E}  & = \frac{\bra{\psi_a}\mc{H}\ket{\psi_a}+\bra{\psi_a}\mc{H}\ket{\psi_b}+\bra{\psi_b}\mc{H}\ket{\psi_a}+\bra{\psi_b}\mc{H}\ket{\psi_b}}{\braket{\psi_a|\psi_a}+\braket{\psi_a|\psi_b}+\braket{\psi_b|\psi_a}+\braket{\psi_b|\psi_b}} ~.
\end{align*}

Introducing the notation $\braket{X}_{pq} = \bra{\psi_p}X\ket{\psi_q}$, we get
\begin{equation*}
     \til{E}  = \frac{\bkt{\mc{H}}{aa}+\bkt{\mc{H}}{ab}+\bkt{\mc{H}}{ba}+\bkt{\mc{H}}{bb}}{\bkt{1}{aa}+\bkt{1}{ab}+\bkt{1}{ba}+\bkt{1}{bb}} ~, 
\end{equation*}
where the Hamiltonian matrix elements are
\begin{align*}
    \bkt{\mc{H}}{aa} & = \bkt{\mc{H}}{a1a1}+\bkt{\mc{H}}{a1a2}+\bkt{\mc{H}}{a2a1}+\bkt{\mc{H}}{a2a2} ~, \\
    \bkt{\mc{H}}{ab} & = \bkt{\mc{H}}{a1b1}+\bkt{\mc{H}}{a1b2}+\bkt{\mc{H}}{a2b1}+\bkt{\mc{H}}{a2b2} ~, \\
    \bkt{\mc{H}}{ba} & = \bkt{\mc{H}}{b1a1}+\bkt{\mc{H}}{b1a2}+\bkt{\mc{H}}{b2a1}+\bkt{\mc{H}}{b2a2} ~, \\
    \bkt{\mc{H}}{bb} & = \bkt{\mc{H}}{b1b1}+\bkt{\mc{H}}{b1b2}+\bkt{\mc{H}}{b2b1}+\bkt{\mc{H}}{b2b2} ~,
\end{align*}
and the normalization constants are
\begin{align*}
    \bkt{1}{aa} & = \bkt{1}{a1a1}+\bkt{1}{a1a2}+\bkt{1}{a2a1}+\bkt{1}{a2a2} ~, \\
    \bkt{1}{ab} & = \bkt{1}{a1b1}+\bkt{1}{a1b2}+\bkt{1}{b2a1}+\bkt{1}{b2a2} ~, \\
    \bkt{1}{ba} & = \bkt{1}{b1a1}+\bkt{1}{b1a2}+\bkt{1}{b2a1}+\bkt{1}{b2a2} ~, \\
    \bkt{1}{bb} & = \bkt{1}{b1b1}+\bkt{1}{b1b2}+\bkt{1}{b2b1}+\bkt{1}{b2b2} ~.
\end{align*}

We can easily evaluate:
\begin{align*}
    \mc{H}\psi_{a1} & = \left( -\frac{1}{2}\nabla_{r1}^2-\frac{1}{2}\nabla_{r2}^2-\frac{Z}{r_1}-\frac{Z}{r_2}-\frac{1}{r_{12}} \right)e^{-\al r_1 - \beta r_2} \\ 
    & = \left( -\frac{1}{2}\left(\al^2-\frac{2\al}{r_1}\right)-\frac{1}{2}\left(\beta^2-\frac{2\beta}{r_2}\right)-\frac{Z}{r_1}-\frac{Z}{r_2}-\frac{1}{r_{12}} \right)e^{-\al r_1 - \beta r_2} \\ 
    & = \left( -\frac{\al^2+\beta^2}{2}+\frac{\al-Z}{r_1}+\frac{\beta-Z}{r_2}+\frac{1}{r_{12}} \right) \psi_{a1} ~.
\end{align*}
Similarly,
\begin{align*}
    \mc{H}\psi_{b1} & = \left( -\frac{1}{2}\nabla_{r1}^2-\frac{1}{2}\nabla_{r2}^2-\frac{Z}{r_1}-\frac{Z}{r_2}-\frac{1}{r_{12}} \right)e^{-\beta r_1 - \al r_2} \\ 
    & = \left( -\frac{1}{2}\left(\beta^2-\frac{2\beta}{r_1}\right)-\frac{1}{2}\left(\al^2-\frac{2\al}{r_2}\right)-\frac{Z}{r_1}-\frac{Z}{r_2}-\frac{1}{r_{12}} \right)e^{-\beta r_1 - \al r_2} \\ 
    & = \left( -\frac{\al^2+\beta^2}{2}+\frac{\beta-Z}{r_1}+\frac{\al-Z}{r_2}+\frac{1}{r_{12}} \right) \psi_{b1} ~.
\end{align*}
Also,
\begin{align*}
    \mc{H}\psi_{a2} = & \left( -\frac{1}{2}\nabla_{r1}^2-\frac{1}{2}\nabla_{r2}^2-\frac{Z}{r_1}-\frac{Z}{r_2}-\frac{1}{r_{12}} \right)c r_{12} e^{-\al r_1 - \beta r_2} \\ 
    = & -\frac{c}{2}\left[\left(\al^2-\frac{2\al}{r_1}\right) r_{12} - 2\al\left( \frac{r_1-r_2cos(\theta)}{r_{12}}\right)+\frac{2}{r_{12}}\right]e^{-\al r_1 - \beta r_2} \\
     & -\frac{c}{2}\left[\left(\beta^2-\frac{2\beta}{r_2}\right) r_{12} - 2\beta\left( \frac{r_2-r_1cos(\theta)}{r_{12}}\right)+\frac{2}{r_{12}}\right]e^{-\al r_1 - \beta r_2} \\
     & -\frac{cZr_{12}e^{-\al r_1 - \beta r_2}}{r_1}-\frac{cZr_{12}e^{-\al r_1 - \beta r_2}}{r_2}-ce^{-\al r_1 - \beta r_2} \\ 
    = & \ c \left[ -\left(\frac{\al^2+\beta^2}{2}\right)r_{12}+\left(\frac{\al-Z}{r_1}\right)r_{12}+\left(\frac{\beta-Z}{r_2}\right)r_{12}+1 +\frac{\al r_1}{r_{12}} +\frac{\beta r_2}{r_{12}}-\frac{\al r_2 cos(\theta)}{r_{12}} -\frac{\beta r_1 cos(\theta)}{r_{12}}-\frac{2}{r_{12}} \right] \psi_{a1} ~.
\end{align*}
Similarly, 
\begin{align*}
    \mc{H}\psi_{b2} = & \left( -\frac{1}{2}\nabla_{r1}^2-\frac{1}{2}\nabla_{r2}^2-\frac{Z}{r_1}-\frac{Z}{r_2}-\frac{1}{r_{12}} \right)c r_{12} e^{-\beta r_1 - \al r_2} \\ 
    = & -\frac{c}{2}\left[\left(\beta^2-\frac{2\beta}{r_1}\right) r_{12} - 2\beta\left( \frac{r_1-r_2cos(\theta)}{r_{12}}\right)+\frac{2}{r_{12}}\right]e^{-\beta r_1 - \al r_2} \\
     & -\frac{c}{2}\left[\left(\al^2-\frac{2\al}{r_2}\right) r_{12} - 2\al\left( \frac{r_2-r_1cos(\theta)}{r_{12}}\right)+\frac{2}{r_{12}}\right]e^{-\beta r_1 - \al r_2} \\
     & -\frac{cZr_{12}e^{-\beta r_1 - \al r_2}}{r_1}-\frac{cZr_{12}e^{-\beta r_1 - \al r_2}}{r_2}-ce^{-\beta r_1 - \al r_2} \\ 
    = & \ c \left[ -\left(\frac{\al^2+\beta^2}{2}\right)r_{12}+\left(\frac{\beta-Z} {r_1}\right)r_{12}+\left(\frac{\al-Z}{r_2}\right)r_{12}+1 +\frac{\beta r_1}{r_{12}} +\frac{\al r_2}{r_{12}}-\frac{\beta r_2 cos(\theta)}{r_{12}} -\frac{\al r_1 cos(\theta)}{r_{12}}-\frac{2}{r_{12}} \right] \psi_{b1} ~.
\end{align*}

\subsubsection*{Calculating $\bkt{\mc{H}}{a1a1}$:}
\begin{align*}
    \bkt{\mc{H}}{a1a1} & = \bra{\psi_{a1}}\mc{H}\ket{\psi_{a2}} = \int dr_1^3 dr_2^3 \psi_{a1}^2 \left( -\frac{\al^2+\beta^2}{2}+\frac{\al-Z}{r_1}+\frac{\beta-Z}{r_2}+\frac{1}{r_{12}} \right) \\
    & = \sum_i H_{a1a1,i} ~, \quad \mr{where} \quad H_{a1a1,i} = \int dr_1^3 dr_2^3 \psi_{a1}^2 h_{a1a1,i}
\end{align*}
The integration is over all space and the integral $\bkt{\mc{H}}{a1a1}$ is written as a sum of individual integrals $H_{a1a1,i}$ of the integrands $h_{a1a1,i}$, where $i$ labels the terms in the sum.

\begin{longtable}{|c|c|c|}
\caption{Calculations for $\bkt{\mc{H}}{a1a1}$} \\

\hline
$i$  & $h_{a1a1,i}$ & $H_{a1a1,i}$ \\ 
\hline
\endfirsthead

\hline
$i$  & $h_{b1a2,i}$ & $H_{b1a2,i}$ \\ 
\hline
\endhead

\hline
\endfoot

\hline
\endlastfoot
          1 & $-\left(\dfrac{\al^2+\beta^2}{2}\right)$ & $-\dfrac{\pi^2}{\al^3\beta^3}\left( \dfrac{\al^2+\beta^2}{2} \right)$ \\ 2 & $\dfrac{\al-Z}{r_1}$ & $\dfrac{\pi^2}{\al^2\beta^3}(\al-Z)$ \\ 
          3 & $\dfrac{\beta-Z}{r_2}$ & $\dfrac{\pi^2}{\al^3\beta^2}(\beta-Z)$ \\ 
          4 & $\dfrac{1}{\rab}$ & $\dfrac{\pi^2(\al^2+3\al \beta +\beta^2)}{(\al^2\beta^2)(\al+\beta)^3}$  \\ 
\end{longtable}

The normalisation constant is
\begin{equation*}
    \bkt{1}{a1a1} = \bkt{\psi_{a1}|\psi_{a1}}{} = \int dr_1^3 dr_2^3 \psi_{a1}^2 = \dfrac{\pi^2}{\al^3\beta^3} ~.
\end{equation*}

\subsubsection*{Calculating $\bkt{\mc{H}}{a2a1}$:}
\begin{align*}
    \bkt{\mc{H}}{a2a1} & = \bra{\psi_{a2}}\mc{H}\ket{\psi_{a2}} = \int dr_1^3 dr_2^3 \psi_{a1}^2 c \rab \left( -\frac{\al^2+\beta^2}{2}+\frac{\al-Z}{r_1}+\frac{\beta-Z}{r_2}+\frac{1}{r_{12}} \right) \\
    & = \sum_i H_{a2a1,i} ~, \quad \mr{where} \quad H_{a2a1,i} = \int dr_1^3 dr_2^3 \psi_{a1}^2 h_{a2a1,i}
\end{align*}

\begin{longtable}{|c|c|c|}
\caption{Calculations for $\bkt{\mc{H}}{a2a1}$} \\

\hline
$i$  & $h_{a2a1,i}$ & $H_{a2a1,i}$ \\ 
\hline
\endfirsthead

\hline
$i$  & $h_{b1a2,i}$ & $H_{b1a2,i}$ \\ 
\hline
\endhead

\hline
\endfoot

\hline
\endlastfoot
          1 & $-c\left(\dfrac{\al^2+\beta^2}{2}\right)\rab$ & $-\dfrac{c\pi^2(\al^2+\beta^2)}{4\beta^5}\left( \dfrac{2}{\al^2}+\dfrac{3\beta^2}{\al^4}+\dfrac{-2\al - 3\beta}{(\al+\beta)^3} \right)$ \\ 
          2 & $c\left(\dfrac{\al-Z}{r_1}\right)\rab$ & $\dfrac{c\pi^2(3\al^3+6\al^2 \beta+4\al \beta^2+2\beta^2)(\al-Z)}{2\al^3\beta^4(\al+\beta)^2}$ \\ 
          3 & $c\left(\dfrac{\beta-Z}{r_2}\right)\rab$ & $\dfrac{c\pi^2(\beta-Z)}{2\beta^4}\left(\dfrac{1}{\al^2}+\dfrac{3\beta^2}{\al^4}-\dfrac{1}{(\al_\beta)^2}\right)$ \\ 
          4 & $c$ & $\dfrac{c\pi^2}{\al^3\beta^3}$  \\ 
         \hline
\end{longtable}

The normalisation constant is
\begin{align*}
    \bkt{1}{a2a1} & = \bkt{\psi_{a2}|\psi_{a1}}{} = \int dr_1^3 dr_2^3 c \rab \psi_{a1}^2 = \dfrac{c \pi^2}{2 \beta^5}\left( \dfrac{2}{\al^2} + \dfrac{3\beta^2}{\al^4} + \dfrac{-2\al-3\beta}{(\al+\beta)^5}\right) ~.
\end{align*}

\subsubsection*{Calculating $\bkt{\mc{H}}{a1a2}$:}
\begin{align*}
    \bkt{\mc{H}}{a1a2}  = & \bra{\psi_{a1}}\mc{H}\ket{\psi_{a2}} \\
     = & \int dr_1^3 dr_2^3 \psi_{a1}^2 c \left[ -\frac{(\al^2+\beta^2)}{2}r_{12}+\frac{(\al-Z)}{r_1}r_{12}+\frac{(\beta-Z)}{r_2}r_{12}+1 +\frac{\al r_1}{r_{12}} +\frac{\beta r_2}{r_{12}}-\frac{\al r_2 cos(\theta)}{r_{12}} -\frac{\beta r_1 cos(\theta)}{r_{12}}-\frac{2}{r_{12}} \right] \\
    = & \sum_i H_{a1a2,i} ~, \quad \mr{where} \quad H_{a1a2,i} = \int dr_1^3 dr_2^3 \psi_{a1}^2 h_{a1a2,i}
\end{align*}

\begin{longtable}{|c|c|c|}
\caption{Calculations for $\bkt{\mc{H}}{a1a2}$} \\

\hline
$i$  & $h_{a1a2,i}$ & $H_{a1a2,i}$ \\ 
\hline
\endfirsthead

\hline
$i$  & $h_{b1a2,i}$ & $H_{b1a2,i}$ \\ 
\hline
\endhead

\hline
\endfoot

\hline
\endlastfoot 
          1 & $-c\left(\dfrac{\al^2+\beta^2}{2}\right)\rab$ & $-\dfrac{c\pi^2(\al^2+\beta^2)}{4\beta^5}\left( \dfrac{2}{\al^2}+\dfrac{3\beta^2}{\al^4}+\dfrac{-2\al - 3\beta}{(\al+\beta)^3} \right)$ \\ 
          2 & $c\left(\dfrac{\al-Z}{r_1}\right)\rab$ & $\dfrac{c\pi^2(3\al^3+6\al^2 \beta+4\al \beta^2+2\beta^2)(\al-Z)}{2\al^3\beta^4(\al+\beta)^2}$ \\ 
          3 & $c\left(\dfrac{\beta-Z}{r_2}\right)\rab$ & $\dfrac{c\pi^2(\beta-Z)}{2\beta^4}\left(\dfrac{1}{\al^2}+\dfrac{3\beta^2}{\al^4}-\dfrac{1}{(\al_\beta)^2}\right)$ \\ 
          4 & $c$ & $\dfrac{c\pi^2}{\al^3\beta^3}$  \\ 
          5 & $\dfrac{c\al \ra}{\rab}$ & $-\dfrac{c\pi^2\al}{2\beta^3}\left( -\dfrac{2}{\al^3}+\dfrac{2\al+5\beta}{(\al+\beta)^4}\right)$ \\ 
          6 & $\dfrac{c\beta \rb}{\rab}$ & $\dfrac{c\pi^2(2\al^3+8\al^2\beta+12\al\beta^2+3\beta^3}{2\al^2\beta^2(\al+\beta)^4}$ \\ 
          7 & $-\dfrac{c \al \rb \cst }{\rab}$ & $-\dfrac{c\pi^2(4\al+\beta)}{2\al^2(\al+\beta)^4} $ \\ 
          8 & $-\dfrac{c \beta \ra \cst }{\rab}$ & $-\dfrac{c\pi^2(\al+4\beta)}{2\beta^2(\al+\beta)^4} $ \\ 
          9 & $-\dfrac{2c}{\rab}$ & $-\dfrac{2c\pi^2(\al^2+3\al\beta+\beta^2)}{\al^2\beta^2(\al+\beta)^3} $ \\ 
         \hline
\end{longtable}

The normalisation constant is
\begin{align*}
    \bkt{1}{a1a2} & = \bkt{\psi_{a1}|\psi_{a2}}{} = \int dr_1^3 dr_2^3 c \rab \psi_{a1}^2 = \bkt{1}{a1a2} ~.
\end{align*}

\subsubsection*{Calculating $\bkt{\mc{H}}{a2a2}$:}
\begin{align*}
    \bkt{\mc{H}}{a2a2}  = & \bra{\psi_{a2}}\mc{H}\ket{\psi_{a2}} \\
     = & \int dr_1^3 dr_2^3 \psi_{a1}^2 c^2 \left[ -\frac{(\al^2+\beta^2)}{2}r_{12}^2+\frac{(\al-Z)}{r_1}r_{12}^2+\frac{(\beta-Z)}{r_2}r_{12}^2+\rab +\al r_1 +\beta r_2-\al r_2 cos(\theta) -\beta r_1 cos(\theta)-2 \right] \\
    = & \sum_i H_{a2a2,i} ~, \quad \mr{where} \quad H_{a2a2,i} = \int dr_1^3 dr_2^3 \psi_{a1}^2 h_{a2a2,i}
\end{align*}

\begin{longtable}{|c|c|c|}
\caption{Calculations for $\bkt{\mc{H}}{a2a2}$} \\

\hline
$i$  & $h_{a2a2,i}$ & $H_{a2a2,i}$ \\ 
\hline
\endfirsthead

\hline
$i$  & $h_{b1a2,i}$ & $H_{b1a2,i}$ \\ 
\hline
\endhead

\hline
\endfoot

\hline
\endlastfoot 
          1 & $-c^2\left(\dfrac{\al^2+\beta^2}{2}\right)\rab^2$ & $-\dfrac{3c^2\pi^2(\al^2+\beta^2)^2}{2\al^5\beta^5}$ \\ 
          2 & $c^2\left(\dfrac{\al-Z}{r_1}\right)\rab^2$ & $\dfrac{3c^2\pi^2(2\al^2+\beta^2)(\al-Z)}{2\al^4\beta^5}$ \\ 
          3 & $c^2\left(\dfrac{\beta-Z}{r_2}\right)\rab^2$ & $\dfrac{3c^2\pi^2(\al^2+2\beta^2)(\beta-Z)}{2\al^5\beta^4}$ \\ 
          4 & $c^2\rab$ & $\dfrac{c^2\pi^2}{2 \beta^5}\left( \dfrac{2}{\al^2}+\dfrac{3\beta^2}{\al^4}+\dfrac{-2\al-3\beta}{(\al+\beta)^2} \right)$  \\ 
          5 & $c^2\al \ra$ & $\dfrac{3c^2\pi^2}{2\al^3\beta^3}$ \\ 
          6 & $c^2\beta \rb$ & $\dfrac{3c^2\pi^2}{2\al^3\beta^3}$  \\ 
          7 & $-\dfrac{c \al \rb \cst }{\rab}$ & $0$ \\ 
          8 & $-\dfrac{c \beta \ra \cst }{\rab}$ & $0$ \\ 
          9 & $-2c^2$ & $-\dfrac{2c^2\pi^2}{\al^6} $ \\ 
         \hline
    \end{longtable}

The normalisation constant is
\begin{align*}
    \bkt{1}{a2a2} & = \bkt{\psi_{a2}|\psi_{a2}}{} = \int dr_1^3 dr_2^3 c^2 \rab^2 \psi_{a1}^2 = \dfrac{3c^2\pi^2(\al^2+\beta^2)}{\al^5\beta^5} ~.
\end{align*}

We thus have $\bkt{\mc{H}}{aa}  = \bkt{\mc{H}}{a1a1}+\bkt{\mc{H}}{a1a2}+\bkt{\mc{H}}{a2a1}+\bkt{\mc{H}}{a2a2}$, and $\bkt{1}{aa}  = \bkt{1}{a1a1}+\bkt{1}{a1a2}+\bkt{1}{a2a1}+\bkt{1}{a2a2}$. 
     
\subsubsection*{Calculating $\bkt{\mc{H}}{b1a1}$:}
\begin{align*}
    \bkt{\mc{H}}{b1a1} & = \bra{\psi_{b1}}\mc{H}\ket{\psi_{a2}} = \int dr_1^3 dr_2^3 \psi_{b1}\psi_{a1} \left( -\frac{\al^2+\beta^2}{2}+\frac{\al-Z}{r_1}+\frac{\beta-Z}{r_2}+\frac{1}{r_{12}} \right) \\
    & = \sum_i H_{b1a1,i} ~, \quad \mr{where} \quad H_{b1a1,i} = \int dr_1^3 dr_2^3 \psi_{a1}\psi_{b1} h_{b1a1,i}
\end{align*}

\begin{longtable}{|c|c|c|}
\caption{Calculations for $\bkt{\mc{H}}{b1a1}$} \\

\hline
$i$  & $h_{b1a1,i}$ & $H_{b1a1,i}$ \\ 
\hline
\endfirsthead

\hline
$i$  & $h_{b1a2,i}$ & $H_{b1a2,i}$ \\ 
\hline
\endhead

\hline
\endfoot

\hline
\endlastfoot
          1 & $-\left(\dfrac{\al^2+\beta^2}{2}\right)$ & $-\dfrac{32\pi^2(\al^2+\beta^2)}{(\al+\beta)^6}$ \\ 
          2 & $\dfrac{\al-Z}{r_1}$ & $\dfrac{32\pi^2(\al-Z)}{(\al+\beta)^5}$ \\ 
          3 & $\dfrac{\beta-Z}{r_2}$ & $\dfrac{32\pi^2(\beta-Z)}{(\al+\beta)^5}$ \\ 
          4 & $\dfrac{1}{\rab}$ & $\dfrac{20\pi^2}{(\al+\beta)^5}$  \\ 
         \hline
\end{longtable}
The normalisation constant is
\begin{align*}
    \bkt{1}{b1a1} & = \bkt{\psi_{b1}|\psi_{a1}}{} = \int dr_1^3 dr_2^3 \psi_{a1}\psi_{b1} = \dfrac{64 \pi^2}{(\al+\beta)^6} ~.
\end{align*}

\subsubsection*{Calculating $\bkt{\mc{H}}{b2a1}$:}
\begin{align*}
    \bkt{\mc{H}}{b2a1} & = \bra{\psi_{b2}}\mc{H}\ket{\psi_{a2}} = \int dr_1^3 dr_2^3 \psi_{b2}\psi_{a1} c \rab \left( -\frac{\al^2+\beta^2}{2}+\frac{\al-Z}{r_1}+\frac{\beta-Z}{r_2}+\frac{1}{r_{12}} \right) \\
    & = \sum_i H_{b2a1,i} ~, \quad \mr{where} \quad H_{b2a1,i} = \int dr_1^3 dr_2^3 \psi_{a1}\psi_{b1} h_{b2a1,i}
\end{align*}

\begin{longtable}{|c|c|c|}
\caption{Calculations for $\bkt{\mc{H}}{b2a1}$} \\

\hline
$i$  & $h_{b2a1,i}$ & $H_{b2a1,i}$ \\ 
\hline
\endfirsthead

\hline
$i$  & $h_{b1a2,i}$ & $H_{b1a2,i}$ \\ 
\hline
\endhead

\hline
\endfoot

\hline
\endlastfoot
          1 & $-c\left(\dfrac{\al^2+\beta^2}{2}\right)\rab$ & $-\dfrac{140c\pi^2(\al^2+\beta^2)}{(\al+\beta)^7}$ \\ 
          2 & $c\left(\dfrac{\al-Z}{r_1}\right)\rab$ & $\dfrac{120 c \pi^2(\al-Z)}{(\al+\beta)^6}$ \\ 
          3 & $c\left(\dfrac{\beta-Z}{r_2}\right)\rab$ & $\dfrac{120 c \pi^2(\beta-Z)}{(\al+\beta)^6}$ \\ 
          4 & $c$ & $\dfrac{64 c\pi^2}{(\al+\beta)^6}$  \\ 
         \hline
    \end{longtable}

The normalisation constant is
\begin{align*}
    \bkt{1}{b2a1} & = \bkt{\psi_{b2}|\psi_{a1}}{} = \int dr_1^3 dr_2^3 c \rab \psi_{a1}\psi_{b1} = \dfrac{280 c \pi^2}{(\al+\beta)^7} ~. 
\end{align*}

\subsubsection*{Calculating $\bkt{\mc{H}}{b1a2}$:}
\begin{align*}
    \bkt{\mc{H}}{b1a2}  = & \bra{\psi_{b1}}\mc{H}\ket{\psi_{a2}} \\
     = & \int dr_1^3 dr_2^3 \psi_{b1}\psi_{a1} c \left[ -\frac{(\al^2+\beta^2)}{2}r_{12}+\frac{(\al-Z)}{r_1}r_{12}+\frac{(\beta-Z)}{r_2}r_{12}+1 +\frac{\al r_1}{r_{12}} +\frac{\beta r_2}{r_{12}}-\frac{\al r_2 cos(\theta)}{r_{12}} -\frac{\beta r_1 cos(\theta)}{r_{12}}-\frac{2}{r_{12}} \right] \\
    = &\sum_i H_{b1a2,i} ~, \quad \mr{where} \quad H_{b1a2,i} = \int dr_1^3 dr_2^3 \psi_{a1}\psi_{b1} h_{b1a2,i}
\end{align*}

\begin{longtable}{|c|c|c|}
\caption{Calculations for $\bkt{\mc{H}}{b1a2}$} \\

\hline
$i$  & $h_{b1a2,i}$ & $H_{b1a2,i}$ \\ 
\hline
\endfirsthead

\hline
$i$  & $h_{b1a2,i}$ & $H_{b1a2,i}$ \\ 
\hline
\endhead

\hline
\endfoot

\hline
\endlastfoot
          1 & $-c\left(\dfrac{\al^2+\beta^2}{2}\right)\rab$ & $-\dfrac{140c\pi^2(\al^2+\beta^2)}{(\al+\beta)^7}$ \\ 
          2 & $c\left(\dfrac{\al-Z}{r_1}\right)\rab$ & $\dfrac{120 c \pi^2(\al-Z)}{(\al+\beta)^6}$ \\ 
          3 & $c\left(\dfrac{\beta-Z}{r_2}\right)\rab$ & $\dfrac{120 c \pi^2(\beta-Z)}{(\al+\beta)^6}$ \\ 
          4 & $c$ & $\dfrac{64 c\pi^2}{(\al+\beta)^6}$  \\ 
          5 & $\dfrac{c\al \ra}{\rab}$ & $\dfrac{50c\pi^2\al}{(\al+\beta)^6}$ \\ 
          6 & $\dfrac{c\beta \rb}{\rab}$ & $\dfrac{50c\pi^2\beta}{(\al+\beta)^6}$ \\ 
          7 & $-\dfrac{c \al \rb \cst }{\rab}$ & $-\dfrac{10c\pi^2\beta}{(\al+\beta)^6}$ \\ 
          8 & $-\dfrac{c \beta \ra \cst }{\rab}$ & $-\dfrac{10c\pi^2\al}{(\al+\beta)^6}$ \\ 
          9 & $-\dfrac{2c}{\rab}$ & $-\dfrac{40c\pi^2}{(\al+\beta)^5}$ \\ 
\end{longtable}

The normalisation constant is
\begin{align*}
    \bkt{1}{b1a2} & = \bkt{\psi_{b1}|\psi_{a2}}{} = \int dr_1^3 dr_2^3 c \rab \psi_{a1}\psi_{b1} = \bkt{1}{b2a1} ~.
\end{align*}

\subsubsection*{Calculating $\bkt{\mc{H}}{b2a2}$:}
\begin{align*}
    \bkt{\mc{H}}{b2a2}  = & \bra{\psi_{b2}}\mc{H}\ket{\psi_{a2}} \\
     = & \int dr_1^3 dr_2^3 \psi_{b1}\psi_{a1} c^2 \left[ -\frac{(\al^2+\beta^2)}{2}r_{12}^2+\frac{(\al-Z)}{r_1}r_{12}^2+\frac{(\beta-Z)}{r_2}r_{12}^2+\rab +\al r_1 +\beta r_2-\al r_2 cos(\theta) -\beta r_1 cos(\theta)-2 \right] \\
    = & \sum_i H_{b2a2,i} ~, \quad \mr{where} \quad H_{b2a2,i} = \int dr_1^3 dr_2^3 \psi_{a1}\psi_{b1} h_{b2a2,i} 
\end{align*}

\begin{longtable}{|c|c|c|}
\caption{Calculations for $\bkt{\mc{H}}{b2a2}$} \\

\hline
$i$  & $h_{b2a2,i}$ & $H_{b2a2,i}$ \\ 
\hline
\endfirsthead

\hline
$i$  & $h_{b2a2,i}$ & $H_{b2a2,i}$ \\ 
\hline
\endhead

\hline
\endfoot

\hline
\endlastfoot
          1 & $-c^2\left(\dfrac{\al^2+\beta^2}{2}\right)\rab^2$ & $-\dfrac{768c^2\pi^2(\al^2+\beta^2)}{(\al+\beta)^8}$ \\ 
          2 & $c^2\left(\dfrac{\al-Z}{r_1}\right)\rab^2$ & $\dfrac{576c^2\pi^2(\al-Z)}{\al+\beta)^7}$ \\ 
          3 & $c^2\left(\dfrac{\beta-Z}{r_2}\right)\rab^2$ & $\dfrac{576c^2\pi^2(\beta-Z)}{\al+\beta)^7}$ \\ 
          4 & $c^2\rab$ & $\dfrac{280c^2\pi^2}{\al+\beta)^7}$  \\ 
          5 & $c^2\al \ra$ & $\dfrac{192c^2\pi^2\al}{\al+\beta)^7}$ \\ 
          6 & $c^2\beta \rb$ & $\dfrac{192c^2\pi^2\beta}{\al+\beta)^7}$  \\ 
          7 & $-\dfrac{c \al \rb \cst }{\rab}$ & $0$ \\ 
          8 & $-\dfrac{c \beta \ra \cst }{\rab}$ & $0$ \\ 
          9 & $-2c^2$ & $-\dfrac{128c^2\pi^2}{(\al+\beta)^6} $ \\ 
         \hline
\end{longtable}

The normalisation constant is
\begin{align*}
    \bkt{1}{b2a2} & = \bkt{\psi_{b2}|\psi_{a2}}{} = \int dr_1^3 dr_2^3 c^2 \rab^2 \psi_{a1}\psi_{b1} = \dfrac{1536 c^2 \pi^2}{(\al+\beta)^6} ~.
\end{align*}

We thus have $\bkt{\mc{H}}{ba}  = \bkt{\mc{H}}{b1a1}+\bkt{\mc{H}}{b1a2}+\bkt{\mc{H}}{b2a1}+\bkt{\mc{H}}{b2a2}$,
and $\bkt{1}{ba}  = \bkt{1}{b1a1}+\bkt{1}{b1a2}+\bkt{1}{b2a1}+\bkt{1}{b2a2}$.

\subsubsection*{Calculating $\bkt{\mc{H}}{ab}$:}

$\bkt{\mc{H}}{ba} = \bra{\psi_b}\mc{H}\ket{\psi_a}$ is $\bkt{\mc{H}}{ab}=\bra{\psi_a}\mc{H}\ket{\psi_b}$ with $\al$ and $\beta$ interchanged:
\begin{equation*}
    \bkt{\mc{H}}{ba}(\al,\beta,c) = \bkt{\mc{H}}{ab}(\beta,\al,c) ~.
\end{equation*}
This is also true for the normalisation constants:
\begin{equation*}
    \bkt{1}{ba}(\al,\beta,c) = \bkt{1}{ab}(\beta,\al,c) ~.
\end{equation*}

\subsubsection*{Calculating $\bkt{\mc{H}}{bb}$:}

$\bkt{\mc{H}}{bb} = \bra{\psi_b}\mc{H}\ket{\psi_b}$ is $\bkt{\mc{H}}{aa}=\bra{\psi_a}\mc{H}\ket{\psi_a}$ with $\al$ and $\beta$ interchanged:
\begin{equation*}
    \bkt{\mc{H}}{bb}(\al,\beta,c) = \bkt{\mc{H}}{aa}(\beta,\al,c) ~.
\end{equation*}
This is also true for the normalisation constants:
\begin{equation*}
    \bkt{1}{bb}(\al,\beta,c) = \bkt{1}{aa}(\beta,\al,c) ~.
\end{equation*}

\subsubsection*{Calculating the upper bounds:}

\begin{equation*}
   \til{E} = \frac{\bra{\til{\psi}}\mc{H}\ket{\til{\psi}}}{\bkt{\til{\psi}|\til{\psi}}{}} = \frac{\bkt{\mc{H}}{aa}+\bkt{\mc{H}}{ab}+\bkt{\mc{H}}{ba}+\bkt{\mc{H}}{bb}}{\bkt{1}{aa}+\bkt{1}{ab}+\bkt{1}{ba}+\bkt{1}{bb}} 
\end{equation*}
This expression in terms of the variational parameters $\al, \beta, c$ and the atomic number $Z$ is:
\begin{align*}
\til{E} = & \frac{1}{\left( \frac{2}{\al^3\beta^3} + \frac{3072c^2}{(\al+\beta)^8} +\frac{1120c}{(\al+\beta)^7}+\frac{128}{(\al+\beta)^6}+\frac{6c^2(\al^2+\beta^2)}{\al^5\beta^5}+\frac{c}{\beta^5}\left( \frac{2}{\al^2} +\frac{3\beta^2}{\al^4}+\frac{-2\al-3\beta}{(\al+\beta)^3}\right) + \frac{c}{\al^5}\left( \frac{2}{\beta^2} +\frac{3\al^2}{\beta^4}+\frac{-3\al-2\beta}{(\al+\beta)^3}\right) \right)} \times \\
 & \frac{1}{\al^5\beta^5(\al+\beta)^8 } \times (\al^2\beta^2 (\al+\beta)^2(\al^8  +6\al^7\beta + \beta^8 + 2\al\beta^6(1+3\beta) + 4\al^2\beta^5(3+4\beta) + 2\al^6 \beta(1+8\beta) \\
 & + 2\al^5\beta^2(6+13\beta) + 2\al^3\beta^4(33+13\beta)+2\al^4 \beta^3(33+79\beta)-2Z(\al^7 +7\al^6 \beta+21\al^5 \beta^2 +99\al^4 \beta^3 +99\al^3 \beta^4  \\
 & + 21\al^2 \beta^5 + 7\al\beta^6 +\beta^7)) + c\al\beta(\al+\beta)(3\al^10 +21\al^9 \beta+3\beta^10 +28\al^7 \beta^2 (1+4\beta)+28\al^3 \beta^6 (3+4\beta) \\
    & + 4\al^8 \beta(1+16\beta) +4\al^2 \beta^7 (7+16\beta)+\al\beta^8 (4+21\beta)+\al^6 \beta^3 (84+53\beta)+ \al^4 \beta^5 (396+53\beta)+2\al^5 \beta^4 (198+547\beta)\\
    & -2Z(3\al^9 +23\al^8 \beta+78\al^7 \beta^2 + 158\al^6 \beta^3 +698\al^5 \beta^4 +698\al^4 \beta^5 +158\al^3 \beta^6 +78\al^2 \beta^7 +23\al\beta^8 +3\beta^9))+ \\
    &c^2 (3\al^12 +24\al^11 \beta+3\beta^12 +8\al^7 \beta^4 (23+6\beta)+3\al\beta^10 (1+8\beta)+8\al^9 \beta^2 (3+23\beta)+ 2\al^2 \beta^9 (12+43\beta) \\
    & + \al^5 \beta^6 (823+48\beta)+\al^10 \beta(3+86\beta)+2\al^3 \beta^8 (43+92\beta)+\al^8 \beta^3 (86+269\beta)+\al^4 \beta^7 (184+269\beta) \\
    & +\al^6 \beta^5(823+2868\beta) -3Z(2\al^11 +17\al^10 \beta+65\al^9 \beta^2 +150\al^8 \beta^3 +
240\al^7 \beta^4 +1062\al^6 \beta^5 +1062\al^5 \beta^6 +240\al^4 \beta^7 \\ & +150\al^3 \beta^8 +65\al^2 \beta^9+17\al\beta^10 +2\beta^11))) ~.
\end{align*}

The upper bounds for the ground state energy $\widehat{E}_0$ are obtained by minimizing $\til{E}$ for different Z.

\newpage
\subsection{Lower Bound Calculations:}

We need the variance of the Hamiltonian $\mc{H}$ to obtain the lower bound on the ground state energy. So, in addition to the already evaluated $\bkt{\mc{H}}{}$, we need to evaluate $\bkt{\mc{H}^2}{}$.

\begin{align*}
\bkt{\mc{H}^2}{} & = \bra{\psi}\mc{H}^2\ket{\psi} \\
& = (\bra{\psi_a}+\bra{\psi_b})\mc{H}^2(\ket{\psi_a}+\ket{\psi_b})\\
& = \bra{\psi_a}\mc{H}^2\ket{\psi_a}+\bra{\psi_a} \mc{H}^2 \ket{\psi_b}+\bra{\psi_b}\mc{H}^2\ket{\psi_a}+\bra{\psi_b}\mc{H}^2\ket{\psi_b} \\
\end{align*}
Using the notation from the previous section, we have
\begin{equation*}
    \bkt{\mc{H}^2}{}  = \bkt{\mc{H}^2}{aa}+\bkt{\mc{H}^2}{ab}+\bkt{\mc{H}^2}{ba}+\bkt{\mc{H}^2}{bb} ~,
\end{equation*}
where
\begin{align*}
    \bkt{\mc{H}^2}{aa} & = \bkt{\mc{H}^2}{a1a1}+\bkt{\mc{H}^2}{a1a2}+\bkt{\mc{H}^2}{a2a1}+\bkt{\mc{H}^2}{a2a2} ~, \\
    \bkt{\mc{H}^2}{ab} & = \bkt{\mc{H}^2}{a1b1}+\bkt{\mc{H}^2}{a1b2}+\bkt{\mc{H}^2}{a2b1}+\bkt{\mc{H}^2}{a2b2} ~, \\
    \bkt{\mc{H}^2}{ba} & = \bkt{\mc{H}^2}{b1a1}+\bkt{\mc{H}^2}{b1a2}+\bkt{\mc{H}^2}{b2a1}+\bkt{\mc{H}^2}{b2a2} ~, \\
    \bkt{\mc{H}^2}{bb} & = \bkt{\mc{H}^2}{b1b1}+\bkt{\mc{H}^2}{b1b2}+\bkt{\mc{H}^2}{b2b1}+\bkt{\mc{H}^2}{b2b2} ~.
\end{align*}

We have already evaluated $\mc{H}\ket{\psi_{a1}}, \mc{H}\ket{\psi_{a2}}, \mc{H}\ket{\psi_{b1}}$ and $\mc{H}\ket{\psi_{b2}}$, and we use them to obtain all the terms required for $\bkt{H^2}{}$ in a manner similar to that in the previous section. Note that the normalisation constants have already been evaluated.

\subsubsection*{Calculating $\bkt{\mc{H}^2}{a1a1}$:}
\begin{align*}
    \bkt{\mc{H}^2}{a1a1} & = (\bra{\psi_{a1}}\mc{H}).(\mc{H}\ket{\psi_{a1}}) = \int dr_1^3 dr_2^3 \psi_{a1}^2 \left( -\frac{\al^2+\beta^2}{2}+\frac{\al-Z}{r_1}+\frac{\beta-Z}{r_2}+\frac{1}{r_{12}} \right)^2 \\
    & = \sum_i Hs_{a1a1,i} ~, \quad \mr{where} \quad Hs_{a1a1,i} = \int dr_1^3 dr_2^3 \psi_{a1}^2 hs_{a1a1,i}
\end{align*}
The integration is over all space and the integral $\bkt{\mc{H}^2}{a1a1}$ is written as a sum of individual integrals $Hs_{a1a1,i}$ of the integrands $hs_{a1a1,i}$, where $i$ labels the terms in the sum.

\begin{longtable}{|c|c|c|}
\caption{Calculations for $\bkt{\mc{H}^2}{a1a1}$} \\

\hline
$i$  & $hs_{a1a1,i}$ & $Hs_{a1a1,i}$ \\ 
\hline
\endfirsthead

\hline
$i$  & $hs_{a1a1,i}$ & $Hs_{a1a1,i}$ \\ 
\hline
\endhead

\hline
\endfoot

\hline
\endlastfoot
 
          1 & $\left(\dfrac{\al^2+\beta^2}{2}\right)^2$ & $\dfrac{\pi^2(\al^2+\beta^2)^2}{4\al^3\beta^3}$ \\ 
          2 & $\left(\dfrac{\al-Z}{r_1}\right)^2$ & $\dfrac{2\pi^2(\al-Z)^2}{\al\beta^3}$ \\ 
          3 & $\left(\dfrac{\beta-Z}{r_2}\right)^2$ & $\dfrac{2\pi^2(\beta-Z)^2}{\al^3\beta}$ \\ 
          4 & $\dfrac{1}{\rab^2}$ & $\dfrac{2\pi^2(\al^4-\beta^4+4\al^2\beta^2 log(\frac{\beta}{\al})}{\al\beta(\al^2-\beta^2)}$  \\ 
          5 & $-\dfrac{(\al^2+\beta^2)(\al-Z)}{\ra}$ & $\dfrac{\pi^2(Z-\al)(\al^2+\beta^2)}{\al^2\beta^3}$ \\ 
          6 & $-\dfrac{(\al^2+\beta^2)(\beta-Z)}{\rb}$ & $\dfrac{\pi^2(Z-\beta)(\al^2+\beta^2)}{\al^3\beta^2}$\\ 
          7 & $-\dfrac{(\al^2+\beta^2)}{\rab}$ & $-\dfrac{\pi^2(\al^2+\beta^2)(\al^2+3\al\beta+\beta^2)}{\al^2\beta^2(\al+\beta)^3}$\\ 
          8 & $2\dfrac{(\al-Z)(\beta-Z)}{\ra \rb}$ & $\dfrac{2 \pi^2(\al-Z)(\beta-Z)}{\al^2\beta^2}$\\ 
          9 & $2\dfrac{(\al-Z)}{\ra \rab}$ & $\dfrac{2 \pi^2(\al-Z)(\al+2\beta)}{\al\beta^2(\al+\beta)^2}$\\ 
          10 & $2\dfrac{(\beta-Z)}{\ra \rab}$ & $\dfrac{2 \pi^2(\beta - Z)}{\beta^2}\left( \dfrac{1}{\al^2}-\dfrac{1}{\al+\beta)^2}\right)$\\
  \end{longtable}

\subsubsection*{Calculating $\bkt{\mc{H}^2}{a2a1}$:}
\begin{align*}
    \bkt{\mc{H}^2}{a2a1} = & (\bra{\psi_{a2}}\mc{H}).(\mc{H}\ket{\psi_{a2}}) = \int dr_1^3 dr_2^3 \psi_{a1}^2 c \left( -\frac{\al^2+\beta^2}{2}+\frac{\al-Z}{r_1}+\frac{\beta-Z}{r_2}+\frac{1}{r_{12}} \right) \times \\
     & \left( -\frac{(\al^2+\beta^2)}{2}r_{12}+\frac{(\al-Z)}{r_1}r_{12}+\frac{(\beta-Z)}{r_2}r_{12}+1 +\frac{\al r_1}{r_{12}} +\frac{\beta r_2}{r_{12}}-\frac{\al r_2 cos(\theta)}{r_{12}} -\frac{\beta r_1 cos(\theta)}{r_{12}}-\frac{2}{r_{12}} \right)\\
    = & \sum_i Hs_{a2a1,i} ~, \quad \mr{where} \quad Hs_{a2a1,i} = \int dr_1^3 dr_2^3 \psi_{a1}^2 hs_{a2a1,i}
\end{align*}  

\begin{longtable}{|c|c|c|}
\caption{Calculations for $\bkt{\mc{H}^2}{a2a1}$} \\

\hline
$i$  & $hs_{a2a1,i}$ & $Hs_{a2a1,i}$ \\ 
\hline
\endfirsthead

\hline
$i$  & $hs_{a2a1,i}$ & $Hs_{a2a1,i}$ \\ 
\hline
\endhead

\hline
\endfoot

\hline
\endlastfoot
          1 & $c\left(\dfrac{\al^2+\beta^2}{2}\right)^2\rab$ & $\dfrac{c\pi^2(\al^2+\beta^2)^2}{8\beta^5}\left(\dfrac{2}{\al^2}+\dfrac{3\beta^2}{\al^4}+\dfrac{-2\al-3\beta}{(\al+\beta)^3}\right)$ \\ 
          2 & $-c\left(\dfrac{\al^2+\beta^2}{2}\right)\left(\dfrac{\al-Z}{r_1}\right)\rab$ & $\dfrac{c\pi^2(Z-\al)(\al^2+\beta^2)(3\al^3+6\al^2\beta+4\al\beta+2\beta^3)}{4\al^3\beta^4(\al+\beta)^2}$ \\ 
          3 & $-c\left(\dfrac{\al^2+\beta^2}{2}\right)\left(\dfrac{\beta-Z}{r_2}\right)\rab$ & $\dfrac{c\pi^2(Z-\beta)(\al^2+\beta^2)}{4\beta^4}\left(\dfrac{1}{\al^2}+\dfrac{3\beta^2}{\al^4}-\dfrac{1}{(\al+\beta)^2}\right)$ \\ 
          4 & $-c\left(\dfrac{\al^2+\beta^2}{2}\right)$ & $-\dfrac{c\pi^2(\al^2+\beta^2)}{2\al^3\beta^3}$  \\ 
          5 & $-c\left(\dfrac{\al^2+\beta^2}{2}\right)\dfrac{\al\ra}{\rab}$ & $\dfrac{c\pi^2\al(\al^2+\beta^2)}{4\beta^3}\left(-\dfrac{2}{\al^3}+\dfrac{2\al+5\beta}{(\al+\beta)^4}\right)$ \\ 
          6 & $-c\left(\dfrac{\al^2+\beta^2}{2}\right)\dfrac{\beta\rb}{\rab}$ & $-\dfrac{c\pi^2(\al^2+\beta)^2(2\al^3+8\al^2\beta+12\al\beta^2+3\beta^3)}{4\al^2\beta^2(\al+\beta)^4}$ \\ 
          7 & $c\left(\dfrac{\al^2+\beta^2}{2}\right)\dfrac{\al\rb\cst}{\rab}$ & $\dfrac{c\pi^2(\al+4\beta)(\al^2+\beta^2)}{4\beta^2(\al+\beta)^4}$\\ 
          8 & $c\left(\dfrac{\al^2+\beta^2}{2}\right)\dfrac{\beta\ra\cst}{\rab}$ & $\dfrac{c\pi^2(4\al+\beta)(\al^2+\beta^2)}{4\al^2(\al+\beta)^4}$ \\ 
          9 & $c\left(\dfrac{\al^2+\beta^2}{\rab}\right)$ & $\dfrac{c\pi^2(\al^2+\beta^2)(3\al^3+6\al^2\beta+4\al\beta^2+2\beta^3)}{4\al^3\beta^4(\al+\beta)^2}$\\ 
          10 & $-c\left(\dfrac{\al-Z}{\ra}\right)\left(\dfrac{\al^2+\beta^2}{2}\right)\rab$ & $\dfrac{c\pi^2(Z-\al)(\al^2+\beta^2)(3\al^3+6\al^2\beta+4\al\beta^2+2\beta^3)}{4\al^3\beta^4(\al+\beta)^2}$\\ 
          11 & $c\left(\dfrac{\al-Z}{\ra}\right)^2\rab$ & $\dfrac{c\pi^2(Z-\al)^2}{\beta^5}\left( -1+\dfrac{\beta^2}{\al^2}+\dfrac{\al}{\al+\beta}-4log\left(\dfrac{\al}{\al+\beta}\right)\right)$ \\ 
          12 & $c\left(\dfrac{\al-Z}{\ra}\right)\left(\dfrac{\beta-Z}{\rb}\right)\rab$ & $\dfrac{c\pi^2(Z-\al)(Z-\beta)(\al^2+\al\beta+\beta^2)}{\al^3\beta^3(\al+\beta)}$ \\ 
          13 & $c\left(\dfrac{\al-Z}{\ra}\right)$ & $\dfrac{c\pi^2(\al-Z)}{\al^2\beta^3}$ \\ 
          14 & $c\left(\dfrac{\al-Z}{\ra}\right)\dfrac{\al\ra}{\rab}$ & $\dfrac{c\pi^2(\al-Z)(\al^2+3\al\beta+\beta^2)}{\al\beta^2(\al+\beta)^3}$ \\ 
          15 & $c\left(\dfrac{\al-Z}{\ra}\right)\dfrac{\beta\rb}{\rab}$ & $\dfrac{c\pi^2(\al-Z)(\al^2+3\al\beta+\beta^2)}{\al\beta^2(\al+\beta)^3}$  \\ 
          16 & $-c\left(\dfrac{\al-Z}{\ra}\right)\dfrac{\al\rb\cst}{\rab}$ & $\dfrac{4 c\pi^2\al(\al-Z)}{\beta^5}\left(\dfrac{\beta(4\al^2+10\al\beta+7\beta^2)}{4(\al+\beta)^3}+log\left(\dfrac{\al}{\al+\beta}\right)\right)$ \\ 
          17 & $-c\left(\dfrac{\al-Z}{\ra}\right)\dfrac{\beta\ra\cst}{\rab}$ & $\dfrac{c\pi^2(Z-\al)}{\al(\al+\beta)^3}$ \\ 
          18 & $-c\left(\dfrac{\al-Z}{\ra}\right)\dfrac{2}{\rab}$ & $\dfrac{2c\pi^2(Z-\al)(\al+2\beta)}{\al\beta^2(\al+\beta)^2}$ \\ 
          19 & $-c\left(\dfrac{\beta-Z}{\rb}\right) \left(\dfrac{\al^2+\beta^2}{2}\right)\rab$ & $\dfrac{c\pi^2(Z-\beta)(\al^2+\beta^2)}{4\beta^4}\left(\dfrac{1}{\al^2}+\dfrac{3\beta^2}{\al^4}-\dfrac{1}{(\al+\beta)^2}\right)$\\ 
          20 & $c\left(\dfrac{\beta-Z}{\rb}\right)^2\rab$ & $\dfrac{c\pi^2(Z-\al)(Z-\beta)(\al^2+\al\beta+\beta^2)}{\al^3\beta^3(\al+\beta)}$ \\ 
          21 & $c\left(\dfrac{\beta-Z}{\rb}\right)\left(\dfrac{\al-Z}{\ra}\right)\rab$ & $\dfrac{c\pi^2(Z-\beta)^2}{\al^5\beta^2}\left(\dfrac{\al(\al^2+\al\beta-\beta^2)}{\al+\beta}+4\beta^2log\left(\dfrac{\al+\beta}{\beta}\right)\right)$ \\ 
          22 & $c\left(\dfrac{\beta-Z}{\rb}\right)$ & $\dfrac{c\pi^2(\beta-Z)}{\al^3\beta^2}$ \\ 
          23 & $c\left(\dfrac{\beta-Z}{\rb}\right)\dfrac{\al\ra}{\rab}$ & $\dfrac{c\pi^2\al(\beta-Z)}{\beta^2}\left(\dfrac{1}{\al^3}-\dfrac{1}{(\al+\beta)^3}\right)$ \\ 
          24 & $c\left(\dfrac{\beta-Z}{\rb}\right)\dfrac{\beta\rb}{\rab}$ & $\dfrac{c\pi^2(\beta-Z)(\al^2+3\al\beta+\beta^2)}{\al^2\beta(\al+\beta)^3}$ \\ 
          25 & $-c\left(\dfrac{\beta-Z}{\rb}\right)\dfrac{\al\rb\cst}{\rab}$ & $-\dfrac{c\pi^2(\beta-Z)}{\beta(\al+\beta)^3}$ \\ 
          26 & $-c\left(\dfrac{\beta-Z}{\rb}\right)\dfrac{\beta\ra\cst}{\rab}$ & $\dfrac{c\pi^2(Z-\beta)}{\beta(\al+\beta)^3}$ \\ 
          27 & $-c\left(\dfrac{\beta-Z}{\rb}\right)\dfrac{2}{\rab}$ & $\dfrac{2c\pi^2(Z-\beta)}{\beta^2}\left(\dfrac{1}{\al^3}-\dfrac{1}{(\al+\beta)^3}\right)$ \\ 
          28 & $-c\left(\dfrac{\al^2+\beta^2}{2}\right)$ & $-\dfrac{c\pi^2(\al^2+\beta^2)}{2\al^3\beta^3}$\\ 
          29 & $c\left(\dfrac{\al-Z}{\ra}\right)$ & $\dfrac{c\pi^2(\al-Z)}{\al^2\beta^3}$ \\ 
          30 & $c\left(\dfrac{\beta-Z}{\rb}\right)$ & $\dfrac{c\pi^2(\beta-Z)}{\al^3\beta^2}$ \\ 
          31 & $\dfrac{c}{\rab}$ & $\dfrac{c\pi^2(\al^2+3\al\beta+\beta^2)}{\al^2\beta^2(\al+\beta)^3}$ \\ 
          32 & $\dfrac{c\al\ra}{\rab^2}$ & $\dfrac{c\pi^2(3\al^6+7\al^4\beta^2-11\al^2\beta^4+\beta^6+4\al^2\beta^2(5\al^2+\beta^2)log(\frac{\beta}{\alpha}))}{\al\beta(\al^2-\beta^2)^4}$ \\ 
          33 & $\dfrac{c\beta\rb}{\rab^2}$ & $\dfrac{c\pi^2(\al^6-11\al^4\beta^2+7\al^2\beta^4+3\beta^6+4\al^2\beta^2(\al^2+5\beta^2)log(\frac{\al}{\beta}))}{\al\beta(\al^2-\beta^2)^4}$ \\ 
          34 & $-\dfrac{c\al\rb\cst}{\rab^2}$ & $\dfrac{c\pi^2(-4\beta^2(\al^4+4\al^2\beta^2-5\beta^4)+16(2\al^2\beta^4+\beta^6)log(\frac{\al}{\beta}))}{2\beta^3(\al^2-\beta^2)^4}$ \\ 
          35 & $-\dfrac{c\beta\ra\cst}{\rab^2}$ & $\dfrac{c\pi^2(2(\al-\beta)\beta^3(\al+\beta)(5\al^2+\beta^2)+8\al^2\beta^3(\al^2+2\beta^2)log(\frac{\beta}{\alpha}))}{\al\beta^2(\al^2-\beta^2)^4}$\\ 
          36 & $-\dfrac{2c}{\rab^2}$ & $-\dfrac{4c\pi^2}{3\al^4}$ \\ 
\end{longtable}

\subsubsection*{Calculating $\bkt{\mc{H}^2}{a2a1}$:}
\begin{align*}
    \bkt{\mc{H}^2}{a2a1} & = (\bra{\psi_{a2}}\mc{H}).(\mc{H}\ket{\psi_{a1}}) = (\bra{\psi_{a1}}\mc{H}).(\mc{H}\ket{\psi_{a2}}) = \bkt{\mc{H}^2}{a1a2} ~, \ \mr{for \ real \ wavefunctions.}
\end{align*}

\subsubsection*{Calculating $\bkt{\mc{H}^2}{a2a2}$:}
\begin{align*}
    \bkt{\mc{H}^2}{a2a2} = & (\bra{\psi_{a2}}\mc{H}).(\mc{H}\ket{\psi_{a2}}) \\
    = & \int dr_1^3 dr_2^3 \psi_{a1}^2 c^2 \left( -\frac{\al^2+\beta^2}{2}r_{12}+\frac{(\al-Z)}{r_1}r_{12}+\frac{(\beta-Z)}{r_2}r_{12}+1 +\frac{\al r_1}{r_{12}} +\frac{\beta r_2}{r_{12}}-\frac{\al r_2 cos(\theta)}{r_{12}} -\frac{\beta r_1 cos(\theta)}{r_{12}}-\frac{2}{r_{12}} \right)^2\\
    = & \sum_i Hs_{a2a2,i} ~, \quad \mr{where} \quad Hs_{a2a2,i} = \int dr_1^3 dr_2^3 \psi_{a1}^2 hs_{a2a2,i}
\end{align*} 

\begin{longtable}{|c|c|c|}
\caption{Calculations for $\bkt{\mc{H}^2}{a2a2}$} \\

\hline
$i$  & $hs_{a2a2,i}$ & $Hs_{a2a2,i}$ \\ 
\hline
\endfirsthead

\hline
$i$  & $hs_{a2a2,i}$ & $Hs_{a2a2,i}$ \\ 
\hline
\endhead

\hline
\endfoot

\hline
\endlastfoot

          1 & $c^2\left(\dfrac{\al^2+\beta^2}{2}\right)^2\rab^2$ & $\dfrac{3c^2\pi^2(\al^2+\beta^2)^3}{4\al^5\beta^5}$ \\ 
          2 & $c^2\left(\dfrac{\al-Z}{r_1}\right)^2\rab^2$ & $\dfrac{c^2\pi^2(\al-Z)^2(6\al^2+\beta^2)}{\al^3\beta^5}$ \\ 
          3 & $c^2\left(\dfrac{\beta-Z}{r_2}\right)^2\rab^2$ & $\dfrac{c^2\pi^2(\beta-Z)^2(\al^2+6\beta^2)}{\al^5\beta^3}$ \\ 
          4 & $c^2$ & $\dfrac{c^2\pi^2}{\al^3\beta^3}$  \\ 
          5 & $\left(\dfrac{c\al\ra}{\rab}\right)^2$ &  $\dfrac{c^2\pi^2(6\al^8+37\al^6\beta^2-45\al^4\beta^4+3\al^2\beta^6-\beta^8+12(5\al^6\beta^2+3\al^4\beta^4) log(\frac{\beta}{\al})}{\al\beta(\al^2-\beta^2)^5}$ \\ 
          6 & $\left(\dfrac{c\beta\rb}{\rab}\right)^2$ &  $\dfrac{c^2\pi^2(2\al^8-6\al^6\beta^2+90\al^4\beta^4-74\al^2\beta^6-12\beta^8+24(3\al^4\beta^4+5\al^2\beta^6) log(\frac{\beta}{\al})}{2\al\beta(\al^2-\beta^2)^5}$\\ 
          7 & $\left(\dfrac{2 c}{\rab}\right)^2$ &  $\dfrac{8c^2\pi^2(\al^4-\beta^4+4\al^2\beta^2 log(\frac{\beta}{\al})}{\al\beta(\al^2-\beta^2)}$\\ 
          8 & $-c^2(\al^2+\beta^2)\left(\dfrac{\al-Z}{\ra}\right)\rab^2$ &  $\dfrac{3c^2\pi^2(Z-\al)(\al^2+\beta^2)(2\al^2+\beta^2)}{2\al^4\beta^5}$\\ 
          9 & $-c^2(\al^2+\beta^2)\left(\dfrac{\beta-Z}{\ra}\right)\rab^2$ &  $\dfrac{3c^2\pi^2(Z-\beta)(\al^2+\beta^2)(\al^2+2\beta^2)}{2\al^5\beta^4}$\\ 
          10 & $-c^2(\al^2+\beta^2)\rab$ &  $-\dfrac{c^2\pi^2(\al^2+\beta^2)}{2\beta^5}\left(\dfrac{2}{\al^2}+\dfrac{3\beta^2}{\al^4}+\dfrac{-2\al-3\beta}{(\al+\beta)^3}\right)$\\ 
          11 & $-c^2(\al^2+\beta^2) \al \ra$ &  $-\dfrac{3 c^2\pi^2(\al^2+\beta^2)}{2\al^3\beta^3}$\\ 
          12 & $-c^2(\al^2+\beta^2) \beta \rb$ &  $-\dfrac{3 c^2\pi^2(\al^2+\beta^2)}{2\al^3\beta^3}$\\ 
          13 & $c^2(\al^2+\beta^2) \al\rb\cst$ &  $0$\\ 
          14 & $c^2(\al^2+\beta^2) \beta\ra\cst$ &  $0$\\ 
          15 & $2c^2(\al^2+\beta^2) $ &  $\dfrac{2c^2\pi^2(\al^2+\beta^2}{\al^3\beta^3}$\\ 
          16 & $2c^2\left(\dfrac{\al-Z}{\ra}\right)\left(\dfrac{\beta-Z}{\rb}\right)\rab^2$ & $\dfrac{3c^2\pi^2(\al-Z)(\beta-Z)(\al^2+\beta^2)}{\al^4\beta^4}$ \\ 
          17 & $2c^2\left(\dfrac{\al-Z}{\ra}\right)\rab$ & $\dfrac{c^2\pi^2(\al-Z)(3\al^3+6\al^2\beta+4\al\beta^2+2\beta^3}{\al^3\beta^4(\al+\beta)^2}$\\ 
          18 & $2c^2\al(\al-Z)$ & $\dfrac{2c^2\pi^2(\al-Z)}{\al^2\beta^3}$\\ 
          19 & $2c^2\left(\dfrac{\al-Z}{\ra}\right)\beta\rb$ & $\dfrac{3c^2\pi^2(\al-Z)}{\al^2\beta^3}$ \\ 
          20 & $-2c^2\left(\dfrac{\al-Z}{\ra}\right)\al\rb\cst$ & $0$ \\ 
          21 & $-2c^2(\al-Z)\beta\cst$ & $0$ \\ 
          22 & $-4c^2\left(\dfrac{\al-Z}{\ra}\right)$ & $\dfrac{4c^2\pi^2(Z-\al)}{\al^2\beta^3}$ \\ 
          23 & $2c^2\left(\dfrac{\beta-Z}{\rb}\right)\rab$ & $\dfrac{c^2\pi^2(\beta-Z)}{\beta^4}\left(\dfrac{1}{\al^2}+\dfrac{3\beta^2}{\al^4}-\dfrac{1}{(\al+\beta)^2}\right)$ \\ 
          24 & $2c^2\left(\dfrac{\beta-Z}{\rb}\right)\al\ra$ & $\dfrac{3c^2\pi^2(\beta-Z)}{\al^3\beta^2}$ \\ 
          25 & $2c^2\beta(\beta-Z)$ & $\dfrac{2c^2\pi^2(\beta-Z)}{\al^3\beta^2}$  \\ 
          26 & $-2c^2(\beta-Z)\al\cst$ & $0$ \\ 
          27 & $-2c^2\left(\dfrac{\beta-Z}{\rb}\right)\beta\ra\cst$ & $0$ \\ 
          28 & $-4c^2\left(\dfrac{\beta-Z}{\rb}\right)$ & $\dfrac{4c^2\pi^2(Z-\beta)}{\al^3\beta^2}$  \\ 
          29 & $\dfrac{2c^2\al\ra}{\rab}$ & $-\dfrac{c^2\pi^2\al}{\beta^3}\left(-\dfrac{2}{\al^3}+\dfrac{2\al+5\beta}{(\al+\beta)^4}\right)$ \\ 
          30 & $\dfrac{2c^2\beta\rb}{\rab}$ & $\dfrac{c^2\pi^2(2\al^3+8\al^2\beta+12\al\beta^2+3\beta^3)}{\al^2\beta^2(\al+\beta)^4}$ \\ 
          31 & $\dfrac{-2c^2\al\rb\cst}{\rab}$ & $-\dfrac{c^2\pi^2(\al+4\beta)}{\beta^2(\al+\beta)^4}$ \\ 
          32 & $\dfrac{-2c^2\beta\ra\cst}{\rab}$ & $-\dfrac{c^2\pi^2(4\al+\beta)}{\al^2(\al+\beta)^4}$ \\ 
          33 & $-\dfrac{4c^2}{\rb^2}$ & $-\dfrac{4c^2\pi^2(\al^2+3\al\beta+\beta^2)}{\al^2\beta^2(\al+\beta)^3}$ \\ 
          34 & $\dfrac{2c^2\al\beta\ra\rb}{\rab^2}$ & $\dfrac{c^2\pi^2(3(\al^8-18\al^6\beta^2+18\al^2\beta^6-\beta^8)+4(5\al^6\beta^2+38\al^4\beta^4+5\al^2\beta^6) log(\frac{\al}{\beta}))}{\al\beta(\al^2-\beta^2)^5}$ \\ 
          35 & $-\dfrac{2c^2\al^2\ra\rb\cst}{\rab^2}$ & $-\dfrac{c^2\pi^2\al(8(\al-\beta)\beta^3(\alpha+\beta)(\al^4+10\al^2\beta^2+\beta^4)+96\al^2\beta^5(\al^2+\beta^2) log(\frac{\beta}{\al}))}{\beta^4(\al^2-\beta^2)^5}$ \\ 
          36 & $-\dfrac{2c^2\al\beta\ra^2\cst}{\rab^2}$ & $\dfrac{2c^2\pi^2\beta(29\al^6-9\al^4\beta^2-21\al^2\beta^4+\beta^6+4(5\al^6+17\al^4\beta^2+2\al^2\beta^4) log(\frac{\beta}{\al}))}{\al(\al-\beta)^5(\al+\beta)^5}$ \\ 
          37 & $-\dfrac{4c^2\al\ra}{\rab^2}$ & $-\dfrac{4c^2\pi^2(3\al^6+7\al^4\beta^2-11\al^2\beta^4+\beta^6+4\al^2
        \beta^2 (5\al^2+\beta^2) log(\frac{\beta}{\al}))}{\al\beta(\al^2-\beta^2)^4}$ \\ 
          38 & $-\dfrac{2c^2\al\beta\rb^2\cst}{\rab^2}$ & $\dfrac{c^2\pi^2\al(-2\al^6\beta^2+42\al^4\beta^4+18\al^2\beta^6-58\beta^8)+8(2\al^4\beta^4+17\al^2\beta^6+5\beta^8) log(\frac{\beta}{\al}))}{\beta^3(\al^2-\beta^2)^5}$ \\ 
          39 & $-\dfrac{2c^2\beta^2\ra\rb\cst}{\rab^2}$ & $-\dfrac{8c^2\pi^2\beta(\al^6+9\al^4\beta^2-9\al^2\beta^4-\beta^6+12\al^2\beta^2(\al^2+\beta^2)log(\frac{\beta}{\al}))}{\al(\al^2-\beta^2)^5}$ \\ 
          40 & $-\dfrac{4c^2\beta\rb}{\rab^2}$ & $-\dfrac{4c^2\pi^2(\al^6-11\al^4\beta^2+7\al^2\beta^4 +3\beta^6 +4\al^2\beta^2(\al^2+5\beta^2)log(\frac{\al}{\beta}))}{\al\beta(\al^2-\beta^2)^4}$ \\ 
          41 & $-\dfrac{4c^2\al\rb\cst}{\rab^2}$ & $\dfrac{2c^2\pi^2\al(4\beta^2(\al^4+4\al^2\beta^2-5\beta^4)+16(2\al^2\beta^4+\beta^6)log(\frac{\beta}{\al}))}{\beta^3(\al^2-\beta^2)^4}$ \\ 
          42 & $-\dfrac{4c^2\beta\ra\cst}{\rab^2}$ & $\dfrac{4c^2\pi^2(2\beta^2(-5\al^4+4\al^2\beta^2+\beta^4) +8\al^2\beta^2(\al^2+2\beta^2)log(\frac{\al}{\beta}))}{\al\beta(\al^2-\beta^2)^4}$ \\ 
  \end{longtable}

This table omits the $hs_{a2a2,i}$ terms containing $\tfrac{\cst^2}{\rab^2} $, such as $\left(\tfrac{c\al\rb\cst}{\rab}\right)^2$, $\left(\tfrac{c\beta\ra\cst}{\rab}\right)^2$ and $\tfrac{2c^2\al\beta\ra\rb\cst^2}{\rab^2}$, because they are difficult to integrate individually. They will be grouped together with similar terms from the following integrals, and then integrated using suitable substitutions.

\subsubsection*{Calculating $\bkt{\mc{H}^2}{b1a1}$:}
\begin{align*}
    \bkt{\mc{H}^2}{b1a1}  = & (\bra{\psi_{b1}}\mc{H}).(\mc{H}\ket{\psi_{a1}}) = \int dr_1^3 dr_2^3 \psi_{a1} \psi_{b1} \left[ \left( -\frac{\al^2+\beta^2}{2}+\frac{\al-Z}{r_1}+\frac{\beta-Z}{r_2}+\frac{1}{r_{12}} \right)  \right. \\
    & \times \left. \left( -\frac{\al^2+\beta^2}{2}+\frac{\beta-Z}{r_1}+\frac{\al-Z}{r_2}+\frac{1}{r_{12}} \right) \right] \\
    = & \sum_i Hs_{b1a1,i} ~, \quad \mr{where} \quad Hs_{b1a1,i} = \int dr_1^3 dr_2^3 \psi_{a1}\psi_{b1} hs_{b1a1,i}
\end{align*}

\begin{longtable}{|c|c|c|}
\caption{Calculations for $\bkt{\mc{H}^2}{b1a1}$} \\

\hline
$i$  & $hs_{b1a1,i}$ & $Hs_{b1a1,i}$ \\ 
\hline
\endfirsthead

\hline
$i$  & $hs_{b1a1,i}$ & $Hs_{b1a1,i}$ \\ 
\hline
\endhead

\hline
\endfoot

\hline
\endlastfoot
          1 & $\left(\dfrac{\al^2+\beta^2}{2}\right)^2$ & $\dfrac{16\pi^2(\al^2+\beta^2)^2}{(\al+\beta)^6}$ \\ 
          2 & $-\left(\dfrac{\al^2+\beta^2}{2}\right)\left(\dfrac{\al-Z}{r_1}\right)$ & $-\dfrac{16\pi^2(\al-Z)(\al^2+\beta^2)}{(\al+\beta)^5}$ \\ 
          3 & $-\left(\dfrac{\al^2+\beta^2}{2}\right)\left(\dfrac{\beta-Z}{r_2}\right)$ & $-\dfrac{16\pi^2(\beta-Z)(\al^2+\beta^2)}{(\al+\beta)^5}$ \\ 
          4 & $-\left(\dfrac{\al^2+\beta^2}{2}\right)\dfrac{1}{\rab}$ & $-\dfrac{10\pi^2(\al^2+\beta^2)}{(\al+\beta)^5}$  \\ 
          5 & $-\left(\dfrac{\al^2+\beta^2}{2}\right)\left(\dfrac{\beta-Z}{\ra}\right)$ & $\dfrac{16\pi^2(Z-\beta)(\al^2+\beta^2)}{(\al+\beta)^5}$ \\ 
          6 & $\dfrac{(\beta-Z)(\al-Z)}{\ra^2}$ & $\dfrac{32\pi^2(\beta-Z)(\al-Z)}{(\al+\beta)^4}$\\ 
          7 & $\dfrac{(\beta-Z)^2}{\ra \rb}$ & $\dfrac{16\pi^2(\beta-Z)^2}{(\al+\beta)^4}$\\ 
          8 & $\dfrac{(\beta-Z)}{\ra \rab}$ & $\dfrac{12 \pi^2(\beta-Z)}{(\al+\beta)^4}$\\ 
          9 & $-\left(\dfrac{\al^2+\beta^2}{2}\right)\left(\dfrac{\al-Z}{\rb}\right)$ & $-\dfrac{16 \pi^2(\al-Z)(\al^2+\beta^2)}{(\al+\beta)^5}$\\ 
          10 & $\dfrac{(\al-Z)^2}{\ra \rb}$ & $\dfrac{16 \pi^2(\al - Z)^2}{(\al+\beta)^4}$\\ 
          11 & $\dfrac{(\al-Z)(\beta-Z)}{\rb^2}$ & $\dfrac{32 \pi^2(\al-Z)(\beta - Z)}{(\al+\beta)^4}$\\ 
          12 & $\dfrac{(\al-Z)}{\rb \rab}$ & $\dfrac{12 \pi^2(\al - Z)}{(\al+\beta)^4}$\\ 
          13 & $-\left(\dfrac{\al^2+\beta^2}{2\rab}\right)$ & $-\dfrac{10 \pi^2(\al^2+\beta^2)}{(\al+\beta)^5}$\\ 
          14 & $\dfrac{(\al-Z)}{\ra \rab}$ & $\dfrac{12 \pi^2(\al - Z)}{(\al+\beta)^4}$\\ 
          15 & $\dfrac{(\beta-Z)}{\rb \rab}$ & $\dfrac{12 \pi^2(\beta - Z)}{(\al+\beta)^4}$\\ 
          16 & $\dfrac{1}{\rab^2}$ & $\dfrac{32 \pi^2}{3(\al+\beta)^4}$
  \end{longtable}

\subsubsection*{Calculating $\bkt{\mc{H}^2}{b2a1}$:}
\begin{align*}
    \bkt{\mc{H}^2}{b2a1}  = & (\bra{\psi_{b2}}\mc{H}).(\mc{H}\ket{\psi_{a1}}) = \int dr_1^3 dr_2^3 \psi_{a1} \psi_{b1} c\left[ \left( -\left( \frac{\al^2+\beta^2}{2}\right)r_{12}+\left(\frac{\beta-Z} {r_1}\right)r_{12}+\left(\frac{\al-Z}{r_2}\right)r_{12}+1 \right. \right. \\ 
    & \left. \left.+ \frac{\beta r_1}{r_{12}} +\frac{\al r_2}{r_{12}}-\frac{\beta r_2 cos(\theta)}{r_{12}} -\frac{\al r_1 cos(\theta)}{r_{12}}-\frac{2}{r_{12}}  \right)
     \times \left( -\frac{\al^2+\beta^2}{2}+\frac{\al-Z}{r_1}+\frac{\beta-Z}{r_2}+\frac{1}{r_{12}} \right) \right] \\
    = & \sum_i Hs_{b2a1,i} , \quad \mr{where} \quad Hs_{b2a1,i} = \int dr_1^3 dr_2^3 \psi_{a1}\psi_{b1} hs_{b2a1,i}
\end{align*}

\begin{longtable}{|c|c|c|}
\caption{Calculations for $\bkt{\mc{H}^2}{b2a1}$} \\

\hline
$i$  & $hs_{b2a1,i}$ & $Hs_{b2a1,i}$ \\ 
\hline
\endfirsthead

\hline
$i$  & $hs_{b2a1,i}$ & $Hs_{b2a1,i}$ \\ 
\hline
\endhead

\hline
\endfoot

\hline
\endlastfoot

          1 & $c\left(\dfrac{\al^2+\beta^2}{2}\right)^2\rab$ & $\dfrac{70 c\pi^2(\al^2+\beta^2)^2}{(\al+\beta)^7}$ \\ 
          2 & $-c\left(\dfrac{\al^2+\beta^2}{2}\right)\left(\dfrac{\al-Z}{r_1}\right)\rab$ & $-\dfrac{60c\pi^2(\al-Z)(\al^2+\beta^2)}{(\al+\beta)^6}$ \\ 
          3 & $-c\left(\dfrac{\al^2+\beta^2}{2}\right)\left(\dfrac{\beta-Z}{r_2}\right)\rab$ & $-\dfrac{60c\pi^2(\beta-Z)(\al^2+\beta^2)}{(\al+\beta)^6}$ \\ 
          4 & $-c\left(\dfrac{\al^2+\beta^2}{2}\right)$ & $-\dfrac{32c\pi^2(\al^2+\beta^2)}{(\al+\beta)^6}$  \\ 
          5 & $-c\left(\dfrac{\al^2+\beta^2}{2}\right)\left(\dfrac{\beta-Z}{\ra}\right)\rab$ & $\dfrac{60c\pi^2(Z-\beta)(\al^2+\beta^2)}{(\al+\beta)^6}$ \\ 
          6 & $c\dfrac{(\beta-Z)(\al-Z)}{\ra^2}\rab$ & $\dfrac{16c\pi^2(\beta-Z)(\al-Z)(1+log(256))}{(\al+\beta)^5}$\\ 
          7 & $c\dfrac{(\beta-Z)^2}{\ra \rb}\rab$ & $\dfrac{48c\pi^2(\beta-Z)^2}{(\al+\beta)^5}$\\ 
          8 & $c\left(\dfrac{\beta-Z}{\ra}\right)$ & $\dfrac{32 c\pi^2(\beta-Z)}{(\al+\beta)^5}$\\ 
          9 & $-c\left(\dfrac{\al^2+\beta^2}{2}\right)\left(\dfrac{\al-Z}{\rb}\right)\rab$ & $-\dfrac{60c \pi^2(\al-Z)(\al^2+\beta^2)}{(\al+\beta)^6}$\\ 
          10 & $c\dfrac{(\al-Z)^2}{\ra \rb}\rab$ & $\dfrac{48c\pi^2(\al - Z)^2}{(\al+\beta)^5}$\\ 
          11 & $c\dfrac{(\al-Z)(\beta-Z)}{\rb^2}\rab$ & $\dfrac{16c\pi^2(\al-Z)(\beta - Z)(1+log(256))}{(\al+\beta)^5}$\\ 
          12 & $c\left(\dfrac{\al-Z}{\rb}\right)$ & $\dfrac{32c\pi^2(\al - Z)}{(\al+\beta)^5}$\\ 
          13 & $-c\left(\dfrac{\al^2+\beta^2}{2}\right)$ & $-\dfrac{32c\pi^2(\al^2+\beta^2)}{(\al+\beta)^6}$\\ 
          14 & $c\left(\dfrac{\al-Z}{\ra}\right)$ & $\dfrac{32c\pi^2(\al - Z)}{(\al+\beta)^5}$\\ 
          15 & $c\left(\dfrac{\beta-Z}{\rb}\right)$ & $\dfrac{32c\pi^2(\beta - Z)}{(\al+\beta)^5}$\\ 
          16 & $\dfrac{c}{\rab}$ & $\dfrac{20c\pi^2}{(\al+\beta)^5}$ \\ 
          17 & $-\dfrac{c\beta\ra}{\rab}\left(\dfrac{\al^2+\beta^2}{2}\right)$ & $-\dfrac{25c\pi^2\beta(\al^2+\beta^2)}{(\al+\beta)^6}$\\ 
          18 & $c\beta\left(\dfrac{\al-Z}{\ra}\right)$ & $\dfrac{20c\pi^2(\al - Z)\beta}{(\al+\beta)^5}$\\ 
          19 & $\dfrac{c\beta\ra}{\rab}\left(\dfrac{\beta-Z}{\rb}\right)$ & $\dfrac{28c\pi^2\beta(\beta - Z)}{(\al+\beta)^5}$\\ 
          20 & $\dfrac{c\beta\ra}{\rab^2}$ & $\dfrac{64c\pi^2\beta}{3(\al+\beta)^5}$ \\ 
          21 & $-\dfrac{c\al\rb}{\rab}\left(\dfrac{\al^2+\beta^2}{2}\right)$ & $-\dfrac{25c\pi^2\al(\al^2+\beta^2)}{(\al+\beta)^6}$\\ 
          22 & $\dfrac{c\al\rb}{\rab}\left(\dfrac{\al-Z}{\ra}\right)$ & $\dfrac{28c\pi^2\al(\al - Z)}{(\al+\beta)^5}$\\ 
          23 & $\dfrac{c\al}{\rab}(\beta-Z)$ & $\dfrac{20c\pi^2\al(\beta - Z)}{(\al+\beta)^5}$\\ 
          24 & $\dfrac{c\al\rb}{\rab}$ & $\dfrac{64c\pi^2}{3(\al+\beta)^5}$ \\ 
          25 & $c\left(\dfrac{\al^2+\beta^2}{2}\right)\dfrac{\beta\rb\cst}{\rab}$ & $\dfrac{5c\pi^2\beta(\al^2+\beta^2)}{(\al+\beta)^6}$\\ 
          26 & $-c\left(\dfrac{\al-Z}{\ra}\right)\dfrac{\beta\rb\cst}{\rab}$ & $\dfrac{4c\pi^2(\al - Z)\beta(21-32log(2))}{(\al+\beta)^5}$\\ 
          27 & $-c(\beta-Z)\dfrac{\beta\cst}{\rab}$ & $-\dfrac{4c\pi^2\beta(\beta - Z)}{(\al+\beta)^5}$\\ 
          28 & $-\dfrac{c\beta\rb\cst}{\rab^2}$ & $-\dfrac{32c\pi^2\beta}{3(\al+\beta)^5}$ \\ 
          29 & $c\left(\dfrac{\al^2+\beta^2}{2}\right)\dfrac{\al\ra\cst}{\rab}$ & $\dfrac{5c\pi^2\al(\al^2+\beta^2)}{(\al+\beta)^6}$\\ 
          30 & $-c(\al-Z)\dfrac{\al\cst}{\rab}$ & $-\dfrac{4c\pi^2\al(\al - Z)}{(\al+\beta)^5}$\\ 
          31 & $-c(\beta-Z)\dfrac{\al\ra\cst}{\rb\rab}$ & $\dfrac{4c\pi^2\al(\beta - Z)(21-32log(2))}{(\al+\beta)^5}$\\ 
          32 & $-\dfrac{c\al\ra\cst}{\rab^2}$ & $-\dfrac{32c\pi^2\al}{(\al+\beta)^5}$ \\ 
          33 & $c\left(\dfrac{\al^2+\beta^2}{\rab}\right)$ & $\dfrac{20c\pi^2(\al^2+\beta^2)}{(\al+\beta)^5}$\\ 
          34 & $-\dfrac{2c}{\rab}\left(\dfrac{\al-Z}{\ra}\right)$ & $-\dfrac{24c\pi^2(\al - Z)}{(\al+\beta)^4}$\\ 
          35 & $-\dfrac{2c}{\rab}\left(\dfrac{\beta-Z}{\rb}\right)$ & $-\dfrac{24c\pi^2(\beta - Z)}{(\al+\beta)^4}$ \\ 
          36 & $-\dfrac{2c}{\rab^2}$ & $-\dfrac{64c\pi^2}{3(\al+\beta)^4}$ \\ 
  \end{longtable}

\subsubsection*{Calculating $\bkt{\mc{H}^2}{b1a2}$:}
    $\bkt{\mc{H}^2}{b1a2} = (\bra{\psi_{b1}}\mc{H}).(\mc{H}\ket{\psi_{a2}})  = (\bra{\psi_{a2}}\mc{H}).(\mc{H}\ket{\psi_{b1}})$ is 
    $\bkt{\mc{H}^2}{b2a1} = (\bra{\psi_{b2}}\mc{H}).(\mc{H}\ket{\psi_{a1}})$ with $\al$ and $\beta$ interchanged:
    \[ \bkt{\mc{H}^2}{b1a2}(\al,\beta,c) = \bkt{\mc{H}^2}{b2a1}(\beta,\al,c) \]

\subsubsection*{Calculating $\bkt{\mc{H}^2}{b2a2}$:}
\begin{align*}
    \bkt{\mc{H}^2}{b2a2}  = & (\bra{\psi_{b2}}\mc{H}).(\mc{H}\ket{\psi_{a2}}) = \int dr_1^3 dr_2^3 \psi_{a1} \psi_{b1} c^2\left[ \left( -\left( \frac{\al^2+\beta^2}{2}\right)r_{12}+\left(\frac{\beta-Z} {r_1}\right)r_{12}+\left(\frac{\al-Z}{r_2}\right)r_{12}+1 \right. \right. \\ 
    & \left. \left. + \frac{\beta r_1}{r_{12}} +\frac{\al r_2}{r_{12}}  -\frac{\beta r_2 cos(\theta)}{r_{12}} -\frac{\al r_1 cos(\theta)}{r_{12}}-\frac{2}{r_{12}}  \right)
     \times \left(-\left(\frac{\al^2+\beta^2}{2}\right)r_{12}+\left(\frac{\al-Z}{r_1}\right)r_{12}+\left(\frac{\beta-Z}{r_2}\right)r_{12} \right. \right. \\ 
    & \left. \left.  +1 +\frac{\al r_1}{r_{12}} +\frac{\beta r_2}{r_{12}}-\frac{\al r_2 cos(\theta)}{r_{12}} -\frac{\beta r_1 cos(\theta)}{r_{12}} -\frac{2}{r_{12}} \right)  \right] \\
    = & \sum_i Hs_{b2a2,i} ~, \quad \mr{where} \quad Hs_{b2a2,i} = \int dr_1^3 dr_2^3 \psi_{a1}\psi_{b1} hs_{b2a2,i}
\end{align*}

\begin{longtable}{|c|c|c|}

\caption{Calculations for $\bkt{\mc{H}^2}{b2a2}$} \\

\hline
$i$  & $hs_{b2a2,i}$ & $Hs_{b2a2,i}$ \\
\hline
\endfirsthead

\hline
$i$  & $hs_{b2a2,i}$ & $Hs_{b2a2,i}$ \\ 
\hline
\endhead

\hline
\endfoot

\hline
\endlastfoot
        
          1 & $c^2\left(\dfrac{\al^2+\beta^2}{2}\right)^2\rab^2$ & $\dfrac{384 c^2\pi^2(\al^2+\beta^2)^2}{(\al+\beta)^8}$ \\ 
          2 & $-c^2\left(\dfrac{\al^2+\beta^2}{2}\right)\left(\dfrac{\beta-Z}{r_1}\right)\rab^2$ & $-\dfrac{288c^2\pi^2(\beta-Z)(\al^2+\beta^2)}{(\al+\beta)^7}$ \\ 
          3 & $-c^2\left(\dfrac{\al^2+\beta^2}{2}\right)\left(\dfrac{\al-Z}{r_2}\right)\rab^2$ & $-\dfrac{288c^2\pi^2(\alpha-Z)(\al^2+\beta^2)}{(\al+\beta)^7}$ \\ 
          4 & $-c^2\left(\dfrac{\al^2+\beta^2}{2}\right)\rab$ & $-\dfrac{140c^2\pi^2(\al^2+\beta^2)}{(\al+\beta)^7}$  \\ 
          5 & $-c^2\left(\dfrac{\al^2+\beta^2}{2}\right)\beta\ra$ & $-\dfrac{96c^2\pi^2\beta(\al^2+\beta^2)}{(\al+\beta)^7}$ \\ 
          6 & $-c^2\left(\dfrac{\al^2+\beta^2}{2}\right)\al\rb$ & $-\dfrac{96c^2\pi^2\al(\al^2+\beta^2)}{(\al+\beta)^7}$\\ 
          7 & $c^2\left(\dfrac{\al^2+\beta^2}{2}\right)\beta\rb\cst$ & $0$\\ 
          8 & $c^2\left(\dfrac{\al^2+\beta^2}{2}\right)\al\ra\cst$ & $0$\\ 
          9 & $c^2(\al^2+\beta^2)$ & $\dfrac{64c^2\pi^2(\al^2+\beta^2)}{(\al+\beta)^6}$\\ 
          10 & $-c^2\left(\dfrac{\al-Z}{\ra}\right)\left(\dfrac{\al^2+\beta^2}{2}\right)\rab^2$ & $-\dfrac{288c^2\pi^2(\al - Z)(\al^2+\beta^2)}{(\al+\beta)^7}$\\ 
          11 & $c^2\left(\dfrac{\al-Z}{\ra}\right)\left(\dfrac{\beta-Z}{\ra}\right)\rab^2$ & $\dfrac{448c^2\pi^2(\al-Z)(\beta - Z)}{(\al+\beta)^6}$\\ 
          12 & $c^2\left(\dfrac{\al-Z}{\ra}\right)\left(\dfrac{\al-Z}{\rb}\right)\rab^2$ & $\dfrac{192c^2\pi^2(\al - Z)^2}{(\al+\beta)^6}$\\ 
          13 & $c^2\left(\dfrac{\al-Z}{\ra}\right)\rab$ & $\dfrac{120c^2\pi^2(\al-Z)}{(\al+\beta)^6}$\\ 
          14 & $c^2(\al-Z)\beta$ & $\dfrac{64c^2\pi^2(\al - Z)\beta}{(\al+\beta)^6}$\\ 
          15 & $c^2\left(\dfrac{\al-Z}{\ra}\right)\al \rb$ & $\dfrac{96c^2\pi^2\al(\al - Z)}{(\al+\beta)^6}$\\ 
          16 & $\dfrac{c}{\rab}$ & $0$ \\ 
          17 & $-c^2\al(\al-Z)\cst$ & $0$\\ 
          18 & $-2c^2\left(\dfrac{\al-Z}{\ra}\right)$ & $-\dfrac{64c^2\pi^2(\al - Z)}{(\al+\beta)^5}$\\ 
          19 & $-c^2\left(\dfrac{\al^2+\beta^2}{2}\right)\left(\dfrac{\beta-Z}{\rb}\right)\rab^2$ & $-\dfrac{288c^2\pi^2(\beta - Z)(\al^2+\beta^2)}{(\al+\beta)^7}$\\ 
          20 & $c^2\left(\dfrac{\beta-Z}{\ra}\right)\left(\dfrac{\beta-Z}{\rb}\right)\rab^2$ & $\dfrac{6192c^2\pi^2(\beta-Z)^2}{(\al+\beta)^6}$ \\ 
          21 & $c^2\left(\dfrac{\al-Z}{\rb}\right)\left(\dfrac{\beta-Z}{\rb}\right)\rab^2$ & $\dfrac{448c^2\pi^2(\al-Z)(\beta-Z)}{(\al+\beta)^6}$\\ 
          22 & $c^2\left(\dfrac{\beta-Z}{\rb}\right)\rab$ & $\dfrac{120c^2\pi^2(\beta - Z)}{(\al+\beta)^6}$\\ 
          23 & $c^2\left(\dfrac{\beta-Z}{\rb}\right)\beta\ra$ & $\dfrac{96c^2\pi^2\beta(\beta - Z)}{(\al+\beta)^6}$\\ 
          24 & $c^2(\beta-Z)\al$ & $\dfrac{64c^2\pi^2\al(\beta-Z)}{(\al+\beta)^6}$ \\ 
          25 & $-c^2\beta(\beta-Z)\cst$ & $0$\\ 
          26 & $-c^2\left(\dfrac{\beta-Z}{\rb}\right)\al\ra\cst$ & $0$\\ 
          27 & $-2c^2\left(\dfrac{\beta-Z}{\rb}\right)$ & $-\dfrac{64c^2\pi^2(\beta - Z)}{(\al+\beta)^5}$\\ 
          28 & $-c^2\left(\dfrac{\al^2+\beta^2}{2}\right)\rab$ & $-\dfrac{140c^2\pi^2(\al^2+\beta^2)}{(\al+\beta)^7}$ \\ 
          29 & $c^2\left(\dfrac{\beta-Z}{\ra}\right)\rab$ & $\dfrac{120c^2\pi^2(\beta-Z)}{(\al+\beta)^6}$\\ 
          30 & $c^2\left(\dfrac{\al-Z}{\rb}\right)\rab$ & $-\dfrac{120c^2\pi^2(\al - Z)}{(\al+\beta)^6}$\\ 
          31 & $c^2$ & $\dfrac{64c^2\pi^2}{(\al+\beta)^6}$\\ 
          32 & $c^2\dfrac{\beta\ra}{\rab}$ & $\dfrac{50c^2\pi^2\beta}{(\al+\beta)^6}$ \\ 
          33 & $c^2\dfrac{\al\rb}{\rab}$ & $\dfrac{50c^2\pi^2\al}{(\al+\beta)^6}$\\ 
          34 & $-c^2\dfrac{\beta\rb\cst}{\rab}$ & $-\dfrac{10c^2\pi^2\beta}{(\al+\beta)^6}$\\ 
          35 & $-c^2\dfrac{\al\ra\cst}{\rab}$ & $-\dfrac{10c^2\pi^2\al}{(\al+\beta)^6}$\\ 
          36 & $-\dfrac{2c^2}{\rab}$ & $-\dfrac{40c^2\pi^2}{(\al+\beta)^5}$ \\ 
          37 & $-c^2\left(\dfrac{\al^2+\beta^2}{2}\right)\al\ra$ & $-\dfrac{96c^2\pi^2\al(\al^2+\beta^2)}{(\al+\beta)^7}$ \\ 
          38 & $c^2(\beta-Z)\al$ & $\dfrac{64c^2\pi^2\al(\beta-Z)}{(\al+\beta)^6}$\\ 
          39 & $c^2\left(\dfrac{\al-Z}{\rb}\right)\al\ra$ & $\dfrac{96c^2\pi^2\al(\al - Z)}{(\al+\beta)^6}$\\ 
          40 & $c^2\dfrac{\al\ra}{\rab}$ & $\dfrac{50c^2\pi^2\al}{(\al+\beta)^6}$\\ 
          41 & $c^2\dfrac{\al\beta\ra^2}{\rab^2}$ & $\dfrac{288c^2\pi^2\al\beta}{5(\al+\beta)^6}$ \\ 
          42 & $c^2\dfrac{\al^2\ra\rb}{\rab^2}$ & $\dfrac{736c^2\pi^2\al^2}{15(\al+\beta)^6}$\\ 
          43 & $-c^2\dfrac{\al\beta\ra\rb\cst}{\rab^2}$ & $-\dfrac{128c^2\pi^2\al\beta}{5(\al+\beta)^6}$\\ 
          44 & $-c^2\dfrac{\al^2\ra^2\cst}{\rab^2}$ & $-\dfrac{416c^2\pi^2\al^2}{15(\al+\beta)^6}$\\ 
          45 & $-\dfrac{2c^2\al\ra}{\rab^2}$ & $-\dfrac{128c^2\pi^2\al}{3(\al+\beta)^5}$ \\ 
          46 & $-c^2\left(\dfrac{\al^2+\beta^2}{2}\right)\beta\rb$ & $-\dfrac{96c^2\pi^2\beta(\al^2+\beta^2)}{(\al+\beta)^7}$ \\ 
          47 & $c^2\left(\dfrac{\beta-Z}{\ra}\right)\beta\rb$ & $\dfrac{96c^2\pi^2\beta(\beta-Z)}{(\al+\beta)^6}$\\ 
          48 & $c^2(\al-Z)\beta$ & $\dfrac{64c^2\pi^2\beta(\al - Z)}{(\al+\beta)^6}$\\ 
          49 & $c^2\dfrac{\beta\rb}{\rab}$ & $\dfrac{50c^2\pi^2\beta}{(\al+\beta)^6}$\\ 
          50 & $c^2\dfrac{\beta^2\ra\rb}{\rab^2}$ & $\dfrac{736c^2\pi^2\beta^2}{15(\al+\beta)^6}$ \\ 
          51 & $c^2\dfrac{\al\beta\rb^2}{\rab^2}$ & $\dfrac{288c^2\pi^2\al\beta}{5(\al+\beta)^6}$\\ 
          52 & $-c^2\dfrac{\beta^2\rb^2\cst}{\rab^2}$ & $-\dfrac{416c^2\pi^2\beta^2}{15(\al+\beta)^6}$\\ 
          53 & $-c^2\dfrac{\al\beta\ra\rb\cst}{\rab^2}$ & $-\dfrac{128c^2\pi^2\al\beta}{5(\al+\beta)^6}$\\ 
          54 & $-\dfrac{2c^2\beta\rb}{\rab^2}$ & $-\dfrac{128c^2\pi^2\beta}{3(\al+\beta)^5}$ \\ 
          55 & $c^2\left(\dfrac{\al^2+\beta^2}{2}\right)\al\rb\cst$ & $0$ \\ 
          56 & $-c^2\left(\dfrac{\beta-Z}{\ra}\right)\al\rb\cst$ & $0$\\ 
          57 & $-c^2(\al-Z)\al\cst$ & $0$\\ 
          58 & $-c^2\dfrac{\al\rb\cst}{\rab}$ & $-\dfrac{10c^2\pi^2\al}{(\al+\beta)^6}$\\ 
          59 & $-c^2\dfrac{\al\beta\ra\rb\cst}{\rab^2}$ & $-\dfrac{128c^2\pi^2\al\beta}{5(\al+\beta)^6}$ \\ 
          60 & $-c^2\dfrac{\al^2\rb^2\cst}{\rab^2}$ & $-\dfrac{416c^2\pi^2\al^2}{15(\al+\beta)^6}$\\ 
          61 & $2c^2\dfrac{\al\rb\cst}{\rab^2}$ & $\dfrac{64c^2\pi^2\al}{3(\al+\beta)^5}$\\ 
          62 & $c^2\left(\dfrac{\al^2+\beta^2}{2}\right)\beta\ra\cst$ & $0$ \\ 
          63 & $-c^2(\beta-Z)\beta\cst$ & $0$\\ 
          64 & $-c^2\left(\dfrac{\al-Z}{\rb}\right)\beta\ra\cst$ & $0$\\ 
          65 & $-c^2\dfrac{\beta\ra\cst}{\rab}$ & $-\dfrac{10c^2\pi^2\beta}{(\al+\beta)^6}$\\ 
          66 & $-c^2\dfrac{\beta^2\ra^2\cst}{\rab^2}$ & $-\dfrac{416c^2\pi^2\beta^2}{15(\al+\beta)^6}$ \\ 
          67 & $-c^2\dfrac{\al\beta\ra\rb\cst}{\rab^2}$ & $-\dfrac{128c^2\pi^2\al\beta}{5(\al+\beta)^6}$\\ 
          68 & $2c^2\dfrac{\beta\ra\cst}{\rab^2}$ & $\dfrac{64c^2\pi^2\beta}{3(\al+\beta)^5}$\\ 
          69 & $c^2(\al^2+\beta^2)$ & $\dfrac{64c^2\pi^2(\al^2+\beta^2)}{(\al+\beta)^6}$ \\ 
          70 & $-2c^2\left(\dfrac{\beta-Z)}{\ra}\right)$ & $\dfrac{64c^2\pi^2(\beta-Z)}{(\al+\beta)^5}$\\ 
          71 & $-2c^2\left(\dfrac{\al-Z}{\rb}\right)$ & $\dfrac{64c^2\pi^2(\al - Z)}{(\al+\beta)^5}$\\ 
          72 & $\dfrac{-2c^2}{\rab}$ & $-\dfrac{40c^2\pi^2}{(\al+\beta)^5}$\\ 
          73 & $-2c^2\dfrac{\beta\ra}{\rab^2}$ & $-\dfrac{128c^2\pi^2\beta}{3(\al+\beta)^5}$ \\ 
          74 & $-2c^2\dfrac{\al\rb}{\rab^2}$ & $-\dfrac{128c^2\pi^2\al}{3(\al+\beta)^5}$\\ 
          75 & $2c^2\dfrac{\beta\rb\cst}{\rab^2}$ & $\dfrac{64c^2\pi^2\beta}{3(\al+\beta)^5}$\\ 
          76 & $2c^2\dfrac{\al\ra\cst}{\rab^2}$ & $\dfrac{64c^2\pi^2\al}{3(\al+\beta)^5}$\\ 
          77 & $\dfrac{4c^2}{\rab^2}$ & $-\dfrac{128c^2\pi^2}{3(\al+\beta)^4}$ \\ 
  \end{longtable}

This table omits the $hs_{b2a2,i}$ terms containing $\tfrac{\cst^2}{\rab^2}$, such as $\left(\tfrac{c^2\al\beta\rb^2\cst^2}{\rab^2}\right)$, $\left(\tfrac{c^2\al^2\ra\rb\cst^2}{\rab^2}\right)$, $\left(\tfrac{c^2\al\beta\ra^2\cst^2}{\rab}\right)^2$ and $\left(\tfrac{c^2\beta^2\ra\rb\cst^2}{\rab^2}\right)$, because they are difficult to integrate individually. They will be grouped with similar terms from $hs_{a2a2,i}$ and the following integrals, and then integrated using suitable substitutions.

 \subsubsection*{Calculating $\bkt{\mc{H}^2}{ab}$:}
    $\bkt{\mc{H}^2}{ab} = (\bra{\psi_{a}}\mc{H}).(\mc{H}\ket{\psi_{b}}) $ is 
    $\bkt{\mc{H}^2}{ba} = (\bra{\psi_{b}}\mc{H}).(\mc{H}\ket{\psi_{a}})$ with $\al$ and $\beta$ interchanged:
    \[ \bkt{\mc{H}^2}{ab}(\al,\beta,c) = \bkt{\mc{H}^2}{ba}(\beta,\al,c) \]

\subsubsection*{Calculating $\bkt{\mc{H}^2}{bb}$:}
    $\bkt{\mc{H}^2}{bb} = (\bra{\psi_{b}}\mc{H}).(\mc{H}\ket{\psi_{b}}) $ is 
    $\bkt{\mc{H}^2}{aa} = (\bra{\psi_{a}}\mc{H}).(\mc{H}\ket{\psi_{a}})$ with $\al$ and $\beta$ interchanged:
    \[ \bkt{\mc{H}^2}{bb}(\al,\beta,c) = \bkt{\mc{H}^2}{aa}(\beta,\al,c) \]

\subsubsection*{Integration of terms  containing $\tfrac{\cst^2}{\rab^2} $:}
All the terms containing $\tfrac{\cst^2}{\rab^2}$ in $\bkt{\mc{H}^2}{aa}$ can be summed and written as:
\[ {hs}_{aa\star} = c^2 (\al^2\rb^2+\beta^2\ra^2+2\al\beta\ra\rb)\dfrac{\cst^2}{\rab^2}  = c^2(\al\rb+\beta\ra^2)\dfrac{\cst^2}{\rab^2} \] 

\begin{equation*}
 Hs_{aa\star} = \int d\ra^3d\rb^3 e^{-2(\al\ra+\beta\rb)}\left(c^2(\al\rb+\beta\ra)^2\dfrac{\cst^2}{\rab^2}\right)   
\end{equation*}
Integration over the $\phi_1$ and $\phi_2$ coordinates from 0 to $2\pi$ gives $2\pi$ each, since the integrand is independent of $\phi$'s. $\theta$ is the relative angle between $\theta_1$ and $\theta_2$, so integrating over one of the $\theta$'s gives a factor of 2, and then we have to carry out the integration with respect to the other $\theta$.

This simplification was used in all the terms containing $\theta$ ($\cst$ and $\rab$) that appeared in the preceding tables, and the integrals were then easy to compute. It is only in the case ($\tfrac{\cst^2}{\rab^2} $ terms) that grouping of terms and further simplification became necessary, since individual terms proved to be difficult to integrate.
\begin{equation*}
    \int_0^\pi d\theta  \dfrac{\snt\cst^2}{\rab^2} = \int_0^\pi d\theta \dfrac{\cst^2\snt}{\ra^2+\rb^2-2\ra\rb\cst}
\end{equation*}
Substituting $t =\ra^2+\rb^2-2\ra\rb\cst$, we get $dt = 2\ra\rb\snt d\theta$ and $\cst = \frac{\ra^2+\rb^2-t}{-2\ra\rb}$. The integral then becomes:
\begin{align*}
    \int_0^\pi d\theta  \dfrac{\snt\cst^2}{\rab^2} & = \int_{(\ra-\rb)^2}^{(\ra+\rb)^2} \dfrac{(\ra^2+\rb^2-t)^2dt}{4\ra^2\rb^2(2\ra\rb)t} \\
    & = \dfrac{1}{8\ra^3\rb^3}\int_{(\ra-\rb)^2}^{(\ra+\rb)^2}\left( \dfrac{(\ra^2+\rb^2)^2}{t} - 2(\ra^2+\rb^2) + t\right)dt \\
    & = \dfrac{1}{8\ra^3\rb^3}\left((\ra^2+\rb^2)^2 log\left(\dfrac{(\ra+\rb)^2}{(\ra-\rb)^2}\right) -4\ra\rb(\ra^2+\rb^2)\right)
\end{align*}

The complete integral is:
\begin{equation*}
    Hs_{aa\star} = c^2\pi^2\int_0^{\infty} \int_0^{\infty} d\ra d\rb \dfrac{e^{-2(\al\ra+\beta\rb)}}{\ra\rb}(\al\rb+\beta\ra)^2(\ra^2+\rb^2)\left(log\left(\dfrac{(\ra+\rb)^2}{(\ra-\rb)^2}\right) -4\ra\rb\right)
\end{equation*}
Let us substitute
\begin{align*}
    \al = \gamma+\delta, \quad \beta = \gamma-\delta ~,
\end{align*}
and
\begin{align*}
    u = \ra+\rb, &\quad 0<u<\infty ~, \\
    v = \rb-\ra, &\quad -u<v<u ~.
\end{align*}
Then we get:
\begin{align*}
    \al\ra+\beta\rb & = \gamma u+\delta v ~, \\
     \al\rb+\beta\ra & = \gamma u-\delta v ~, \\
     \ra^2 + \rb^2 & = \dfrac{u^2+v^2}{2} ~, \\
     \ra\rb & = \dfrac{u^2-v^2}{4} ~,
\end{align*}
and
\begin{align*}
    d\ra d\rb & = |J|dudv = \begin{vmatrix}
        \pdv{\ra}{u} & \pdv{\ra}{v} \\
        \pdv{\rb}{u} & \pdv{\rb}{v}
    \end{vmatrix}du dv = \begin{vmatrix}
        \frac{1}{2} & -\frac{1}{2} \\
        \frac{1}{2} & \frac{1}{2}
    \end{vmatrix}du dv = du dv ~.
\end{align*}
The transformed integral becomes:
\begin{align*}
   Hs_{aa\star} &= c^2\pi^2\int_0^{\infty}du  \int_{-u}^{u} dv \left\{\dfrac{4e^{-2(\gamma u-\delta v)}}{(u^2-v^2)}(\gamma u+\delta v)^2\left(\dfrac{u^2+v^2}{2}\right)\left[log\left(\dfrac{u^2}{|v|^2}\right)\left(\dfrac{u^2+v^2}{2}\right)-(u^2-v^2)\right] \right\} \\
   &= 2c^2\pi^2\int_0^{\infty}du  \int_{-u}^{u} dv \left\{ e^{-2(\gamma u-\delta v)}(\gamma u+\delta v)^2\left(\dfrac{u^2+v^2}{u^2-v^2}\right)\left[log\left(\dfrac{u}{|v|}\right)(u^2+v^2)-(u^2-v^2)\right] \right\} ~.
\end{align*}

We first carry out the integral over the variable $v$, by expanding $e^{2\delta v}$ as a power series, and separating the $v>0$ and $v<0$ contributions. The individual terms in the expansion are:
\begin{align*}
  hs_{aa\star,vn}  = & \int_{-u}^{u} dv \left\{ \dfrac{(2\delta v)^n}{n!}(\gamma u+\delta v)^2\left(\dfrac{u^2+v^2}{u^2-v^2}\right)\left[log\left(\dfrac{u}{|v|}\right)(u^2+v^2)-(u^2-v^2)\right] \right\} ~, \\
  = & \int_{-u}^{0} dv \left\{ \dfrac{(2\delta v)^n}{n!}(\gamma u+\delta v)^2\left(\dfrac{u^2+v^2}{u^2-v^2}\right)\left[log\left(\dfrac{u}{-v}\right)(u^2+v^2)-(u^2-v^2)\right] \right\} \\
  + & \int_{0}^{u} dv \left\{ \dfrac{(2\delta v)^n}{n!}(\gamma u+\delta v)^2\left(\dfrac{u^2+v^2}{u^2-v^2}\right)\left[log\left(\dfrac{u}{v}\right)(u^2+v^2)-(u^2-v^2)\right] \right\} ~.
\end{align*}
Evaluating this integral, followed by integration over the variable $u$, we obtain
\begin{align*}
hs_{aa\star,n} & = 2 c^2\pi^2 \int_0^{\infty}du \ e^{-2\gamma u}hs_{aa\star,vn} ~, \\
hs_{aa\star,n} & = \dfrac{1}{32n!}c^2\pi^2\gamma^{-(n+6)}\Gamma(n+6)\left[ -(-\delta)^n\left( \left( \dfrac{3}{(n+1)^2}+\dfrac{1}{(n+3)^2}\right)\gamma^2 -\dfrac{8\gamma\delta(13+7n+n^2)}{(8+6n+n^2)^2} + \left( \dfrac{4}{(n+1)^2} \right. \right. \right. \\ 
& \left. \left. \left. + \dfrac{3}{(n+3)^2} + \dfrac{1}{(n+5)^2}  \right)\delta^2 + 2\gamma\delta \ \Upsilon\left(1+\dfrac{n}{2}\right) +(\gamma^2+\delta^2) \ \Upsilon\left(\dfrac{1+n}{2}\right) \right) + \right.  (\delta)^n\left( \left(- \dfrac{3}{(n+1)^2}-\dfrac{1}{(3+n)^2}\right)\gamma^2 \right. \\
&\left. \left. -\dfrac{8\gamma\delta(13+7n+n^2)}{(8+6n+n^2)^2} + \left( -\dfrac{4}{(n+1)^2}-\dfrac{3}{(n+3)^2} - \dfrac{1}{(n+5)^2}  \right)\delta^2 + 2\gamma\delta \ \Upsilon\left(1+\dfrac{n}{2} \right)+(\gamma^2+\delta^2) \  \Upsilon\left(\dfrac{1+n}{2}\right) \right) \right] ,
\end{align*}
where $\Gamma(n) = \int_0^{\infty} t^{n-1}e^{-t}dt$ is the Euler Gamma function, and $\Upsilon(n) = \frac{d}{dn} \left(\frac{\Gamma'(n)}{\Gamma(n)}\right)$ in the trigamma function.

Similarly, the terms containing $\tfrac{\cst^2}{\rab^2} $ in $\bkt{\mc{H}^2}{bb}$ can be summed and written as: 
\[ {hs}_{bb\star} = c^2 (\al^2\ra^2+\beta^2\rb^2+2\al\beta\ra\rb)\dfrac{\cst^2}{\rab^2}  = c^2(\al\ra+\beta\rb^2)\dfrac{\cst^2}{\rab^2} ~. \]
The same change of variables, power series expansion and integration give:
\begin{align*}
hs_{bb\star,n} & = \dfrac{1}{32n!}c^2\pi^2\gamma^{-(n+6)}\Gamma(n+6)\left[ (-\delta)^n\left( \left(- \dfrac{3}{(n+1)^2}-\dfrac{1}{(n+3)^2}\right)\gamma^2 +\dfrac{8\gamma\delta(13+7n+n^2)}{(8+6n+n^2)^2} + \left(- \dfrac{4}{(n+1)^2} \right. \right. \right. \\ 
& \left. \left. \left. - \dfrac{3}{(n+3)^2} -\dfrac{1}{(n+5)^2}  \right)\delta^2 + 2\gamma\delta \ \Upsilon\left(1+\dfrac{n}{2}\right) +(\gamma^2+\delta^2) \ \Upsilon\left(\dfrac{1+n}{2}\right) \right) + \right.  (\delta)^n\left( \left(- \dfrac{3}{(n+1)^2}-\dfrac{1}{(3+n)^2}\right)\gamma^2 \right. \\
& \left. \left. -\dfrac{8\gamma\delta(13+7n+n^2)}{(8+6n+n^2)^2} + \left( -\dfrac{4}{(n+1)^2}-\dfrac{3}{(n+3)^2} - \dfrac{1}{(n+5)^2}  \right)\delta^2 + 2\gamma\delta \ \Upsilon\left(1+\dfrac{n}{2} \right)+(\gamma^2+\delta^2) \  \Upsilon\left(\dfrac{1+n}{2}\right) \right) \right] ~.
\end{align*}

We now add $hs_{aa\star,n}$ and $hs_{bb\star,n}$, and sum over all $n$ the terms that do not contain the trigamma function, to obtain the first part of the total integral $Hs_{aa\star,1}$:
\begin{align*}
    Hs_{aabb\star,1} & = \dfrac{c^2\pi^2}{\gamma^5(\gamma^2-\delta^2)^4\delta^3}\left( 2\gamma\delta(\gamma^{10}-24\gamma^8\delta^2-32\gamma^6\delta^4+8\gamma^4\delta^6-19\gamma^2\delta^8+6\delta^{10})  \right. \\
    & \left. - (\gamma^2-\delta^2)^4(\gamma^4+3\gamma^2\delta^2+6\delta^4)log\left(\frac{\gamma+\delta}{\gamma}\right)+(\gamma^2-\delta^2)(\gamma^4+3\gamma^2\delta^2+6\delta^4)log\left( 1-\frac{\delta}{\gamma}\right)\right) ~.
\end{align*}

We next consider the terms in $hs_{aa\star,n}$ and $hs_{bb\star,n}$ that contain the trigamma function. They cannot be summed for all $n$ directly. We first sum over all odd $n = 2l+1$ with $l = 0,1,2,...\infty$, to obtain
\begin{align*}
    Hs_{aabb\star,2} & = \dfrac{c^2\pi^2\delta}{\gamma^3(\gamma^2-\delta^2)^6}\biggl[ 2 \gamma \delta ((-163 + 45 \pi^2) \gamma^8 + 
   6 (-11 + 25 \pi^2) \gamma^6 \delta^2 + 
   15 (16 + 3 \pi^2) \gamma^4 \delta^4 - 
   14 \gamma^2 \delta^6 + 3 \delta^8) \\
 & + \delta (137 \gamma^{10} + 555 \gamma^8 \delta^2 - 
   570 \gamma^6 \delta^4 - 110 \gamma^4 \delta^6 - 
   15 \gamma^2 \delta^8 + 3 \delta^{10}) \left(log\left(\dfrac{\gamma - \delta}{\gamma + \delta}\right)\right) \\
 & + 60 \gamma^4 (\gamma^6 + 
   15 \gamma^4 \delta^2 + 15 \gamma^2 \delta^4 + \delta^6)\left(\dilog\left(-\dfrac{\delta}{\gamma}\right) - \dilog\left(\dfrac{\delta}{\gamma}\right) \right) \biggr] ~,
\end{align*}
where $\dilog(x) = \sum_{k=0}^{\infty} \frac{x^k}{k^2}$ is the dilogarithmic function

We then sum over all even $n = 2l$ with $l = 0,1,2,...\infty$, to obtain 
\begin{align*}
    Hs_{aabb\star,3} & = \dfrac{c^2 \pi^2 (\gamma^2 + \delta^2)}{2 \gamma^5 (\gamma^2 - \delta^2)^6 }\biggl[ 3 \gamma (5 \pi^2 \gamma^{10} + 
   3 (-84 + 25 \pi^2) \gamma^8 \delta^2 + 
   3 (92 + 25 \pi^2) \gamma^6 \delta^4 + 
   5 (-12 + \pi^2) \gamma^4 \delta^6 + 
   44 \gamma^2 \delta^8 - 8 \delta^{10}) \\
   &  -2 \delta (-261 \gamma^{10} + 80 \gamma^8 \delta^2 + 
   120 \gamma^6 \delta^4 + 90 \gamma^4 \delta^6 - 
   35 \gamma^2 \delta^8 + 
   6 \delta^{10}) \left(log\left(\dfrac{\gamma - \delta}{\gamma + \delta}\right)\right) \\
   & +120 \gamma^6 \delta (3 \gamma^4 + 10 \gamma^2 \delta^2 + 
   3 \delta^4) \left(\dilog\left( -\dfrac{\delta}{\gamma}\right) - 
   \dilog\left( \dfrac{\delta}{\gamma}\right)\right)\biggr] ~.
\end{align*}

The complete integral of the terms containing $\tfrac{\cst^2}{\rab^2}$ in $\bkt{\mc{H}^2}{aa}$ and $\bkt{\mc{H}^2}{bb}$ is then given by:
\begin{equation*}
    Hs_{aa\star}+ Hs_{bb\star} = Hs_{aabb\star,1} + Hs_{aabb\star,2} + Hs_{aabb\star,3} ~.
\end{equation*}

The terms containing $\tfrac{\cst^2}{\rab^2} $ in $\bkt{\mc{H}^2}{ba}$ can be summed and written as: 
\[ {hs}_{ba\star} = c^2(\al\beta(\ra^2+\rb^2)+(\al^2+\beta^2)\ra\rb)\dfrac{\cst^2}{\rab^2} ~. \]
After the change of variables and integration, we get:
\begin{align*}
   Hs_{ba\star} & = 2c^2\pi^2\int_0^{\infty}du  \int_{-u}^{u} dv \left\{ e^{-2\gamma^2 u^2}(\gamma^2 u^2-\delta^2 v^2)\left(\dfrac{u^2+v^2}{u^2-v^2}\right)\left[(u^2+v^2)log\left(\dfrac{u}{|v|}\right)(u^2+v^2)-(u^2-v^2)\right] \right\} ~, \\
   & = \dfrac{c^2\pi^2(25\gamma^2(9\pi^2-80)+3\delta^2(736 - 75\pi^2))}{60\gamma^6} ~, \\
   & = \dfrac{16c^2\pi^2(52\al^2+(225\pi^2-2104)\al\beta+52\beta^2)}{15(\al+\beta)^6} ~.
\end{align*}
The terms containing $\tfrac{\cst^2}{\rab^2}$ in $\bkt{\mc{H}^2}{ab}$ give the same result, $Hs_{ab\star} = Hs_{ba\star}$.

We obtain $\bkt{\mc{H}^2}{}$ by combining all the integrals that have been computed.

The variance is then given by: 
$ (\Delta \mc{H})^2 = \braket{(\mc{H}-\braket{\mc{H}})^2} = \braket{\mc{H}^2}-\braket{\mc{H}}^2 $.

Finally, the lower bound on the ground state energy, $\widecheck{E}_0$, is computed by maximising (for each Z):
\begin{equation*}
\til{E} = \braket{\mc{H}} - \Delta \mc{H} ~.
\end{equation*}

%% file: appendix2.tex
\newpage 
\section{Hydrogen Molecular Ion}

Here we list the errors incurred in the energy, when truncating the exact solution for the hydrogen molecular ion by series approximations. The accurate ground state electronic energy eigenvalue for $R=2$, obtained by solving the differential equations is $E=-1.10263462$.

\subsection{Power Series Approximation}

\begin{table}[htbp]
    \centering
    \caption{Power series approximation: The expectation value of the electronic energy $E_{\mr{approx}}$ (rounded to 6 decimal places), and its error (rounded to 6 decimal places, and to 4 significant digits when smaller)}
    \begin{tabular}{|M{2.5em}|M{3.5em}|M{5.5em}|M{5.5em}|M{5.5em}|M{5.5em}|M{5.5em}|M{5.5em}|M{5.5em}|M{5.5em}|}
    \hline
     $N_{\mr{rad}} \ \downarrow$ & $N_{\mr{ang}} \ \rightarrow $ &  0 & 2 & 4 & 6 & 8 & 10 \\
     \hline
      \multirow{2}{3em}{0} & $E_{\mr{approx}}$ & -1.072822 & -1.096952 & -1.097024 & -1.097024 & -1.097024  & -1.097024\\ 
      \cline{2-8}
       & Error & 0.029813 & 0.005682 & 0.005610 & 0.005610 & 0.005610 & 0.005610 \\
     \hline
     \multirow{2}{3em}{1} & $E_{\mr{approx}}$ & -1.078376 & -1.101705 & -1.101774 & -1.101774 &  -1.101774 &  -1.101774 \\ 
     \cline{2-8}
      & Error  & 0.024258 & 0.000929 & 0.000860 & 0.000860 &  0.000860 &  0.000860 \\
     \hline
     \multirow{2}{3em}{2} & $E_{\mr{approx}}$ & -1.079576 & -1.102485 & -1.102553 & -1.102553 & -1.102553 & -1.102553 \\ 
     \cline{2-8}
      & Error  & 0.023059 & 0.000149 & 0.000081 & 0.000081 & 0.000081 & 0.000081 \\
     \hline
     \multirow{2}{3em}{3} & $E_{\mr{approx}}$ & -1.079724 & -1.102546 & -1.102613 & -1.102613 & -1.102613 & -1.102613   \\ 
     \cline{2-8}
      & Error  & 0.022910 & 0.000088 & 0.000021 & 0.000021 & 0.000021 & 0.000021  \\
     \hline
     \multirow{2}{3em}{4} & $E_{\mr{approx}}$ & -1.079819 & -1.102567 & -1.102634 & -1.102634  & -1.102634  & -1.102634   \\ 
     \cline{2-8}
      & Error & 0.022816 & 0.000067 & $1.397 \times 10^{-7}$ & $1.144 \times 10^{-7}$ & $1.144 \times 10^{-7}$ & $1.144 \times 10^{-7}$  \\
     \hline
     \multirow{2}{3em}{5} & $E_{\mr{approx}}$ & -1.079805 & -1.102566 & -1.102633 & -1.102633 & -1.102633 & -1.102633  \\ 
     \cline{2-8}
      & Error  & 0.022829 & 0.000069 & $1.414 \times 10^{-6}$ & $1.388 \times 10^{-6}$ & $1.388 \times 10^{-6}$ & $1.388 \times 10^{-6}$  \\
     \hline
     \multirow{2}{3em}{6} & $E_{\mr{approx}}$ & -1.079817 & -1.102567 & -1.102634 & -1.102634 & -1.102634 & -1.102634\\ 
     \cline{2-8}
      & Error  & 0.022817 & 0.000067 & $1.701 \times 10^{-7}$ & $1.448 \times 10^{-7}$ & $1.448 \times 10^{-7}$ & $1.448 \times 10^{-7}$ \\
     \hline
     \multirow{2}{3em}{7} & $E_{\mr{approx}}$ & -1.079814 & -1.102567 & -1.102634 & -1.102634 & -1.102634 & -1.102634  \\ 
     \cline{2-8}
      & Error  & 0.022820 & 0.000067 & $1.335 \times 10^{-7}$ & $1.082 \times 10^{-7}$ & $1.082 \times 10^{-7}$ & $1.082 \times 10^{-7}$\\
     \hline
     \multirow{2}{3em}{8} & $E_{\mr{approx}}$ & -1.079815 & -1.102567 & -1.102634 & -1.102634 & -1.102634 & -1.102634  \\ 
     \cline{2-8}
      & Error & 0.022819 & 0.000067 & $4.037 \times 10^{-8}$ & $1.505 \times 10^{-8}$ &  $1.505 \times 10^{-8}$ & $1.505 \times 10^{-8}$ \\
     \hline   
    \end{tabular}
    \label{tab:approx_pow}
\end{table}

\newpage
\subsection{Chebyshev Polynomial Approximation}

\begin{table}[htbp]
    \centering
    \caption{Chebyshev polynomial approximation: The expectation value of the electronic energy $E_{\mr{approx}}$ (rounded to 6 decimal places), and its error (rounded to 6 decimal places, and to 4 significant digits when smaller)}
\begin{tabular}{|M{2.5em}|M{3.5em}|M{5.5em}|M{5.5em}|M{5.5em}|M{5.5em}|M{5.5em}|M{5.5em}|M{5.5em}|}
    \hline
     $N_{\mr{rad}} \ \downarrow$ & $N_{\mr{ang}} \ \rightarrow $ &  0 & 2 & 4 & 6 & 8 & 10 \\
    \hline
     \multirow{2}{3em}{0} & $E_{\mr{approx}}$ & -1.072822 & -1.096952 & -1.097024 & -1.097024 & -1.097024 & -1.097024 \\ 
     \cline{2-8}
      & Error & 0.029813 & 0.005682 & 0.005610 & 0.005610 & -0.005610 & -0.005610 \\
     \hline
     \multirow{2}{3em}{1} & $E_{\mr{approx}}$ & -1.078376 & -1.101706 &-1.101774 & -1.101775 &  -1.101775 & -1.101775 \\ 
     \cline{2-8}
      & Error  & 0.024258 & -0.000929 & -0.000860 &  -0.000860 & -0.000860 & -0.000860\\
     \hline
     \multirow{2}{3em}{2} & $E_{\mr{approx}}$ & -1.079576 & -1.102486 & -1.102553 & -1.102553 & -1.102553 & -1.102553 \\ 
     \cline{2-8}
      & Error  & 0.023058 & 0.000149 & 0.000081 & 0.000081 & 0.000081 & 0.000081 \\
     \hline
     \multirow{2}{3em}{3} & $E_{\mr{approx}}$ & -1.079724 &-1.102546 & -1.102613 & -1.102613 & -1.102613 & -1.102613 \\ 
     \cline{2-8}
      & Error  & 0.022910 & 0.000088 & 0.000021 & 0.000021 & 0.000021 & 0.000021 \\
     \hline
     \multirow{2}{3em}{4} & $E_{\mr{approx}}$ & -1.079819 & -1.102567 & -1.102634 &-1.102634 & -1.102634 & -1.102634 \\ 
     \cline{2-8}
      & Error  & 0.022816 & 0.000067 & 1.398 $\times 10^{-7}$ & $1.145 \times 10^{-7}$ & $1.145 \times 10^{-7}$ & $1.145 \times 10^{-7}$ \\
     \hline
     \multirow{2}{3em}{5} & $E_{\mr{approx}}$ & -1.079805 & -1.102566  & -1.102633 & -1.102633 & -1.102633 & -1.102633 \\ 
     \cline{2-8}
      & Error & 0.022829 & 0.000069 & 1.410 $\times 10^{-6}$ & 1.385 $\times 10^{-6}$ & 1.385 $\times 10^{-6}$ & 1.385 $\times 10^{-6}$ \\
     \hline
     \multirow{2}{3em}{6} & $E_{\mr{approx}}$ &-1.079817 & -1.102567 & -1.102634 & -1.102634 & -1.102634 & -1.102634 \\ 
     \cline{2-8}
      & Error  & 0.022817 & 0.000067 & $1.7 \times 10^{-7}$ & $1.447 \times 10^{-7}$ & $1.447 \times 10^{-7}$ & $1.447 \times 10^{-7}$  \\
     \hline
     \multirow{2}{3em}{7} & $E_{\mr{approx}}$ & -1.079814 & -1.102567 & -1.102634 & -1.1026341 & -1.1026341 & -1.1026341\\ 
     \cline{2-8}
      & Error & 0.022820 & 0.000067 & $1.329 \times 10^{-7}$ & $1.076 \times 10^{-7}$ & $1.076 \times 10^{-7}$ & $1.078 \times 10^{-7}$ \\
      \hline
     \multirow{2}{3em}{8} & $E_{\mr{approx}}$ & -1.079815 & -1.102567 & -1.102634 & -1.1026341 & -1.1026341 & -1.1026341\\ 
     \cline{2-8}
      & Error & 0.022819 & 0.000067 & $4.028 \times 10^{-8}$ & $1.496 \times 10^{-8}$ & $1.495 \times 10^{-8}$ & $1.495 \times 10^{-8}$ \\
     \hline   
    \end{tabular}
    \label{tab:approx_cheb}
\end{table}